\documentclass[onecolumn,showpacs,showkeys,letterpaper]{revtex4}

\usepackage{amsmath}            
\usepackage{amstext}            
\usepackage{amssymb}            
\usepackage{amsfonts}           
\usepackage{graphicx}           
\usepackage{pict2e}             
\usepackage{color}              
\usepackage{hyperref}           

\newcommand{\C}[2]{{#1 \choose #2}}
\renewcommand{\O}[1]{\mathcal{O}\left(#1\right)}
\newcommand{\arcsinh}{\operatorname{arcsinh}}

\newcommand{\diameter}{2}
\newcommand{\node}[1]{\put #1{\circle*{\diameter}}}
\newcommand{\rnode}[1]{\put #1{\circle{\diameter}}}
\newcommand{\cnode}[2]{\put #1{\circle{4}}\put #1{\!\raisebox{-1\unitlength}{#2}}}
\newcommand{\smallcircle}[1]{\put #1{\circle{4}}}
\newcommand{\mediumcircle}[1]{\put #1{\circle{7}}}
\newcommand{\largecircle}[1]{\put #1{\circle{10}}}
\newcommand{\edge}[3]{\put #1{\line #2{#3}}}
\newcommand{\arrow}[4]{\put #1{\line #2{#3}}\put #1{\vector #2{#4}}}
\newcommand{\cadre}[2]{}

\newcommand{\elementaryRoot}{\begin{picture}(2,2)\cadre{2}{2}\rnode{(1,1)}\end{picture}}
\newcommand{\elementaryNode}{\begin{picture}(2,2)\cadre{2}{2}\node{(1,1)}\end{picture}}
\newcommand{\compositeNode}[1]{\begin{picture}(4,3)\cadre{4}{3}\put(2,1){\circle{4}}\put(2,1){\!\raisebox{-1\unitlength}{#1}}\end{picture}}

\newcommand{\Ca}{\begin{picture}(4,3)\cadre{4}{3}\put(2,1){\circle{4}}\node{(2,1)}\end{picture}}
\newcommand{\CCCa}{\begin{picture}(7,5)\cadre{7}{5}\put(3.5,1){\circle{7}}\node{(1.8,1)}\node{(4.5,2.5)}\node{(4.5,-0.5)}\end{picture}}
\newcommand{\CCCCCa}{\begin{picture}(10,6)\cadre{10}{6}\put(5,1){\circle{10}}\node{(2,1)}\node{(4.3,3.5)}\node{(4.3,-1.5)}\node{(7.5,2.7)}\node{(7.5,-0.7)}\end{picture}}
\newcommand{\CCCCCCCa}{\begin{picture}(13,7.5)\cadre{13}{7.5}\put(6.5,1){\circle{13}}\node{(2.5,1)}\node{(4.3,4)}\node{(4.3,-2)}\node{(7.5,5)}\node{(7.5,-3)}\node{(10.2,2.7)}\node{(10.2,-0.7)}\end{picture}}

\newcommand{\rootedGa}{\begin{picture}(4,3)\cadre{4}{3}\put(2,1){\circle{4}}\rnode{(2,1)}\end{picture}}
\newcommand{\rootedGGa}{\begin{picture}(4,7)\cadre{4}{7}\put(2,5){\circle{4}}\rnode{(2,5)}\edge{(2,3.1)}{(0,-1)}{4.1}\put(2,-3){\circle{4}}\node{(2,-3)}\end{picture}}
\newcommand{\rootedGGGa}{\begin{picture}(4,11)\cadre{4}{11}\put(2,9){\circle{4}}\rnode{(2,9)}\edge{(2,7.1)}{(0,-1)}{4.1}\put(2,1){\circle{4}}\node{(2,1)}\edge{(2,-0.9)}{(0,-1)}{4.1}\put(2,-7){\circle{4}}\node{(2,-7)}\end{picture}}
\newcommand{\rootedGGGb}{\begin{picture}(12,7)\cadre{12}{7}\put(6,5){\circle{4}}\rnode{(6,5)}\edge{(5.1,3.2)}{(-1,-2)}{2.2}\edge{(6.9,3.2)}{(1,-2)}{2.2}\put(2,-3){\circle{4}}\node{(2,-3)}\put(10,-3){\circle{4}}\node{(10,-3)}\end{picture}}
\newcommand{\rootedGGGc}{\begin{picture}(7,5)\cadre{7}{5}\put(3.5,1){\circle{7}}\rnode{(1.8,1)}\node{(4.5,2.5)}\node{(4.5,-0.5)}\end{picture}}
\newcommand{\rootedGGGGa}{\begin{picture}(4,15)\cadre{4}{15}\put(2,13){\circle{4}}\rnode{(2,13)}\edge{(2,11.1)}{(0,-1)}{4.1}\put(2,5){\circle{4}}\node{(2,5)}\edge{(2,3.1)}{(0,-1)}{4.1}\put(2,-3){\circle{4}}\node{(2,-3)}\edge{(2,-4.9)}{(0,-1)}{4.1}\put(2,-11){\circle{4}}\node{(2,-11)}\end{picture}}
\newcommand{\rootedGGGGb}{\begin{picture}(12,11)\cadre{12}{11}\put(6,9){\circle{4}}\rnode{(6,9)}\edge{(5.1,7.2)}{(-1,-2)}{2.2}\edge{(6.9,7.2)}{(1,-2)}{2.2}\put(2,1){\circle{4}}\node{(2,1)}\put(10,1){\circle{4}}\node{(10,1)}\put(2,1){\circle{4}}\node{(2,1)}\edge{(2,-0.9)}{(0,-1)}{4.1}\put(2,-7){\circle{4}}\node{(2,-7)}\end{picture}}
\newcommand{\rootedGGGGc}{\begin{picture}(12,11)\cadre{12}{11}\put(6,9){\circle{4}}\rnode{(6,9)}\edge{(6,7.1)}{(0,-1)}{4.1}\put(6,1){\circle{4}}\node{(6,1)}\edge{(5.1,-0.8)}{(-1,-2)}{2.2}\edge{(6.9,-0.8)}{(1,-2)}{2.2}\put(2,-7){\circle{4}}\node{(2,-7)}\put(10,-7){\circle{4}}\node{(10,-7)}\end{picture}}
\newcommand{\rootedGGGGd}{\begin{picture}(20,7)\cadre{20}{7}\put(10,5){\circle{4}}\rnode{(10,5)}\edge{(8.6,3.6)}{(-1,-1)}{5.2}\edge{(10,3.1)}{(0,-1)}{4.1}\edge{(11.4,3.6)}{(1,-1)}{5.2}\put(2,-3){\circle{4}}\node{(2,-3)}\put(10,-3){\circle{4}}\node{(10,-3)}\put(18,-3){\circle{4}}\node{(18,-3)}\end{picture}}
\newcommand{\rootedGGGGe}{\begin{picture}(7,9)\cadre{7}{9}\put(3.5,7){\circle{4}}\rnode{(3.5,7)}\edge{(3.5,5)}{(0,-1)}{3.5}\put(3.5,-2){\circle{7}}\node{(1.8,-2)}\node{(4.5,-0.5)}\node{(4.5,-3.5)}\end{picture}}
\newcommand{\rootedGGGGf}{\begin{picture}(7,9)\cadre{7}{9}\put(3.5,5.5){\circle{7}}\rnode{(1.8,5.5)}\node{(4.5,7)}\node{(4.5,4)}\edge{(3.5,2)}{(0,-1)}{3.6}\put(3.5,-3.5){\circle{4}}\node{(3.5,-3.5)}\end{picture}}
\newcommand{\rootedGGGGGa}{\begin{picture}(4,19)\cadre{4}{19}\put(2,17){\circle{4}}\rnode{(2,17)}\edge{(2,15.1)}{(0,-1)}{4.1}\put(2,9){\circle{4}}\node{(2,9)}\edge{(2,7.1)}{(0,-1)}{4.1}\put(2,1){\circle{4}}\node{(2,1)}\edge{(2,-0.9)}{(0,-1)}{4.1}\put(2,-7){\circle{4}}\node{(2,-7)}\edge{(2,-8.9)}{(0,-1)}{4.1}\put(2,-15){\circle{4}}\node{(2,-15)}\end{picture}}
\newcommand{\rootedGGGGGb}{\begin{picture}(12,15)\cadre{12}{15}\put(6,13){\circle{4}}\rnode{(6,13)}\edge{(5.1,11.2)}{(-1,-2)}{2.2}\edge{(6.9,11.2)}{(1,-2)}{2.2}\put(2,5){\circle{4}}\node{(2,5)}\put(10,5){\circle{4}}\node{(10,5)}\edge{(2,3.1)}{(0,-1)}{4.1}\put(2,-3){\circle{4}}\node{(2,-3)}\edge{(2,-4.9)}{(0,-1)}{4.1}\put(2,-11){\circle{4}}\node{(2,-11)}\end{picture}}
\newcommand{\rootedGGGGGc}{\begin{picture}(12,11)\cadre{12}{11}\put(6,9){\circle{4}}\rnode{(6,9)}\edge{(5.1,7.2)}{(-1,-2)}{2.2}\edge{(6.9,7.2)}{(1,-2)}{2.2}\put(2,1){\circle{4}}\node{(2,1)}\put(10,1){\circle{4}}\node{(10,1)}\put(2,1){\circle{4}}\node{(2,1)}\edge{(2,-0.9)}{(0,-1)}{4.1}\put(2,-7){\circle{4}}\node{(2,-7)}\edge{(10,-0.9)}{(0,-1)}{4.1}\put(10,-7){\circle{4}}\node{(10,-7)}\end{picture}}
\newcommand{\rootedGGGGGd}{\begin{picture}(12,15)\cadre{12}{15}\put(6,13){\circle{4}}\rnode{(6,13)}\edge{(6,11.1)}{(0,-1)}{4.1}\put(6,5){\circle{4}}\node{(6,5)}\edge{(6,3.1)}{(0,-1)}{4.1}\put(6,-3){\circle{4}}\node{(6,-3)}\edge{(5.1,-4.8)}{(-1,-2)}{2.2}\edge{(6.9,-4.8)}{(1,-2)}{2.2}\put(2,-11){\circle{4}}\node{(2,-11)}\put(10,-11){\circle{4}}\node{(10,-11)}\end{picture}}
\newcommand{\rootedGGGGGe}{\begin{picture}(16,11)\cadre{16}{11}\put(10,9){\circle{4}}\rnode{(10,9)}\edge{(9.1,7.2)}{(-1,-2)}{2.2}\edge{(10.9,7.2)}{(1,-2)}{2.2}\put(6,1){\circle{4}}\node{(6,1)}\put(14,1){\circle{4}}\node{(14,1)}\edge{(5.1,-0.8)}{(-1,-2)}{2.2}\edge{(6.9,-0.8)}{(1,-2)}{2.2}\put(2,-7){\circle{4}}\node{(2,-7)}\put(10,-7){\circle{4}}\node{(10,-7)}\end{picture}}
\newcommand{\rootedGGGGGf}{\begin{picture}(20,11)\cadre{20}{11}\put(10,9){\circle{4}}\rnode{(10,9)}\edge{(8.6,7.6)}{(-1,-1)}{5.2}\edge{(10,7.1)}{(0,-1)}{4.1}\edge{(11.4,7.6)}{(1,-1)}{5.2}\put(2,1){\circle{4}}\node{(2,1)}\put(10,1){\circle{4}}\node{(10,1)}\put(18,1){\circle{4}}\node{(18,1)}\edge{(2,-0.9)}{(0,-1)}{4.1}\put(2,-7){\circle{4}}\node{(2,-7)}\end{picture}}
\newcommand{\rootedGGGGGg}{\begin{picture}(12,15)\cadre{12}{15}\put(6,13){\circle{4}}\rnode{(6,13)}\edge{(6,11.1)}{(0,-1)}{4.1}\put(6,5){\circle{4}}\node{(6,5)}\edge{(5.1,3.2)}{(-1,-2)}{2.2}\edge{(6.9,3.2)}{(1,-2)}{2.2}\put(2,-3){\circle{4}}\node{(2,-3)}\put(10,-3){\circle{4}}\node{(10,-3)}\edge{(2,-4.9)}{(0,-1)}{4.1}\put(2,-11){\circle{4}}\node{(2,-11)}\end{picture}}
\newcommand{\rootedGGGGGh}{\begin{picture}(20,11)\cadre{20}{11}\put(10,9){\circle{4}}\rnode{(10,9)}\edge{(10,7.1)}{(0,-1)}{4.1}\put(10,1){\circle{4}}\node{(10,1)}\edge{(8.6,-0.4)}{(-1,-1)}{5.2}\edge{(10,-0.9)}{(0,-1)}{4.1}\edge{(11.4,-0.4)}{(1,-1)}{5.2}\put(2,-7){\circle{4}}\node{(2,-7)}\put(10,-7){\circle{4}}\node{(10,-7)}\put(18,-7){\circle{4}}\node{(18,-7)}\end{picture}}
\newcommand{\rootedGGGGGi}{\begin{picture}(28,7)\cadre{28}{7}\put(14,5){\circle{4}}\rnode{(14,5)}\edge{(12.35,3.9)}{(-3,-2)}{8.7}\edge{(13.1,3.2)}{(-1,-2)}{2.2}\edge{(14.9,3.2)}{(1,-2)}{2.2}\edge{(15.65,3.9)}{(3,-2)}{8.7}\put(2,-3){\circle{4}}\node{(2,-3)}\put(10,-3){\circle{4}}\node{(10,-3)}\put(18,-3){\circle{4}}\node{(18,-3)}\put(26,-3){\circle{4}}\node{(26,-3)}\end{picture}}
\newcommand{\rootedGGGGGj}{\begin{picture}(7,13)\cadre{7}{13}\put(3.5,11){\circle{4}}\rnode{(3.5,11)}\edge{(3.5,9.1)}{(0,-1)}{4.1}
\put(3.5,-6){\circle{7}}\node{(1.8,-6)}\node{(4.5,-4.5)}\node{(4.5,-7.5)}\edge{(3.5,-2.5)}{(0,1)}{3.6}\put(3.5,3){\circle{4}}\node{(3.5,3)}\end{picture}}
\newcommand{\rootedGGGGGk}{\begin{picture}(7,13)\cadre{7}{13}\put(3.5,9.5){\circle{7}}\rnode{(1.8,9.5)}\node{(4.5,11)}\node{(4.5,8)}\edge{(3.5,6)}{(0,-1)}{3.6}\put(3.5,0.5){\circle{4}}\node{(3.5,0.5)}\edge{(3.5,-1.4)}{(0,-1)}{4.1}\put(3.5,-7.5){\circle{4}}\node{(3.5,-7.5)}\end{picture}}
\newcommand{\rootedGGGGGl}{\begin{picture}(13.5,9)\cadre{13.5}{9}\put(7.5,6){\circle{4}}\rnode{(7.5,6)}\edge{(6.6,4.2)}{(-1,-2)}{1.5}\edge{(8.4,4.2)}{(1,-2)}{2.2}\put(3.5,-2){\circle{7}}\node{(1.8,-2)}\node{(4.5,-0.5)}\node{(4.5,-3.5)}\put(11.5,-2){\circle{4}}\node{(11.5,-2)}\end{picture}}
\newcommand{\rootedGGGGGm}{\begin{picture}(7,13)\cadre{7}{13}\put(3.5,11){\circle{4}}\rnode{(3.5,11)}\edge{(3.5,9)}{(0,-1)}{3.5}\put(3.5,2){\circle{7}}\node{(1.8,2)}\node{(4.5,3.5)}\node{(4.5,0.5)}\edge{(3.5,-1.5)}{(0,-1)}{3.5}\put(3.5,-7){\circle{4}}\node{(3.5,-7)}\end{picture}}
\newcommand{\rootedGGGGGn}{\begin{picture}(12,10)\cadre{12}{10}\put(6,5.5){\circle{7}}\rnode{(4.3,5.5)}\node{(7,7)}\node{(7,4)}\edge{(4.4,2.3)}{(-1,-2)}{1.5}\edge{(7.6,2.3)}{(1,-2)}{1.5}\put(2,-2.5){\circle{4}}\node{(2,-2.5)}\put(10,-2.5){\circle{4}}\node{(10,-2.5)}\end{picture}}
\newcommand{\rootedGGGGGo}{\begin{picture}(10,6)\cadre{10}{6}\put(5,1){\circle{10}}\node{(2,1)}\rnode{(4.3,3.5)}\node{(4.3,-1.5)}\node{(7.5,2.7)}\node{(7.5,-0.7)}\end{picture}}

\newcommand{\Ga}{\begin{picture}(4,3)\cadre{4}{3}\smallcircle{(2,1)}\node{(2,1)}\end{picture}}
\newcommand{\GGa}{\begin{picture}(11,3)\cadre{11}{3}\smallcircle{(2,1)}\node{(2,1)}\edge{(3.9,1)}{(1,0)}{3.1}\smallcircle{(9,1)}\node{(9,1)}\end{picture}}
\newcommand{\GGGa}{\begin{picture}(18,3)\cadre{18}{3}\smallcircle{(2,1)}\node{(2,1)}\edge{(3.9,1)}{(1,0)}{3.1}\smallcircle{(9,1)}\node{(9,1)}\edge{(10.9,1)}{(1,0)}{3.1}\smallcircle{(16,1)}\node{(16,1)}\end{picture}}
\newcommand{\GGGb}{\begin{picture}(7,5)\cadre{7}{5}\mediumcircle{(3.5,1)}\node{(1.8,1)}\node{(4.5,2.5)}\node{(4.5,-0.5)}\end{picture}}
\newcommand{\GGGGa}{\begin{picture}(25,3)\cadre{25}{3}\smallcircle{(2,1)}\node{(2,1)}\edge{(3.9,1)}{(1,0)}{3.1}\smallcircle{(9,1)}\node{(9,1)}\edge{(10.9,1)}{(1,0)}{3.1}\smallcircle{(16,1)}\node{(16,1)}\edge{(17.9,1)}{(1,0)}{3.1}\smallcircle{(23,1)}\node{(23,1)}\end{picture}}
\newcommand{\GGGGb}{\begin{picture}(18.5,7)\cadre{18.5}{7}\smallcircle{(2,1)}\node{(2,1)}\edge{(3.9,1)}{(1,0)}{3.1}\smallcircle{(9,1)}\node{(9,1)}\edge{(10.7889,1.89443)}{(2,1)}{3.6}\smallcircle{(16.2,4.6)}\node{(16.2,4.6)}\edge{(10.7889,0.105573)}{(2,-1)}{3.6}\smallcircle{(16.2,-2.6)}\node{(16.2,-2.6)}\end{picture}}
\newcommand{\GGGGc}{\begin{picture}(14,5)\cadre{14}{5}\mediumcircle{(3.5,1)}\node{(1.8,1)}\node{(4.5,2.5)}\node{(4.5,-0.5)}\edge{(6.9,1)}{(1,0)}{3.1}\smallcircle{(12,1)}\node{(12,1)}\end{picture}}
\newcommand{\GGGGGa}{\begin{picture}(32,3)\cadre{25}{3}\smallcircle{(2,1)}\node{(2,1)}\edge{(3.9,1)}{(1,0)}{3.1}\smallcircle{(9,1)}\node{(9,1)}\edge{(10.9,1)}{(1,0)}{3.1}\smallcircle{(16,1)}\node{(16,1)}\edge{(17.9,1)}{(1,0)}{3.1}\smallcircle{(23,1)}\node{(23,1)}\edge{(24.9,1)}{(1,0)}{3.1}\smallcircle{(30,1)}\node{(30,1)}\end{picture}}
\newcommand{\GGGGGb}{\begin{picture}(25.5,7)\cadre{25.5}{7}\smallcircle{(2,1)}\node{(2,1)}\edge{(3.9,1)}{(1,0)}{3.1}\smallcircle{(9,1)}\node{(9,1)}\edge{(10.9,1)}{(1,0)}{3.1}\smallcircle{(16,1)}\node{(16,1)}\edge{(17.7889,1.89443)}{(2,1)}{3.6}\smallcircle{(23.2,4.6)}\node{(23.2,4.6)}\edge{(17.7889,0.105573)}{(2,-1)}{3.6}\smallcircle{(23.2,-2.6)}\node{(23.2,-2.6)}\end{picture}}
\newcommand{\GGGGGc}{\begin{picture}(18,10)\cadre{18}{10}\smallcircle{(2,1)}\node{(2,1)}\edge{(3.9,1)}{(1,0)}{3.1}\smallcircle{(9,1)}\node{(9,1)}\edge{(9,-0.9)}{(0,-1)}{3.1}\smallcircle{(9,-6)}\node{(9,-6)}\edge{(10.9,1)}{(1,0)}{3.1}\smallcircle{(16,1)}\node{(16,1)}\edge{(9,2.9)}{(0,1)}{3.1}\smallcircle{(9,8)}\node{(9,8)}\end{picture}}
\newcommand{\GGGGGd}{\begin{picture}(21,5)\cadre{21}{5}\mediumcircle{(3.5,1)}\node{(1.8,1)}\node{(4.5,2.5)}\node{(4.5,-0.5)}\edge{(6.9,1)}{(1,0)}{3.1}\smallcircle{(12,1)}\node{(12,1)}\edge{(13.9,1)}{(1,0)}{3.1}\smallcircle{(19,1)}\node{(19,1)}\end{picture}}
\newcommand{\GGGGGe}{\begin{picture}(21,5)\cadre{21}{5}\smallcircle{(2,1)}\node{(2,1)}\edge{(3.9,1)}{(1,0)}{3.1}\mediumcircle{(10.5,1)}\node{(8.8,1)}\node{(11.5,2.5)}\node{(11.5,-0.5)}\edge{(13.9,1)}{(1,0)}{3.1}\smallcircle{(19,1)}\node{(19,1)}\end{picture}}
\newcommand{\GGGGGf}{\begin{picture}(10,6)\cadre{10}{6}\largecircle{(5,1)}\node{(2,1)}\node{(4.3,3.5)}\node{(4.3,-1.5)}\node{(7.5,2.7)}\node{(7.5,-0.7)}\end{picture}}

\newcommand{\sGa}{\begin{picture}(2,2)\cadre{2}{2}\node{(1,1)}\end{picture}}
\newcommand{\sGGa}{\begin{picture}(7,2)\cadre{7}{2}\node{(1,1)}\edge{(1,1)}{(1,0)}{5}\node{(6,1)}\end{picture}}
\newcommand{\sGGGa}{\begin{picture}(12,2)\cadre{12}{2}\node{(1,1)}\edge{(1,1)}{(1,0)}{5}\node{(6,1)}\edge{(6,1)}{(1,0)}{5}\node{(11,1)}\end{picture}}
\newcommand{\sGGGb}{\begin{picture}(4,3)\cadre{4}{3}\cnode{(2,1)}{3}\end{picture}}
\newcommand{\sGGGGa}{\begin{picture}(17,2)\cadre{17}{2}\node{(1,1)}\edge{(1,1)}{(1,0)}{5}\node{(6,1)}\edge{(6,1)}{(1,0)}{5}\node{(11,1)}\edge{(11,1)}{(1,0)}{5}\node{(16,1)}\end{picture}}
\newcommand{\sGGGGb}{\begin{picture}(10,5)\cadre{10}{5}\node{(1,1)}\edge{(1,1)}{(1,0)}{5}\node{(6,1)}\edge{(6,1)}{(1,1)}{3.7}\node{(9,4)}\edge{(6,1)}{(1,-1)}{3.7}\node{(9,-2)}\end{picture}}
\newcommand{\sGGGGc}{\begin{picture}(9,3)\cadre{9}{3}\cnode{(2,1)}{3}\edge{(3.9,1)}{(1,0)}{4}\node{(8,1)}\end{picture}}
\newcommand{\sGGGGGa}{\begin{picture}(22,2)\cadre{22}{2}\node{(1,1)}\edge{(1,1)}{(1,0)}{5}\node{(6,1)}\edge{(6,1)}{(1,0)}{5}\node{(11,1)}\edge{(11,1)}{(1,0)}{5}\node{(16,1)}\edge{(16,1)}{(1,0)}{5}\node{(21,1)}\end{picture}}
\newcommand{\sGGGGGb}{\begin{picture}(15,5)\cadre{15}{5}\node{(1,1)}\edge{(1,1)}{(1,0)}{5}\node{(6,1)}\edge{(6,1)}{(1,0)}{5}\node{(11,1)}\edge{(11,1)}{(1,1)}{3.7}\node{(14,4)}\edge{(11,1)}{(1,-1)}{3.7}\node{(14,-2)}\end{picture}}
\newcommand{\sGGGGGc}{\begin{picture}(12,7)\cadre{12}{7}\node{(1,1)}\edge{(1,1)}{(1,0)}{5}\node{(6,1)}\edge{(6,1)}{(1,0)}{5}\node{(11,1)}\edge{(6,1)}{(0,1)}{5}\node{(6,6)}\edge{(6,1)}{(0,-1)}{5}\node{(6,-4)}\end{picture}}
\newcommand{\sGGGGGd}{\begin{picture}(14,3)\cadre{14}{3}\cnode{(2,1)}{3}\edge{(3.9,1)}{(1,0)}{4}\node{(8,1)}\edge{(8,1)}{(1,0)}{5}\node{(13,1)}\end{picture}}
\newcommand{\sGGGGGe}{\begin{picture}(14,3)\cadre{14}{3}\node{(1,1)}\edge{(1,1)}{(1,0)}{4}\cnode{(7,1)}{3}\edge{(8.9,1)}{(1,0)}{4}\node{(13,1)}\end{picture}}
\newcommand{\sGGGGGf}{\begin{picture}(4,3)\cadre{4}{3}\cnode{(2,1)}{5}\end{picture}}
\newcommand{\sGGGGGGa}{\begin{picture}(27,2)\cadre{27}{2}\node{(1,1)}\edge{(1,1)}{(1,0)}{5}\node{(6,1)}\edge{(6,1)}{(1,0)}{5}\node{(11,1)}\edge{(11,1)}{(1,0)}{5}\node{(16,1)}\edge{(16,1)}{(1,0)}{5}\node{(21,1)}\edge{(21,1)}{(1,0)}{5}\node{(26,1)}\end{picture}}
\newcommand{\sGGGGGGb}{\begin{picture}(20,5)\cadre{20}{5}\node{(1,1)}\edge{(1,1)}{(1,0)}{5}\node{(6,1)}\edge{(6,1)}{(1,0)}{5}\node{(11,1)}\edge{(11,1)}{(1,0)}{5}\node{(16,1)}\edge{(16,1)}{(1,1)}{3.7}\node{(19,4)}\edge{(16,1)}{(1,-1)}{3.7}\node{(19,-2)}\end{picture}}
\newcommand{\sGGGGGGc}{\begin{picture}(13,5)\cadre{13}{5}\node{(1,-2)}\edge{(1,-2)}{(1,1)}{3.7}\node{(1,4)}\edge{(1,4)}{(1,-1)}{3.7}\node{(4,1)}\edge{(4,1)}{(1,0)}{5}\node{(9,1)}\edge{(9,1)}{(1,1)}{3.7}\node{(12,4)}\edge{(9,1)}{(1,-1)}{3.7}\node{(12,-2)}\end{picture}}
\newcommand{\sGGGGGGd}{\begin{picture}(18,8)\cadre{18}{8}\node{(1,1)}\edge{(1,1)}{(1,0)}{5}\node{(6,1)}\edge{(6,1)}{(1,0)}{5}\node{(11,1)}\edge{(11,1)}{(1,1)}{3.7}\node{(14,4)}\edge{(14,4)}{(1,1)}{3.7}\node{(17,7)}\edge{(11,1)}{(1,-1)}{3.7}\node{(14,-2)}\end{picture}}
\newcommand{\sGGGGGGe}{\begin{picture}(17,7)\cadre{17}{7}\node{(1,1)}\edge{(1,1)}{(1,0)}{5}\node{(6,1)}\edge{(6,1)}{(1,0)}{5}\node{(11,1)}\edge{(11,1)}{(1,0)}{5}\node{(16,1)}\edge{(11,1)}{(0,1)}{5}\node{(11,6)}\edge{(11,1)}{(0,-1)}{5}\node{(11,-4)}\end{picture}}
\newcommand{\sGGGGGGf}{\begin{picture}(10,6)\cadre{10}{6}\node{(4,1)}\edge{(4,1)}{(1,0)}{5}\node{(9,1)}\edge{(4,1)}{(1,2)}{2}\node{(6,5)}\edge{(4,1)}{(-1,1)}{3.6}\node{(1,4)}\edge{(4,1)}{(-1,-1)}{3.6}\node{(1,-2)}\edge{(4,1)}{(1,-2)}{2}\node{(6,-3)}\end{picture}}
\newcommand{\sGGGGGGg}{\begin{picture}(19,3)\cadre{19}{3}\cnode{(2,1)}{3}\edge{(3.9,1)}{(1,0)}{4}\node{(8,1)}\edge{(8,1)}{(1,0)}{5}\node{(13,1)}\edge{(13,1)}{(1,0)}{5}\node{(18,1)}\end{picture}}
\newcommand{\sGGGGGGh}{\begin{picture}(19,3)\cadre{19}{3}\node{(1,1)}\edge{(1,1)}{(1,0)}{4}\cnode{(7,1)}{3}\edge{(8.9,1)}{(1,0)}{4}\node{(13,1)}\edge{(13,1)}{(1,0)}{5}\node{(18,1)}\end{picture}}
\newcommand{\sGGGGGGi}{\begin{picture}(12,3)\cadre{12}{3}\cnode{(2,1)}{3}\edge{(3.9,1)}{(1,0)}{4}\node{(8,1)}\edge{(8,1)}{(1,1)}{3.7}\node{(11,4)}\edge{(8,1)}{(1,-1)}{3.7}\node{(11,-2)}\end{picture}}
\newcommand{\sGGGGGGj}{\begin{picture}(12.5,6.5)\cadre{12.5}{6.5}\node{(1,1)}\edge{(1,1)}{(1,0)}{4}\cnode{(7,1)}{3}\edge{(8.4,2.4)}{(1,1)}{3.6}\node{(11.3,5.3)}\edge{(8.4,-0.4)}{(1,-1)}{3.6}\node{(11.5,-3.3)}\end{picture}}
\newcommand{\sGGGGGGk}{\begin{picture}(9,3)\cadre{9}{3}\cnode{(2,1)}{5}\edge{(3.9,1)}{(1,0)}{4}\node{(8,1)}\end{picture}}
\newcommand{\sGGGGGGl}{\begin{picture}(11,3)\cadre{11}{3}\cnode{(2,1)}{3}\edge{(3.9,1)}{(1,0)}{3}\cnode{(8.9,1)}{3}\end{picture}}
\newcommand{\sGGGGGGGGGd}{\begin{picture}(19.5,6.5)\cadre{19.5}{6.5}\node{(1,1)}\edge{(1,1)}{(1,0)}{4}\cnode{(7,1)}{3}\edge{(9,1)}{(1,0)}{3}\cnode{(14,1)}{3}\edge{(15.4,2.4)}{(1,1)}{3.6}\node{(18.3,5.3)}\edge{(15.4,-0.4)}{(1,-1)}{3.6}\node{(18.5,-3.3)}\end{picture}}

\newcommand{\tildeCa}{\begin{picture}(4,3)\cadre{4}{3}\put(2,1){\circle{4}}\node{(2,1)}\end{picture}}
\newcommand{\tildeCCCa}{\begin{picture}(14,3)\cadre{14}{3}\put(7,1){\oval(14,4)}\node{(2,1)}\edge{(2,1)}{(1,0)}{5}\node{(7,1)}\edge{(7,1)}{(1,0)}{5}\node{(12,1)}\end{picture}}
\newcommand{\tildeCCCCCa}{\begin{picture}(24,3)\cadre{24}{3}\put(12,1){\oval(24,4)}\node{(2,1)}\edge{(2,1)}{(1,0)}{5}\node{(7,1)}\edge{(7,1)}{(1,0)}{5}\node{(12,1)}\edge{(12,1)}{(1,0)}{5}\node{(17,1)}\edge{(17,1)}{(1,0)}{5}\node{(22,1)}\end{picture}}
\newcommand{\tildeCCCCCb}{\begin{picture}(19,6)\cadre{19}{6}\put(9.5,1){\oval(19,10)}\node{(2,1)}\edge{(2,1)}{(1,0)}{5}\node{(7,1)}\edge{(7,1)}{(1,0)}{5}\node{(12,1)}\edge{(12,1)}{(1,1)}{3.7}\node{(15,4)}\edge{(12,1)}{(1,-1)}{3.7}\node{(15,-2)}\end{picture}}
\newcommand{\tildeCCCCCc}{\begin{picture}(14,8)\cadre{14}{8}\put(7,1){\oval(14,14)}\node{(2,1)}\edge{(2,1)}{(1,0)}{5}\node{(7,1)}\edge{(7,1)}{(1,0)}{5}\node{(12,1)}\edge{(7,1)}{(0,1)}{5}\node{(7,6)}\edge{(7,1)}{(0,-1)}{5}\node{(7,-4)}\end{picture}}

\newcommand{\tildeGa}{\begin{picture}(4,3)\cadre{4}{3}\put(2,1){\circle{4}}\rnode{(2,1)}\end{picture}}
\newcommand{\tildeGGa}{\begin{picture}(5,7)\cadre{5}{7}\put(2,5){\circle{4}}\rnode{(2,5)}\arrow{(2,3.1)}{(0,-1)}{4.1}{3}\put(2.5,0){\small$i_{1}$}\put(2,-3){\circle{4}}\node{(2,-3)}\end{picture}}
\newcommand{\tildeGGaa}{\begin{picture}(5,7)\cadre{5}{7}\put(2,5){\circle{4}}\rnode{(2,5)}\arrow{(2,-1)}{(0,+1)}{4.1}{3}\put(2.5,0){\small$i_{1}$}\put(2,-3){\circle{4}}\node{(2,-3)}\end{picture}}
\newcommand{\tildeGGGa}{\begin{picture}(12,7)\cadre{12}{7}\put(6,5){\circle{4}}\rnode{(6,5)}\arrow{(5.1,3.2)}{(-1,-2)}{2.2}{1.7}\arrow{(6.9,3.2)}{(1,-2)}{2.2}{1.7}\put(1,1){\small$i_{1}$}\put(8.5,1){\small$i_{2}$}\put(2,-3){\circle{4}}\node{(2,-3)}\put(10,-3){\circle{4}}\node{(10,-3)}\end{picture}}
\newcommand{\tildeGGGaa}{\begin{picture}(12,7)\cadre{12}{7}\put(6,5){\circle{4}}\rnode{(6,5)}\arrow{(2.9,-1.2)}{(+1,+2)}{2.2}{1.7}\arrow{(6.9,3.2)}{(1,-2)}{2.2}{1.7}\put(1,1){\small$i_{1}$}\put(8.5,1){\small$i_{2}$}\put(2,-3){\circle{4}}\node{(2,-3)}\put(10,-3){\circle{4}}\node{(10,-3)}\end{picture}}
\newcommand{\tildeGGGaaa}{\begin{picture}(12,7)\cadre{12}{7}\put(6,5){\circle{4}}\rnode{(6,5)}\arrow{(5.1,3.2)}{(-1,-2)}{2.2}{1.7}\arrow{(9.1,-1.2)}{(-1,+2)}{2.2}{1.7}\put(1,1){\small$i_{1}$}\put(8.5,1){\small$i_{2}$}\put(2,-3){\circle{4}}\node{(2,-3)}\put(10,-3){\circle{4}}\node{(10,-3)}\end{picture}}
\newcommand{\tildeGGGaaaa}{\begin{picture}(12,7)\cadre{12}{7}\put(6,5){\circle{4}}\rnode{(6,5)}\arrow{(2.9,-1.2)}{(+1,+2)}{2.2}{1.7}\arrow{(9.1,-1.2)}{(-1,+2)}{2.2}{1.7}\put(1,1){\small$i_{1}$}\put(8.5,1){\small$i_{2}$}\put(2,-3){\circle{4}}\node{(2,-3)}\put(10,-3){\circle{4}}\node{(10,-3)}\end{picture}}
\newcommand{\tildeGGGb}{\begin{picture}(5,10)\cadre{5}{10}\put(2,8){\circle{4}}\rnode{(2,8)}\arrow{(2,6.1)}{(0,-1)}{4.1}{3}\put(2.5,3){\small$i_{1}$}\put(2,0){\circle{4}}\node{(2,0)}\arrow{(2,-1.9)}{(0,-1)}{4.1}{3}\put(2.5,-5){\small$i_{2}$}\put(2,-8){\circle{4}}\node{(2,-8)}\end{picture}}
\newcommand{\tildeGGGbb}{\begin{picture}(5,10)\cadre{5}{10}\put(2,8){\circle{4}}\rnode{(2,8)}\arrow{(2,2)}{(0,+1)}{4.1}{3}\put(2.5,3){\small$i_{1}$}\put(2,0){\circle{4}}\node{(2,0)}\arrow{(2,-1.9)}{(0,-1)}{4.1}{3}\put(2.5,-5){\small$i_{2}$}\put(2,-8){\circle{4}}\node{(2,-8)}\end{picture}}
\newcommand{\tildeGGGbbb}{\begin{picture}(5,10)\cadre{5}{10}\put(2,8){\circle{4}}\rnode{(2,8)}\arrow{(2,6.1)}{(0,-1)}{4.1}{3}\put(2.5,3){\small$i_{1}$}\put(2,0){\circle{4}}\node{(2,0)}\arrow{(2,-6)}{(0,+1)}{4.1}{3}\put(2.5,-5){\small$i_{2}$}\put(2,-8){\circle{4}}\node{(2,-8)}\end{picture}}
\newcommand{\tildeGGGbbbb}{\begin{picture}(5,10)\cadre{5}{10}\put(2,8){\circle{4}}\rnode{(2,8)}\arrow{(2,2)}{(0,+1)}{4.1}{3}\put(2.5,3){\small$i_{1}$}\put(2,0){\circle{4}}\node{(2,0)}\arrow{(2,-6)}{(0,+1)}{4.1}{3}\put(2.5,-5){\small$i_{2}$}\put(2,-8){\circle{4}}\node{(2,-8)}\end{picture}}
\newcommand{\tildeGGGbbbbb}{\begin{picture}(5,10)\cadre{5}{10}\put(2,8){\circle{4}}\rnode{(2,8)}\arrow{(2,6.1)}{(0,-1)}{4.1}{3}\put(2.5,3){\small$i_{2}$}\put(2,0){\circle{4}}\node{(2,0)}\arrow{(2,-1.9)}{(0,-1)}{4.1}{3}\put(2.5,-5){\small$i_{1}$}\put(2,-8){\circle{4}}\node{(2,-8)}\end{picture}}
\newcommand{\tildeGGGbbbbbb}{\begin{picture}(5,10)\cadre{5}{10}\put(2,8){\circle{4}}\rnode{(2,8)}\arrow{(2,2)}{(0,+1)}{4.1}{3}\put(2.5,3){\small$i_{2}$}\put(2,0){\circle{4}}\node{(2,0)}\arrow{(2,-1.9)}{(0,-1)}{4.1}{3}\put(2.5,-5){\small$i_{1}$}\put(2,-8){\circle{4}}\node{(2,-8)}\end{picture}}
\newcommand{\tildeGGGbbbbbbb}{\begin{picture}(5,10)\cadre{5}{10}\put(2,8){\circle{4}}\rnode{(2,8)}\arrow{(2,6.1)}{(0,-1)}{4.1}{3}\put(2.5,3){\small$i_{2}$}\put(2,0){\circle{4}}\node{(2,0)}\arrow{(2,-6)}{(0,+1)}{4.1}{3}\put(2.5,-5){\small$i_{1}$}\put(2,-8){\circle{4}}\node{(2,-8)}\end{picture}}
\newcommand{\tildeGGGbbbbbbbb}{\begin{picture}(5,10)\cadre{5}{10}\put(2,8){\circle{4}}\rnode{(2,8)}\arrow{(2,2)}{(0,+1)}{4.1}{3}\put(2.5,3){\small$i_{2}$}\put(2,0){\circle{4}}\node{(2,0)}\arrow{(2,-6)}{(0,+1)}{4.1}{3}\put(2.5,-5){\small$i_{1}$}\put(2,-8){\circle{4}}\node{(2,-8)}\end{picture}}
\newcommand{\tildeGGGc}{\begin{picture}(11,6)\cadre{11}{6}\put(5.5,0.8){\circle{10.6}}\rnode{(5.5,4.5)}\arrow{(5,3.5)}{(-1,-2)}{2.2}{1.7}\arrow{(6,3.5)}{(1,-2)}{2.2}{1.7}\put(0.5,0.5){\small$i_{1}$}\put(8,0.5){\small$i_{2}$}\node{(2.5,-1.5)}\node{(8.5,-1.5)}\end{picture}}
\newcommand{\tildeGGGcc}{\begin{picture}(11,6)\cadre{11}{6}\put(5.5,0.8){\circle{10.6}}\rnode{(5.5,4.5)}\arrow{(2.8,-0.9)}{(+1,+2)}{2.2}{1.7}\arrow{(6,3.5)}{(1,-2)}{2.2}{1.7}\put(0.5,0.5){\small$i_{1}$}\put(8,0.5){\small$i_{2}$}\node{(2.5,-1.5)}\node{(8.5,-1.5)}\end{picture}}
\newcommand{\tildeGGGccc}{\begin{picture}(11,6)\cadre{11}{6}\put(5.5,0.8){\circle{10.6}}\rnode{(5.5,4.5)}\arrow{(5,3.5)}{(-1,-2)}{2.2}{1.7}\arrow{(8.2,-0.9)}{(-1,+2)}{2.2}{1.7}\put(0.5,0.5){\small$i_{1}$}\put(8,0.5){\small$i_{2}$}\node{(2.5,-1.5)}\node{(8.5,-1.5)}\end{picture}}
\newcommand{\tildeGGGcccc}{\begin{picture}(11,6)\cadre{11}{6}\put(5.5,0.8){\circle{10.6}}\rnode{(5.5,4.5)}\arrow{(2.8,-0.9)}{(+1,+2)}{2.2}{1.7}\arrow{(8.2,-0.9)}{(-1,+2)}{2.2}{1.7}\put(0.5,0.5){\small$i_{1}$}\put(8,0.5){\small$i_{2}$}\node{(2.5,-1.5)}\node{(8.5,-1.5)}\end{picture}}
\newcommand{\tildeGGGd}{\begin{picture}(6,9)\cadre{6}{9}\put(3,1){\oval(6,16)}\rnode{(2,7)}\arrow{(2,6)}{(0,-1)}{4}{3}\put(2.5,3){\small$i_{1}$}\node{(2,1)}\arrow{(2,0)}{(0,-1)}{4}{3}\put(2.5,-3){\small$i_{2}$}\node{(2,-5)}\end{picture}}
\newcommand{\tildeGGGdd}{\begin{picture}(6,9)\cadre{6}{9}\put(3,1){\oval(6,16)}\rnode{(2,7)}\arrow{(2,2)}{(0,+1)}{4}{3}\put(2.5,3){\small$i_{1}$}\node{(2,1)}\arrow{(2,0)}{(0,-1)}{4}{3}\put(2.5,-3){\small$i_{2}$}\node{(2,-5)}\end{picture}}
\newcommand{\tildeGGGddd}{\begin{picture}(6,9)\cadre{6}{9}\put(3,1){\oval(6,16)}\rnode{(2,7)}\arrow{(2,6)}{(0,-1)}{4}{3}\put(2.5,3){\small$i_{1}$}\node{(2,1)}\arrow{(2,-4)}{(0,+1)}{4}{3}\put(2.5,-3){\small$i_{2}$}\node{(2,-5)}\end{picture}}
\newcommand{\tildeGGGdddd}{\begin{picture}(6,9)\cadre{6}{9}\put(3,1){\oval(6,16)}\rnode{(2,7)}\arrow{(2,2)}{(0,+1)}{4}{3}\put(2.5,3){\small$i_{1}$}\node{(2,1)}\arrow{(2,-4)}{(0,+1)}{4}{3}\put(2.5,-3){\small$i_{2}$}\node{(2,-5)}\end{picture}}
\newcommand{\tildeGGGddddd}{\begin{picture}(6,9)\cadre{6}{9}\put(3,1){\oval(6,16)}\rnode{(2,7)}\arrow{(2,6)}{(0,-1)}{4}{3}\put(2.5,3){\small$i_{2}$}\node{(2,1)}\arrow{(2,0)}{(0,-1)}{4}{3}\put(2.5,-3){\small$i_{1}$}\node{(2,-5)}\end{picture}}
\newcommand{\tildeGGGdddddd}{\begin{picture}(6,9)\cadre{6}{9}\put(3,1){\oval(6,16)}\rnode{(2,7)}\arrow{(2,2)}{(0,+1)}{4}{3}\put(2.5,3){\small$i_{2}$}\node{(2,1)}\arrow{(2,0)}{(0,-1)}{4}{3}\put(2.5,-3){\small$i_{1}$}\node{(2,-5)}\end{picture}}
\newcommand{\tildeGGGddddddd}{\begin{picture}(6,9)\cadre{6}{9}\put(3,1){\oval(6,16)}\rnode{(2,7)}\arrow{(2,6)}{(0,-1)}{4}{3}\put(2.5,3){\small$i_{2}$}\node{(2,1)}\arrow{(2,-4)}{(0,+1)}{4}{3}\put(2.5,-3){\small$i_{1}$}\node{(2,-5)}\end{picture}}
\newcommand{\tildeGGGdddddddd}{\begin{picture}(6,9)\cadre{6}{9}\put(3,1){\oval(6,16)}\rnode{(2,7)}\arrow{(2,2)}{(0,+1)}{4}{3}\put(2.5,3){\small$i_{2}$}\node{(2,1)}\arrow{(2,-4)}{(0,+1)}{4}{3}\put(2.5,-3){\small$i_{1}$}\node{(2,-5)}\end{picture}}

\newcommand{\sTildeGa}{\begin{picture}(2,2)\cadre{2}{2}\rnode{(1,1)}\end{picture}}
\newcommand{\sTildeGGa}{\begin{picture}(4,5)\cadre{4}{5}\rnode{(1,4)}\arrow{(1,3)}{(0,-1)}{5}{3}\put(1.5,0){\small$i_{1}$}\node{(1,-2)}\end{picture}}
\newcommand{\sTildeGGaa}{\begin{picture}(4,5)\cadre{4}{5}\rnode{(1,4)}\arrow{(1,3)}{(0,-1)}{5}{3}\put(1.5,0){\small$i_{2}$}\node{(1,-2)}\end{picture}}
\newcommand{\sTildeGGGa}{\begin{picture}(10,5)\cadre{10}{5}\rnode{(5,4)}\arrow{(4.55279,3.10557)}{(-1,-2)}{3}{1.7}\arrow{(5.44721,3.10557)}{(1,-2)}{3}{1.7}\put(0,0){\small$i_{1}$}\put(7.5,0){\small$i_{2}$}\node{(2,-2)}\node{(8,-2)}\end{picture}}
\newcommand{\sTildeGGGb}{\begin{picture}(4,8)\cadre{4}{8}\rnode{(1,7)}\arrow{(1,6)}{(0,-1)}{5}{3}\put(1.5,3){\small$i_{1}$}\node{(1,1)}\arrow{(1,0)}{(0,-1)}{5}{3}\put(1.5,-3){\small$i_{2}$}\node{(1,-5)}\end{picture}}
\newcommand{\sTildeGGGc}{\begin{picture}(11,6)\cadre{11}{6}\put(5.5,0.8){\circle{10.6}}\rnode{(5.5,4.5)}\arrow{(5.05279,3.60557)}{(-1,-2)}{3}{1.7}\arrow{(5.94721,3.60557)}{(1,-2)}{3}{1.7}\put(0.5,0.5){\small$i_{1}$}\put(8,0.5){\small$i_{2}$}\node{(2.5,-1.5)}\node{(8.5,-1.5)}\end{picture}}
\newcommand{\sTildeGGGd}{\begin{picture}(6,9)\cadre{6}{9}\put(3,1){\oval(6,16)}\rnode{(2,7)}\arrow{(2,6)}{(0,-1)}{5}{3}\put(2.5,3){\small$i_{1}$}\node{(2,1)}\arrow{(2,0)}{(0,-1)}{5}{3}\put(2.5,-3){\small$i_{2}$}\node{(2,-5)}\end{picture}}
\newcommand{\sTildeGGGGa}{\begin{picture}(11,8)\cadre{11}{8}\rnode{(5,7)}\arrow{(4.55279,6.10557)}{(-1,-2)}{3}{1.7}\arrow{(5.44721,6.10557)}{(1,-2)}{3}{1.7}\put(0,3){\small$i_{1}$}\put(7.5,3){\small$i_{2}$}\node{(2,1)}\node{(8,1)}\arrow{(8,1)}{(0,-1)}{6}{4}\put(8.5,-3){\small$i_{3}$}\node{(8,-5)}\end{picture}}
\newcommand{\sTildeGGGGb}{\begin{picture}(4,11)\cadre{4}{11}\rnode{(1,10)}\arrow{(1,9)}{(0,-1)}{5}{3}\put(1.5,6){\small$i_{1}$}\node{(1,4)}\arrow{(1,3)}{(0,-1)}{5}{3}\put(1.5,0){\small$i_{2}$}\node{(1,-2)}\arrow{(1,-3)}{(0,-1)}{5}{3}\put(1.5,-6){\small$i_{3}$}\node{(1,-8)}\end{picture}}
\newcommand{\sTildeGGGGc}{\begin{picture}(10,8)\cadre{10}{8}\rnode{(5,7)}\arrow{(5,6)}{(0,-1)}{5}{3}\put(5.5,3){\small$i_{1}$}\node{(5,1)}\arrow{(5,1)}{(-1,-2)}{3}{2}\arrow{(5,1)}{(1,-2)}{3}{2}\put(0,-3){\small$i_{2}$}\put(7.5,-3){\small$i_{3}$}\node{(2,-5)}\node{(8,-5)}\end{picture}}
\newcommand{\sTildeGGGGd}{\begin{picture}(14,5)\cadre{14}{5}\rnode{(7,4)}\arrow{(6.2,3.2)}{(-1,-1)}{5}{3}\put(0.5,1){\small$i_{1}$}\node{(1,-2)}\arrow{(7,3)}{(0,-1)}{5}{3}\put(7.5,-0.8){\small$i_{2}$}\node{(7,-2)}\arrow{(7.8,3.2)}{(1,-1)}{5}{3}\put(11,1){\small$i_{3}$}\node{(13,-2)}\end{picture}}
\newcommand{\sTildeGGGGe}{\begin{picture}(12.5,9.5)\cadre{12.5}{9.5}\qbezier(4,4)(-1,14)(10,6)\qbezier(10,6)(13.0152,3.80711)(12,0)\qbezier(12,0)(8,-15)(5,0)\qbezier(5,0)(4.33333,3.33333)(4,4)\rnode{(5,7)}\arrow{(4.55279,6.10557)}{(-1,-2)}{3}{1.7}\arrow{(5.44721,6.10557)}{(1,-2)}{3}{1.7}\put(0,3){\small$i_{1}$}\put(7.5,3){\small$i_{2}$}\node{(2,1)}\node{(8,1)}\arrow{(8,1)}{(0,-1)}{6}{4}\put(8.5,-3){\small$i_{3}$}\node{(8,-5)}\end{picture}}
\newcommand{\sTildeGGGGf}{\begin{picture}(11.5,9)\cadre{11.5}{9}\put(5.5,3.3){\circle{10.6}}\rnode{(5.5,7)}\arrow{(5.05279,6.10557)}{(-1,-2)}{3}{1.7}\arrow{(5.94721,6.10557)}{(1,-2)}{3}{1.7}\put(0.5,3){\small$i_{1}$}\put(8,3){\small$i_{2}$}\node{(2.5,1)}\node{(8.5,1)}\arrow{(8.5,1)}{(0,-1)}{6}{4}\put(9,-3){\small$i_{3}$}\node{(8.5,-5)}\end{picture}}
\newcommand{\sTildeGGGGg}{\begin{picture}(6,12)\cadre{6}{12}\put(3,-1){\oval(6,16)}\rnode{(2,11)}\arrow{(2,10)}{(0,-1)}{5}{3}\put(2.5,8){\small$i_{1}$}\node{(2,5)}\arrow{(2,4)}{(0,-1)}{5}{3}\put(2.5,1){\small$i_{2}$}\node{(2,-1)}\arrow{(2,-2)}{(0,-1)}{5}{3}\put(2.5,-5){\small$i_{3}$}\node{(2,-7)}\end{picture}}
\newcommand{\sTildeGGGGh}{\begin{picture}(6,12)\cadre{6}{12}\put(3,4){\oval(6,16)}\rnode{(2,10)}\arrow{(2,9)}{(0,-1)}{5}{3}\put(2.5,6){\small$i_{1}$}\node{(2,4)}\arrow{(2,3)}{(0,-1)}{5}{3}\put(2.5,0){\small$i_{2}$}\node{(2,-2)}\arrow{(2,-3)}{(0,-1)}{5}{3}\put(2.5,-7){\small$i_{3}$}\node{(2,-8)}\end{picture}}
\newcommand{\sTildeGGGGi}{\begin{picture}(11,8)\cadre{11}{8}\put(5.5,-2.7){\circle{10.6}}\rnode{(5.5,7)}\arrow{(5.5,6)}{(0,-1)}{6}{3}\put(6.2,3.5){\small$i_{1}$}\node{(5.5,1)}\arrow{(5.5,1)}{(-1,-2)}{3}{2}\arrow{(5.5,1)}{(1,-2)}{3}{2}\put(0.5,-3){\small$i_{2}$}\put(8,-3){\small$i_{3}$}\node{(2.5,-5)}\node{(8.5,-5)}\end{picture}}
\newcommand{\sTildeGGGGj}{\begin{picture}(11,10)\cadre{11}{10}\qbezier(3,7)(3,12)(8,7)\qbezier(8,7)(11,4)(11,-5)\qbezier(11,-5)(11,-13.33333)(6,-5)\qbezier(6,-5)(3,0)(3,7)\rnode{(5.5,7)}\arrow{(5.5,6)}{(0,-1)}{6}{3}\put(6.2,3.5){\small$i_{1}$}\node{(5.5,1)}\arrow{(5.5,1)}{(-1,-2)}{3}{2}\arrow{(5.5,1)}{(1,-2)}{3}{2}\put(0.5,-3){\small$i_{2}$}\put(8,-3){\small$i_{3}$}\node{(2.5,-5)}\node{(8.5,-5)}\end{picture}}
\newcommand{\sTildeGGGGk}{\begin{picture}(17,8)\cadre{17}{8}\qbezier(4.5,4)(4.5,12)(15.5,0.5)\qbezier(15.5,0.5)(20.5,-4.5)(7,-4.5)\qbezier(7,-4.5)(4.5,-4.5)(4.5,4)\rnode{(7,4)}\arrow{(6.2,3.2)}{(-1,-1)}{5}{3}\put(0.5,1){\small$i_{1}$}\node{(1,-2)}\arrow{(7,3)}{(0,-1)}{5}{3}\put(7.5,-0.8){\small$i_{2}$}\node{(7,-2)}\arrow{(7.8,3.2)}{(1,-1)}{5}{3}\put(11,1){\small$i_{3}$}\node{(13,-2)}\end{picture}}
\newcommand{\sTildeGGGGGGa}{\begin{picture}(4,17)\cadre{4}{17}\rnode{(1,16)}\arrow{(1,15)}{(0,-1)}{5}{3}\put(1.5,12){\small$i_{5}$}\node{(1,10)}\arrow{(1,9)}{(0,-1)}{5}{3}\put(1.5,6){\small$i_{4}$}\node{(1,4)}\arrow{(1,3)}{(0,-1)}{5}{3}\put(1.5,0){\small$i_{3}$}\node{(1,-2)}\arrow{(1,-3)}{(0,-1)}{5}{3}\put(1.5,-6){\small$i_{2}$}\node{(1,-8)}\arrow{(1,-9)}{(0,-1)}{5}{3}\put(1.5,-12){\small$i_{1}$}\node{(1,-14)}\end{picture}}
\newcommand{\sTildeGGGGGGaa}{\begin{picture}(14,14)\cadre{14}{14}\rnode{(7,13)}\arrow{(7,12)}{(0,-1)}{5}{3}\put(7.5,9){\small$i_{5}$}\node{(7,7)}\arrow{(7,6)}{(0,-1)}{5}{3}\put(7.5,3){\small$i_{4}$}\node{(7,1)}\arrow{(7,0)}{(0,-1)}{5}{3}\put(7.5,-3){\small$i_{3}$}\node{(7,-5)}\arrow{(6.2,-5.8)}{(-1,-1)}{5}{3}\put(0.5,-8){\small$i_{1}$}\node{(1,-11)}\arrow{(7.8,-5.8)}{(1,-1)}{5}{3}\put(10.5,-8){\small$i_{2}$}\node{(13,-11)}\end{picture}}
\newcommand{\sTildeGGGGGGaaa}{\begin{picture}(14,11)\cadre{14}{11}\rnode{(7,10)}\arrow{(7,9)}{(0,-1)}{5}{3}\put(7.5,6){\small$i_{5}$}\node{(7,4)}\arrow{(7,3)}{(0,-1)}{5}{3}\put(7.5,0){\small$i_{4}$}\node{(7,-2)}\arrow{(6,-2)}{(-1,0)}{5}{3}\put(2,-5){\small$i_{1}$}\node{(1,-2)}\arrow{(7,-3)}{(0,-1)}{5}{3}\put(7.5,-6){\small$i_{2}$}\node{(7,-8)}\arrow{(8,-2)}{(1,0)}{5}{3}\put(11,-5){\small$i_{3}$}\node{(13,-2)}\end{picture}}
\newcommand{\sTildeGGGGGGaaaa}{\begin{picture}(16,9)\cadre{16}{9}\node{(8,1)}\arrow{(8.8,0.2)}{(1,-1)}{5}{3}\arrow{(7.2,0.2)}{(-1,-1)}{5}{3}\arrow{(7,1.5)}{(-2,1)}{5}{3}\arrow{(8,7)}{(0,-1)}{5}{3}\arrow{(9,1.5)}{(2,1)}{5}{3}\put(9.5,-5){\small$i_{3}$}\put(4,-5){\small$i_{2}$}\put(2,0.5){\small$i_{1}$}\put(8.5,5){\small$i_{5}$}\put(12,0.5){\small$i_{4}$}\node{(14,-5)}\node{(2,-5)}\node{(1,4.5)}\rnode{(8,8)}\node{(15,4.5)}\end{picture}}
\newcommand{\sTildeGGGGGGb}{\begin{picture}(16,8)\cadre{16}{8}\rnode{(8,1)}\arrow{(8.8,1.8)}{(1,1)}{5}{3}\arrow{(7.2,1.8)}{(-1,1)}{5}{3}\arrow{(7,0.5)}{(-2,-1)}{5}{3}\arrow{(8,0)}{(0,-1)}{5}{3}\arrow{(9,0.5)}{(2,-1)}{5}{3}\put(9,5.5){\small$i_{1}$}\put(4,5.5){\small$i_{2}$}\put(2,0){\small$i_{3}$}\put(8.5,-4){\small$i_{4}$}\put(11,0){\small$i_{5}$}\node{(14,7)}\node{(2,7)}\node{(1,-2.5)}\node{(8,-6)}\node{(15,-2.5)}\end{picture}}
\newcommand{\sTildeGGGGGGbb}{\begin{picture}(14,11)\cadre{14}{11}\rnode{(7,4)}\arrow{(8,4)}{(1,0)}{5}{3}\put(9,5){\small$i_{5}$}\node{(13,4)}\arrow{(7,5)}{(0,1)}{5}{3}\put(3.5,7){\small$i_{4}$}\node{(7,10)}\arrow{(6,4)}{(-1,0)}{5}{3}\put(1.5,1){\small$i_{3}$}\node{(1,4)}\arrow{(7,3)}{(0,-1)}{5}{3}\put(7.5,0){\small$i_{2}$}\node{(7,-2)}\arrow{(7,-3)}{(0,-1)}{5}{3}\put(7.5,-6){\small$i_{1}$}\node{(7,-8)}\end{picture}}
\newcommand{\sTildeGGGGGGbbb}{\begin{picture}(14,14)\cadre{14}{14}\rnode{(7,7)}\arrow{(7.8,7.8)}{(1,1)}{5}{3}\put(10,8){\small$i_{5}$}\node{(13,13)}\arrow{(6.2,7.8)}{(-1,1)}{5}{3}\put(2,8){\small$i_{4}$}\node{(1,13)}\arrow{(7,6)}{(0,-1)}{5}{3}\put(7.5,3){\small$i_{3}$}\node{(7,1)}\arrow{(7,0)}{(0,-1)}{5}{3}\put(7.5,-3){\small$i_{2}$}\node{(7,-5)}\arrow{(7,-6)}{(0,-1)}{5}{3}\put(7.5,-9){\small$i_{1}$}\node{(7,-11)}\end{picture}}
\newcommand{\sTildeGGGGGGbbbb}{\begin{picture}(16,14)\cadre{16}{14}\rnode{(7,13)}\arrow{(6.2,12.2)}{(-1,-1)}{5}{3}\put(0,9.5){\small$i_{5}$}\node{(1,7)}\arrow{(7.8,12.2)}{(1,-1)}{5}{3}\put(10.5,9.5){\small$i_{4}$}\node{(13,7)}\arrow{(13,6)}{(0,-1)}{5}{3}\put(13.5,3){\small$i_{3}$}\node{(13,1)}\arrow{(13,0)}{(0,-1)}{5}{3}\put(13.5,-3){\small$i_{2}$}\node{(13,-5)}\arrow{(13,-6)}{(0,-1)}{5}{3}\put(13.5,-9){\small$i_{1}$}\node{(13,-11)}\end{picture}}
\newcommand{\sTildeGGGGGGGGGy}{\begin{picture}(15,14)\cadre{15}{14}\put(5.5,8.8){\circle{10.6}}\rnode{(5.5,12.5)}\arrow{(5.05279,11.60557)}{(-1,-2)}{3}{1.7}\arrow{(5.94721,11.60557)}{(1,-2)}{3}{1.7}\put(0.5,8.5){\small$i_{1}$}\put(8,8.5){\small$i_{2}$}\node{(2.5,6.5)}\node{(8.5,6.5)}\arrow{(2.5,6.5)}{(0,-1)}{6}{4}\arrow{(8.5,6.5)}{(0,-1)}{6}{4}\put(-0.4,2.5){\small$i_{3}$}\put(9.3,2.6){\small$i_{4}$}\node{(2.5,0.5)}\put(8.5,-3.2){\circle{10.6}}\node{(8.5,0.5)}\arrow{(5.5,-5.5)}{(+1,+2)}{3}{2}\arrow{(8.5,0.5)}{(1,-2)}{3}{2}\put(3.5,-3.3){\small$i_{5}$}\put(10.8,-3.3){\small$i_{6}$}\node{(5.5,-5.5)}\node{(11.5,-5.5)}\arrow{(5.5,-5.5)}{(0,-1)}{6}{4}\arrow{(11.5,-11.5)}{(0,+1)}{6}{4}\put(2.4,-9.5){\small$i_{7}$}\put(12.1,-9.6){\small$i_{8}$}\node{(5.5,-11.5)}\node{(11.5,-11.5)}\end{picture}}
\newcommand{\sTildeGGGGGGGGGyy}{\begin{picture}(26,14)\cadre{26}{14}\qbezier(4.5,6.5)(4.5,14)(14,14)\qbezier(14,14)(24,14)(10,0)\qbezier(10,0)(8,-2)(7,-2)\qbezier(7,-2)(4.5,-2)(4.5,6.5)\put(18.5,-3.3){\circle{10.6}}\rnode{(7,6.5)}\arrow{(6.2,5.7)}{(-1,-1)}{5}{3.7}\arrow{(7,5.4)}{(0,-1)}{5}{3.7}\arrow{(7.9,5.8)}{(2,-1)}{10}{7.5}\arrow{(7.8,7.3)}{(1,1)}{5}{3.7}\put(0,3){\small$i_{3}$}\put(7.5,2){\small$i_{1}$}\put(15,3){\small$i_{4}$}\put(6,9){\small$i_{2}$}\node{(1,0.5)}\node{(7,0.5)}\node{(18.5,0.5)}\node{(13,12.5)}\arrow{(15.5,-5.5)}{(+1,+2)}{3}{2}\arrow{(18.5,0.5)}{(1,-2)}{3}{2}\put(13.5,-3.7){\small$i_{5}$}\put(20.8,-3.7){\small$i_{6}$}\node{(15.5,-5.5)}\node{(21.5,-5.5)}\arrow{(15.5,-6.5)}{(0,-1)}{5}{3.7}\arrow{(21.5,-11.5)}{(0,+1)}{5}{3.7}\put(12,-9.5){\small$i_{7}$}\put(22,-9.5){\small$i_{8}$}\node{(15.5,-11.5)}\node{(21.5,-11.5)}\end{picture}}
\newcommand{\sTildeGGGGGGGGGyyy}{\begin{picture}(23,12.5)\cadre{23}{12.5}\put(11.5,3.5){\oval(12,18)}\rnode{(11.5,3.5)}\arrow{(11,4.5)}{(-1,+2)}{2.5}{1.5}\arrow{(12,4.5)}{(+1,+2)}{2.5}{1.5}\arrow{(8.5,-2.5)}{(+1,+2)}{2.5}{2}\arrow{(12,2.5)}{(1,-2)}{2.5}{1.5}\arrow{(10.5,3.5)}{(-1,0)}{10}{8}\arrow{(12.5,3.5)}{(1,0)}{10}{8}\put(6.5,6){\small$i_{1}$}\put(13.8,6){\small$i_{2}$}\put(6.5,-0.7){\small$i_{5}$}\put(13.8,-0.7){\small$i_{6}$}\put(2.5,4.5){\small$i_{3}$}\put(18,4.5){\small$i_{4}$}\node{(8.5,9.5)}\node{(14.5,9.5)}\node{(8.5,-2.5)}\node{(14.5,-2.5)}\node{(1,3.5)}\node{(22,3.5)}\arrow{(8.5,-3.5)}{(0,-1)}{5}{3.7}\arrow{(14.5,-8.5)}{(0,+1)}{5}{3.7}\put(5,-6.5){\small$i_{7}$}\put(15,-6.5){\small$i_{8}$}\node{(8.5,-8.5)}\node{(14.5,-8.5)}\end{picture}}
\newcommand{\sTildeGGGGGGGGGz}{\begin{picture}(17,10)\cadre{17}{10}\qbezier(1.5,7)(8.5,12)(15.5,7)\qbezier(15.5,7)(19,4.5)(13.5,1.2)\qbezier(13.5,1.2)(8.5,-1.8)(3.5,1.2)\qbezier(3.5,1.2)(-2,4.5)(1.5,7)\rnode{(8.5,1)}\arrow{(9.39443,1.44721)}{(2,1)}{6}{4}\arrow{(11.94721,7.89443)}{(-1,-2)}{3}{2.5}\arrow{(8.05279,1.89443)}{(-1,2)}{3}{2}\arrow{(7.60557,1.44721)}{(-2,1)}{6}{4}\arrow{(7.60557,0.55279)}{(-2,-1)}{6}{4}\arrow{(8.05279,0.10557)}{(-1,-2)}{3}{2}\arrow{(8.94721,0.10557)}{(1,-2)}{3}{2}\arrow{(15.39443,-2.44721)}{(-2,+1)}{6}{5}\put(4,3.6){\tiny$i_{1}$}\put(6.3,5.5){\tiny$i_{2}$}\put(8.4,5.9){\tiny$i_{5}$}\put(11.1,4.1){\tiny$i_{6}$}\put(3.6,-2.7){\tiny$i_{3}$}\put(6.3,-4.5){\tiny$i_{4}$}\put(8.6,-4.5){\tiny$i_{7}$}\put(10.8,-2.2){\tiny$i_{8}$}\node{(14.5,4)}\node{(11.5,7)}\node{(5.5,7)}\node{(2.5,4)}\node{(2.5,-2)}\node{(5.5,-5)}\node{(11.5,-5)}\node{(14.5,-2)}\end{picture}}
\newcommand{\sTildeGGGGGGGGGzz}{\begin{picture}(27.5,12)\cadre{27.5}{12}\qbezier(5,7.5)(5,12)(9.5,12)\qbezier(9.5,12)(15.5,12)(22.5,-2)\qbezier(22.5,-2)(25,-9)(16.5,-9)\qbezier(16.5,-9)(6.5,-9)(8.5,-5)\qbezier(8.5,-5)(12.5,3)(9.5,6)\qbezier(9.5,6)(8.5,7)(7.5,6)\qbezier(7.5,6)(5,3.5)(5,7.5)\rnode{(13,7.5)}\arrow{(12,7.5)}{(-1,0)}{5}{3}\put(9,8.5){\small$i_{1}$}\node{(7,7.5)}\arrow{(6,7.5)}{(-1,0)}{5}{3}\put(1.5,8.5){\small$i_{3}$}\node{(1,7.5)}\arrow{(13.5,6.5)}{(1,-2)}{2.5}{1.5}\put(11,4){\small$i_{2}$}\node{(16,1.5)}\arrow{(16.5,2.5)}{(+1,+1)}{6}{4}\arrow{(13,-4.5)}{(+1,+2)}{2.5}{2}\arrow{(16.5,0.5)}{(1,-2)}{2.5}{1.5}\arrow{(15,1.5)}{(-1,0)}{10}{8}\arrow{(27,1.5)}{(-1,0)}{10}{4}\put(21,4.5){\small$i_{4}$}\put(11,-2.7){\small$i_{5}$}\put(18.3,-2.7){\small$i_{6}$}\put(7,2.5){\small$i_{7}$}\put(22.5,-1.5){\small$i_{8}$}\node{(22.5,8.5)}\node{(13,-4.5)}\node{(19,-4.5)}\node{(5.5,1.5)}\node{(26.5,1.5)}\end{picture}}
\newcommand{\sTildeGGGGGGGGGzzz}{\begin{picture}(24,12)\cadre{24}{12}\put(10,6.8){\circle{10.6}}\rnode{(10,10.5)}\arrow{(9.5,9.5)}{(-1,-2)}{2.7}{1.7}\arrow{(10.5,9.5)}{(1,-2)}{2.7}{1.7}\put(5,6.5){\small$i_{1}$}\put(12.5,6.5){\small$i_{2}$}\node{(7,4.5)}\node{(13,4.5)}\arrow{(7,4.5)}{(-1,0)}{6}{4}\arrow{(13,4.5)}{(0,-1)}{6}{4}\put(2.3,1.5){\small$i_{3}$}\put(13.8,0.6){\small$i_{4}$}\node{(1,4.5)}\put(13,-5.2){\circle{10.6}}\node{(13,-1.5)}\arrow{(13,-1.5)}{(-1,0)}{10}{8}\arrow{(10,-7.5)}{(+1,+2)}{3}{2}\arrow{(13,-1.5)}{(1,-2)}{3}{2}\arrow{(23,-1.5)}{(-1,0)}{10}{4}\put(4,-4.5){\small$i_{7}$}\put(8,-5.3){\small$i_{5}$}\put(15.3,-5.3){\small$i_{6}$}\put(20,-4.5){\small$i_{8}$}\node{(3,-1.5)}\node{(10,-7.5)}\node{(16,-7.5)}\node{(23,-1.5)}\end{picture}}

\begin{document}

\title{Tree structures for the current fluctuations in the exclusion process}
\author{Sylvain Prolhac}
\email[]{prolhac@ma.tum.de}
\affiliation{Institut de Physique Th\'eorique,\\
CEA, IPhT, F-91191 Gif-sur-Yvette, France\\
CNRS, URA 2306, F-91191 Gif-sur-Yvette, France\\
Zentrum Mathematik,\\
Technische Universit\"at M\"unchen,\\
D-85747 Garching, Germany}
\date{January 27, 2010}

\begin{abstract}
We consider the asymmetric simple exclusion process on a ring, with an arbitrary asymmetry between the hopping rates of the particles. Using a functional formulation of the Bethe equations of the model, we derive exact expressions for all the cumulants of the current in the stationary state. These expressions involve tree structures with composite nodes. In the thermodynamic limit, three regimes can be observed for the current fluctuations depending on how the asymmetry scales with the size of the system.

\end{abstract}

\maketitle

\begin{section}{Introduction}
The one-dimensional asymmetric simple exclusion process (ASEP) is a stochastic model featuring classical hard-core particles hopping between neighboring sites of a one-dimensional lattice, with an asymmetry between the hopping rates forcing a current of particles to flow through the system. The ASEP thus belongs to the large class of driven diffusive systems \cite{S91.1,SZ95.1,SZ98.1}, which have been playing an important role in the understanding of of out of equilibrium systems. In particular, it is a special case of the Katz-Lebowitz-Spohn model \cite{KLS83.1,KLS84.1} describing a lattice gas with particles subject to nearest-neighbor interactions and driven by an external field. Since it is one of the simplest examples of an interacting particles model with a non equilibrium steady state, the ASEP has been studied much in the past \cite{S70.1,L85.1,D98.1,S01.1,GM06.1}.\\\indent
Various boundary conditions have been considered in the study of the ASEP. Open boundary conditions \cite{DEHP93.1,SD93.1,USW04.1,dGE08.1} have been used to model the coupling of the system to reservoirs of particles. If the particles are interpreted as quanta of energy, the ASEP models heat transport between two heat baths at different temperatures \cite{D07.1}. The ASEP has also been considered on an infinite line \cite{PS00.1,S07.1,F08.2} and on a finite ring \cite{GS92.2,K95.1,P03.1,GM05.1,GM05.2}. Even though the model defined on the infinite line can be seen as the infinite system size limit of the periodic model, these two models behave differently in their respective stationary state. Indeed, the stationary state corresponds to taking the infinite time limit, which does not always commute with the infinite system size limit. In this paper, we will consider only the periodic model on a ring.\\\indent
The ASEP can be related to several other models of statistical physics, in particular models of a fluctuating interface growing through the deposition and evaporation of particles. The deposition of a particle in the growth model corresponds in the ASEP to the move of a particle in the forward direction, while the evaporation of a particle corresponds in the ASEP to the move of a particle in the backward direction. Thus, the fluctuations of the integrated current in the ASEP are related in the corresponding growth model to the fluctuations of the height of the growing interface. The symmetric exclusion process (SSEP), for which the forward and backward hopping rates are equal, corresponds to a fluctuating interface with the same deposition and evaporation rates. It is described at large scales by the Edwards-Wilkinson equation \cite{EW82.1}. On the other hand, the asymmetric exclusion process, with different forward and backward hopping rates, corresponds to a growing interface described at large scales by the Kardar-Parisi-Zhang equation \cite{KPZ86.1}. The SSEP can then be seen as discrete version of a system evolving by the Edwards-Wilkinson equation, while the ASEP is a discrete version of the Kardar-Parisi-Zhang equation.\\\indent
The fluctuations of the current of particles is an important quantity for the exclusion process. In the case of the periodic system on a ring, these fluctuations have been calculated using mainly two methods: the matrix Ansatz \cite{BE07.1}, first introduced for the ASEP in \cite{DEHP93.1} for the calculation of the stationary measure of the open system, and the Bethe Ansatz, which relies on the underlying \textit{quantum-integrability} of the model. The diffusion constant, which characterizes the average deviation of the current from its mean value, has initially been calculated using the matrix Ansatz: first in \cite{DEM93.1} for the the totally asymmetric model (TASEP), for which the particles hop only in the forward direction, and then in \cite{DM97.1} for the more general case of the partially asymmetric model. Higher cumulants of the current have been obtained subsequently using the Bethe Ansatz. In \cite{ADLvW08.1}, the thermodynamic limit of all the cumulants was obtained for the symmetric model. Before that, in \cite{DL98.1,DA99.1}, finite size expressions have been calculated for all the cumulants of the current in the totally asymmetric case. In \cite{LK99.1}, this result was generalized in the thermodynamic limit to the partially asymmetric case with non-vanishing asymmetry. However, this result did not allow to study precisely the transition between the symmetric and the totally asymmetric regimes, as this transition occurs in a scaling where the asymmetry vanishes in the thermodynamic limit. It was thus needed to find finite size expressions for the cumulants of the current as this would allow to study all the possible scalings for the asymmetry. A first step in this direction was to recover the exact expression for the partially asymmetric diffusion constant from the Bethe Ansatz, using a functional formulation of the Bethe equations \cite{PM08.1}. This calculation was then extended to the third cumulant of the current in \cite{P08.1}, and all the cumulants of the current were obtained in \cite{PM09.1} for the weakly asymmetric model (WASEP), for which the asymmetry scales as the inverse of the size of the system.\\\indent
In this article, we obtain finite size expressions for all the cumulants of the current in the model with arbitrary asymmetry, generalizing the known results for the three first cumulants, as well as the result for all the cumulants in the totally asymmetric model. We write the cumulants of the current in terms of sums over tree sets. These expressions for the cumulants, that we conjectured earlier in \cite{P09.1}, are proved here using functional Bethe Ansatz equations. From these exact expressions, we study the thermodynamic limit of the cumulants of the current in various scalings for the asymmetry. We observe in particular that the Edwards-Wilkinson and the Kardar-Parisi-Zhang regimes are separated by an intermediate regime corresponding to an asymmetry large with respect to the inverse of the size of the system but small with respect to the square root of the inverse of the size of the system.\\\indent
The article is organized as follows. In section \ref{Section current fluctuations}, we summarize the main results of this paper for the cumulants of the current. In section \ref{Section Bethe Ansatz}, we recall the Bethe Ansatz for the exclusion process and prove that the solution of the functional Bethe equation corresponding to the stationary state is solution of a simpler functional equation. Then, in section \ref{Section tree structures for w(t)}, we show that this simpler functional equation gives rise to tree structures. In section \ref{Section expansion t} we express the solution of the functional Bethe equation in terms of these tree structures. In section \ref{Section E(gamma) parametric}, we write a parametric expression for the generating function of the cumulants of the current. Finally, in section \ref{Section E(gamma) explicit}, we reduce this parametric expression to an explicit expression for the cumulants of the current involving forest structures. A few technical calculations are relegated to the appendices.
\end{section}

\begin{section}{Current fluctuations in the asymmetric exclusion process}
\label{Section current fluctuations}
We consider the asymmetric simple exclusion process on a ring of size $L$ with $n$ particles hopping locally both one site forward (with rate $p$, \textit{i.e.} the probability of hopping is $p\,dt$ in any infinitesimal time interval of length $dt$) and backward (with rate $q$). By the exclusion rule, the particles are only allowed to hop if the destination site is empty. This dynamics is represented in fig. \ref{fig rates ASEP ring} for a configuration with three particles on sixteen sites. In all this article, we will use the notation
\begin{equation}
x=\frac{q}{p}\;
\end{equation}
for the ratio of the hopping rates. The symmetric model then corresponds to $x=1$, while the totally asymmetric model corresponds to $x=0$. In the following, we will call $1-x$ the \textit{asymmetry} between the hopping rates.\\\indent
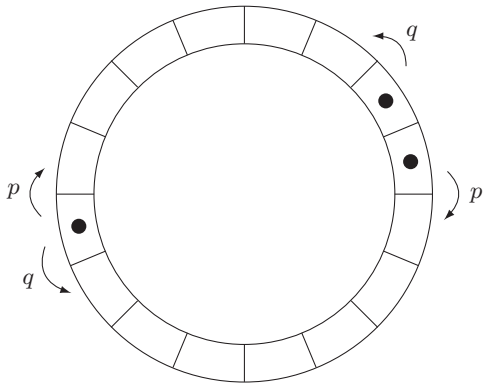
\begin{figure}
\begin{center}
\begin{picture}(70,50)
\put(35,25){\circle{40}}\put(35,25){\circle{50}}
\put(35,45){\line(0,1){5}}\put(42.6,43.48){\line(5,12){1.92}}\put(49.2,39.2){\line(1,1){3.54}}\put(53.48,32.6){\line(12,5){4.62}}\put(55,25){\line(1,0){5}}\put(53.48,17.4){\line(12,-5){4.62}}\put(49.2,10.8){\line(1,-1){3.54}}\put(42.6,6.52){\line(5,-12){1.92}}\put(35,5){\line(0,-1){5}}\put(27.4,6.52){\line(-5,-12){1.92}}\put(20.8,10.8){\line(-1,-1){3.54}}\put(16.52,17.4){\line(-12,-5){4.62}}\put(15,25){\line(-1,0){5}}\put(16.52,32.6){\line(-12,5){4.62}}\put(20.8,39.2){\line(-1,1){3.54}}\put(27.4,43.48){\line(-5,12){1.92}}
\put(53.8,37.4){\circle*{2}}\put(57.1,29.3){\circle*{2}}\put(13,20.7){\circle*{2}}
\qbezier(56.5,42)(56.5,46)(52.5,46)\put(52.5,46){\vector(-1,0){0.5}}\qbezier(62,28)(65,25)(62,22)\put(62,22){\vector(-1,-1){0.5}}\qbezier(8.5,18)(7,13.5)(11.5,12)\put(11.5,12){\vector(9,-3){0.5}}\qbezier(8,22)(5,25)(8,28)\put(8,28){\vector(1,1){0.5}}
\put(56.5,46){$q$}\put(65,24.5){$p$}\put(5.5,13){$q$}\put(3.5,25){$p$}
\end{picture}
\caption{Transition rates for the asymmetric exclusion process on a ring. The particles move one site forward with rate $p$ and one site backward with rate $q$ if the destination site is empty.}
\label{fig rates ASEP ring}
\end{center}
\end{figure}
In this section, we will recall known results about the current fluctuations in the stationary state of the asymmetric exclusion process on a ring. We will then summarize the results that will be obtained in the rest of the article.

\begin{subsection}{Cumulants of the stationary state current}
We call $Y_{t}$ the total distance covered by all the particles between time $0$ and time $t$. Each time a particle moves forward, $Y_{t}$ increases by $1$, while each time a particle moves backward, $Y_{t}$ decreases by $1$. The quantity $Y_{t}$ is thus the integrated current of all the particles. In the long time limit, the probability to observe a value $j$ of the current different from its mean value vanishes exponentially fast \cite{D07.1} as
\begin{equation}
P\left(\frac{Y_{t}}{t}=j\right)\sim e^{-tG(j)}\;.
\end{equation}
The function $G$ is called the large deviations function of the current. Introducing a fugacity $\gamma$ associated to $Y_{t}$, we define the Legendre transform $E(\gamma)$ of $G(j)$ by
\begin{equation}
E(\gamma)=\max_{j\in\mathbb{R}}[j\gamma-G(j)]\;.
\end{equation}
The long time behavior of the mean value of $e^{\gamma Y_{t}}$ is then given in terms of $E(\gamma)$ by the relation
\begin{equation}
\langle e^{\gamma Y_{t}}\rangle\sim e^{E(\gamma)t}\;.
\end{equation}
The previous equation means that $E(\gamma)$ is the exponential generating function of the cumulants of the current in the stationary state. If we write its perturbative expansion near $\gamma=0$ as
\begin{equation}
E(\gamma)=J(x)\gamma+\frac{D(x)}{2!}\gamma^{2}+\frac{E_{3}(x)}{3!}\gamma^{3}+\frac{E_{4}(x)}{4!}\gamma^{4}+\ldots\;,
\end{equation}
then the derivatives of $E(\gamma)$ at $\gamma=0$ give
\begin{align}
\label{J[Y]}
J(x)=&\lim_{t\to\infty}\frac{\langle Y_{t}\rangle}{t}\\
\label{D[Y]}
D(x)=&\lim_{t\to\infty}\frac{\langle Y_{t}^{2}\rangle-\langle Y_{t}\rangle^{2}}{t}=\lim_{t\to\infty}\frac{\left\langle(Y_{t}-\langle Y_{t}\rangle)^{2}\right\rangle}{t}\\
\label{E3[Y]}
E_{3}(x)=&\lim_{t\to\infty}\frac{\langle Y_{t}^{3}\rangle-3\langle Y_{t}\rangle\langle Y_{t}^{2}\rangle+2\langle Y_{t}\rangle^{3}}{t}=\lim_{t\to\infty}\frac{\left\langle(Y_{t}-\langle Y_{t}\rangle)^{3}\right\rangle}{t}\\
\label{E4[Y]}
E_{4}(x)=&\lim_{t\to\infty}\frac{\langle Y_{t}^{4}\rangle-4\langle Y_{t}\rangle\langle Y_{t}^{3}\rangle-3\langle Y_{t}^{2}\rangle^{2}+12\langle Y_{t}\rangle^{2}\langle Y_{t}^{2}\rangle-6\langle Y_{t}\rangle^{4}}{t}\\
&=\lim_{t\to\infty}\frac{\left\langle(Y_{t}-\langle Y_{t}\rangle)^{4}\right\rangle-3\left\langle(Y_{t}-\langle Y_{t}\rangle)^{2}\right\rangle^{2}}{t}\;.\nonumber
\end{align}
Here, $J(x)$ is the mean value of the current, $D(x)$ the diffusion constant, and $E_{3}(x)$ and $E_{4}(x)$ respectively the third and the fourth cumulant of the current.\\\indent
The mean value of the current $J(x)$ can be calculated using the stationary measure of the exclusion process on a ring \cite{D98.1} $P_{\text{stat}}(\mathcal{C})=1/\C{L}{n}$ for any configuration $\mathcal{C}$ of the system. This leads to
\begin{equation}
\label{J(x)}
\frac{J(x)}{p}=(1-x)\frac{n(L-n)}{L-1}\;.
\end{equation}
The higher cumulants of the current are more difficult to obtain, as they involve correlation functions at different times and not just equal-time correlation functions. In particular, they can not be obtained knowing only the stationary measure. The diffusion constant $D(x)$ was first calculated by Derrida and Mallick in \cite{DM97.1} using an extension of the matrix Ansatz for the stationary state. It was then recovered by Bethe Ansatz in \cite{PM08.1}. It is given by
\begin{equation}
\label{D(x)}
\frac{D(x)}{p}=\frac{2(1-x)L}{L-1}\sum_{k=1}^{\infty}k^{2}\frac{1+x^{k}}{1-x^{k}}\frac{\C{L}{n+k}\C{L}{n-k}}{\C{L}{n}^{2}}\;.
\end{equation}
The third cumulant of the current was then obtained in \cite{P08.1} using again Bethe Ansatz. The exact result for $E_{3}(x)$ can be written as
\begin{align}
\label{E3(x)}
\frac{(L-1)E_{3}(x)}{p(1-x)L^{2}}=&\frac{1}{6}\sum_{i\in\mathbb{Z}}\sum_{j\in\mathbb{Z}}\left(i^{2}+ij+j^{2}\right)\frac{\C{L}{n+i}\C{L}{n+j}\C{L}{n-i-j}}{\C{L}{n}^{3}}\\
&-\frac{3}{2}\sum_{i\in\mathbb{Z}}\sum_{j\in\mathbb{Z}}\left(i^{2}+ij+j^{2}\right)\frac{\C{L}{n+i}\C{L}{n+j}\C{L}{n-i-j}}{\C{L}{n}^{3}}\frac{1+x^{|i|}}{1-x^{|i|}}\frac{1+x^{|j|}}{1-x^{|j|}}\nonumber\\
&+\frac{3}{2}\sum_{i\in\mathbb{Z}}\sum_{j\in\mathbb{Z}}\left(i^{2}+j^{2}\right)\frac{\C{L}{n+i}\C{L}{n-i}\C{L}{n+j}\C{L}{n-j}}{\C{L}{n}^{4}}\frac{1+x^{|i|}}{1-x^{|i|}}\frac{1+x^{|j|}}{1-x^{|j|}}\;,\nonumber
\end{align}
where all the $(1+x^{|0|})/(1-x^{|0|})$ must be replaced by an arbitrary constant $\lambda$ which cancels from the result.
\end{subsection}

\begin{subsection}{Combinatorial formula for the cumulants of the current}
In this article, we will obtain finite size expressions (\ref{E(gamma) forests}) for all the cumulants of the current, generalizing the previous expressions (\ref{D(x)}) and (\ref{E3(x)}) of the diffusion constant and the third cumulant. After solving the functional Bethe equation at all order in the fugacity $\gamma$, we will prove that the $k$-th cumulant $E_{k}(x)$ has the following structure:
\begin{equation}
\label{Ek structure}
\frac{E_{k}(x)}{p}=\frac{1-x}{L-1}\left(-\frac{L}{2}\right)^{k-1}\sum_{h\in\mathcal{H}_{k-1}}\frac{W(h)}{S_{f}(h)}\;.
\end{equation}
In this expression, $\mathcal{H}_{k-1}$ is a particular set of forests (a forest being a set of trees) that will be defined in section \ref{Section def forests}. The first sets $\mathcal{H}_{k}$ are given by
\begin{align}
&\mathcal{H}_{1}=\left\{\left[\GGa\right]\right\}\\
&\mathcal{H}_{2}=\left\{\left[\GGGa\right],\left[\GGGb\right],\left[\begin{array}{c}\GGa\vspace{1mm}\\\GGa\end{array}\right]\right\}\\
&\mathcal{H}_{3}=\left\{\left[\GGGGa\right],\left[\GGGGb\right],\left[\GGGGc\right],\left[\begin{array}{c}\GGGa\vspace{1mm}\\\GGa\end{array}\right],\left[\begin{array}{c}\GGGb\vspace{2mm}\\\GGa\end{array}\right]\right\}\\
&\mathcal{H}_{4}=\left\{\left[\GGGGGa\right],\left[\GGGGGb\right],\left[\GGGGGc\right],\left[\GGGGGd\right],\left[\GGGGGe\right],\right.\nonumber\\
&\qquad\qquad\left.\begin{picture}(0,10)\end{picture}\left[\GGGGGf\right],\left[\begin{array}{c}\GGGGa\vspace{1mm}\\\GGa\end{array}\right],\left[\begin{array}{l}\GGGGb\vspace{2mm}\\\GGa\end{array}\right],\left[\begin{array}{c}\GGGGc\vspace{3mm}\\\GGa\end{array}\right],\left[\begin{array}{c}\GGGa\vspace{1mm}\\\GGGa\end{array}\right],\right.\nonumber\\
&\qquad\left.\begin{picture}(0,10)\end{picture}\left[\begin{array}{c}\GGGb\vspace{3mm}\\\GGGa\end{array}\right],\left[\GGGb\;\;\GGGb\right],\left[\begin{array}{c}\GGGa\vspace{1mm}\\\GGa\quad\GGa\end{array}\right],\left[\GGGb\;\begin{array}{c}\GGa\vspace{1mm}\\\GGa\end{array}\right],\left[\begin{array}{c}\GGa\;\;\GGa\vspace{1mm}\\\GGa\;\;\GGa\end{array}\right]\right\}\;.
\end{align}
In equation (\ref{Ek structure}), the rational number $S_{f}(h)$ is a symmetry factor associated to the forest $h$, while $W(h)$ is equal to $k-1$ nested sums over the integers of product of binomial coefficients and factors of the form
\begin{equation}
\frac{1+x^{|i|}}{1-x^{|i|}}\;.
\end{equation}
For example, for the fourth cumulant of the current, we will find
\begin{align}
\label{E4(x)}
\frac{(L-1)E_{4}(x)}{p(1-x)L^{3}}=\sum_{i\in\mathbb{Z}}&\sum_{j\in\mathbb{Z}}\sum_{k\in\mathbb{Z}}\left(3\frac{1+x^{|i|}}{1-x^{|i|}}\frac{1+x^{|j|}}{1-x^{|j|}}\frac{1+x^{|k|}}{1-x^{|k|}}-\frac{1+x^{|i|}}{1-x^{|i|}}\right)\nonumber\\
&\times\left(i^{2}+ij+j^{2}+jk+k^{2}\right)\frac{\C{L}{n+i}\C{L}{n-i-j}\C{L}{n+j+k}\C{L}{n-k}}{\C{L}{n}^{4}}\nonumber\\
+\sum_{i\in\mathbb{Z}}&\sum_{j\in\mathbb{Z}}\sum_{k\in\mathbb{Z}}\left(\frac{1+x^{|i|}}{1-x^{|i|}}\frac{1+x^{|j|}}{1-x^{|j|}}\frac{1+x^{|k|}}{1-x^{|k|}}\right)\nonumber\\
&\times\left(i^{2}+j^{2}+k^{2}+ij+ik+jk\right)\frac{\C{L}{n+i}\C{L}{n+j}\C{L}{n+k}\C{L}{n-i-j-k}}{\C{L}{n}^{4}}\nonumber\\
+\sum_{i\in\mathbb{Z}}&\sum_{j\in\mathbb{Z}}\sum_{k\in\mathbb{Z}}\left(\frac{1+x^{|k|}}{1-x^{|k|}}-9\frac{1+x^{|i|}}{1-x^{|i|}}\frac{1+x^{|j|}}{1-x^{|j|}}\frac{1+x^{|k|}}{1-x^{|k|}}\right)\nonumber\\
&\times\left(i^{2}+ij+j^{2}+k^{2}\right)\frac{\C{L}{n+i}\C{L}{n+j}\C{L}{n-i-j}\C{L}{n+k}\C{L}{n-k}}{\C{L}{n}^{5}}\nonumber\\
+\sum_{i\in\mathbb{Z}}&\sum_{j\in\mathbb{Z}}\sum_{k\in\mathbb{Z}}\left(5\frac{1+x^{|i|}}{1-x^{|i|}}\frac{1+x^{|j|}}{1-x^{|j|}}\frac{1+x^{|k|}}{1-x^{|k|}}\right)\nonumber\\
&\times\left(i^{2}+j^{2}+k^{2}\right)\frac{\C{L}{n+i}\C{L}{n-i}\C{L}{n+j}\C{L}{n-j}\C{L}{n+k}\C{L}{n-k}}{\C{L}{n}^{6}}\;.
\end{align}
In the totally asymmetric limit $x=0$, this expression leads to (using some binomial coefficient formulas, see \textit{e.g.} the appendix A of \cite{P08.1})
\begin{equation}
\frac{E_{4}(x=0)}{p}=\frac{n(L-n)L^{3}}{L-1}\left(\frac{18}{4L-1}\,\frac{\C{4L}{4n}}{\C{L}{n}^{4}}-\left(\frac{24}{3L-1}+\frac{8}{2L-1}\right)\frac{\C{2L}{2n}\C{3L}{3n}}{\C{L}{n}^{5}}+\frac{15}{2L-1}\,\frac{\C{2L}{2n}^{3}}{\C{L}{n}^{6}}\right)\;,
\end{equation}
which agrees with the exact solution \cite{DL98.1} of Derrida and Lebowitz.
\end{subsection}

\begin{subsection}{Three regimes for the current fluctuations}
\label{Section 3 regimes}
In the symmetric exclusion process ($x=1$), the forward and backward hopping rates are equal and the system on a ring reaches an equilibrium stationary state in the long time limit. On the contrary, the partially asymmetric exclusion process ($x\neq1$) reaches in the long time limit a non equilibrium stationary state. These two system thus belong to different universality classes. It is known \cite{GS92.2} that at large scales, the dynamics of the symmetric model can be described by the Edwards-Wilkinson (EW) equation \cite{EW82.1}, while the dynamics of the partially asymmetric model can be described by the Kardar-Parisi-Zhang (KPZ) equation \cite{KPZ86.1}.\\\indent
The crossover between the EW and the KPZ regimes can be studied more precisely by looking at models for which the asymmetry $1-x$ scales in the large system size limit as $1-x\sim1/L^{r}$ for a positive real number $r$. It can be expected that for large values of $r$, the model lies in the EW regime, while for small values of $r$, the model lies in the KPZ regime. Finding the values of $r$ for which the system belongs to the EW or KPZ regimes is thus a natural question.\\\indent
A heuristic argument \cite{PM09.1} indicates that both values $r=1$ and $r=1/2$ correspond to a natural separation between a weakly and a strongly asymmetric model. This tends to show that there are in fact three distinct regimes for the current fluctuations: a regime $1-x\ll1/L$ corresponding to the EW equation, a regime $1-x\gg1/\sqrt{L}$ corresponding to the KPZ equation, and an intermediate regime (I) for which the current fluctuations are neither described by the EW equation nor by the KPZ equation. The scaling $1-x\sim1/L$ is usually called the \textit{weakly asymmetric} scaling. In the following, we will call the scaling $1-x\sim1/\sqrt{L}$ the \textit{strongly asymmetric} scaling.\\\indent
The weakly asymmetric scaling has received much attention recently. In this scaling, a phase transition has been observed by Derrida and Bodineau \cite{BD05.1,BD07.1} using a hydrodynamical approach to the current fluctuations based on the ``macroscopic fluctuations theory'' of Bertini, De Sole, Gabrielli, Jona-Lasinio and Landim \cite{BDSGJLL09.1}. There is a critical value $\nu_{c}$ of the asymmetry such that for $\nu<\nu_{c}$, the generating function of the cumulants of the current $E(\gamma)$ is quadratic when $L\to\infty$, while for $\nu>\nu_{c}$, the function $E(\gamma)$ is not quadratic even when $L\to\infty$. This phase transition corresponds to a change in the density profile adopted by the system when a current different from the mean value of the current $J(x)$ is forced to flow through the system. The phase transition can also be seen from a non analyticity in the generating function of the cumulants of the current in the weakly asymmetric scaling \cite{PM09.1}.\\\indent
The existence of the three regimes EW, I and KPZ can be justified by taking the thermodynamic limit $L,n\to\infty$ with particle density $\rho=n/L$ fixed in the the exact formulas for the cumulants of the current. For the diffusion constant, it was shown in \cite{DM97.1} that the exact formula (\ref{D(x)}) becomes for large systems
\begin{equation}
\label{D regimes}
\left\{\begin{array}{lllll}
\frac{D}{p}\sim\frac{\sqrt{\pi}}{2}(1-x)\rho^{3/2}(1-\rho)^{3/2}L^{3/2}&\;&\text{if $\frac{1}{\sqrt{L}}\ll1-x$}&\;&\begin{array}{c}\text{KPZ regime}\\\text{contains TASEP}\end{array}\\
&&\\
\frac{D}{p}\sim 4\Phi\rho(1-\rho)L\int_{0}^{\infty}du\frac{u^{2}e^{-u^{2}}}{\tanh(\Phi u)}&\;&\text{if $1-x\sim\frac{2\Phi}{\sqrt{\rho(1-\rho)L}}$}&\;&\\
&&\\
\frac{D}{p}\sim 2\rho(1-\rho)L&\;&\text{if $1-x\ll\frac{1}{\sqrt{L}}$}&\;&\begin{array}{c}\text{EW and I regimes}\\\text{contains SSEP}\end{array}
\end{array}\right.\;.
\end{equation}
We observe on these expressions that the behavior of the system is separated in two distinct regimes, depending on whether $1-x$ is large or small with respect to $1/\sqrt{L}$. The regime $1/\sqrt{L}\ll1-x$ is the KPZ regime, which contains in particular the totally asymmetric model. The regime $1-x\ll1/\sqrt{L}$ correspond to the reunion of the EW and I regimes, and contains the symmetric model. We note that the diffusion constant keeps the same value in the EW and I regimes. It becomes larger in the KPZ regime, where it depends on the asymmetry.\\\indent
For the third cumulant of the current, it was shown in \cite{P08.1} that in the thermodynamic limit
\begin{equation}
\label{E3 regimes}
\left\{\begin{array}{lll}
\frac{E_{3}}{p}\sim-(1-x)\left(\frac{8\pi}{3\sqrt{3}}-\frac{3\pi}{2}\right)\rho^{2}(1-\rho)^{2}L^{3}&\;&\text{if $\frac{1}{\sqrt{L}}\ll1-x$ (KPZ regime)}\\
&&\\
\frac{E_{3}}{p}\sim2\Phi h_{3}(\Phi)\rho^{3/2}(1-\rho)^{3/2}L^{5/2}&\;&\text{if $1-x\sim\frac{2\Phi}{\sqrt{\rho(1-\rho)L}}$}\\
&&\\
\frac{E_{3}}{p}\sim-\frac{1}{60}(1-x)^{3}\rho^{3}(1-\rho)^{3}L^{4}&\;&\text{if $\frac{1}{L}\ll1-x\ll\frac{1}{\sqrt{L}}$ (I regime)}\\
&&\\
\frac{E_{3}}{p}\sim\nu\rho^{2}(1-\rho)^{2}\left(1-\frac{\nu^{2}}{60}\rho(1-\rho)\right)L&\;&\text{if $1-x\sim\frac{\nu}{L}$}\\
&&\\
\frac{E_{3}}{p}\sim(1-x)\rho^{2}(1-\rho)^{2}L^{2}&\;&\text{if $1-x\ll\frac{1}{L}$ (EW regime)}
\end{array}\right.\;.
\end{equation}
Here, $h_{3}(\Phi)$ is given by the double integral
\begin{equation}
h_{3}(\Phi)=-\frac{2\pi}{3\sqrt{3}}+6\int_{0}^{\infty}\!\!\int_{0}^{\infty}du\,dv\,\frac{(u^{2}+v^{2})e^{-u^{2}-v^{2}}-(u^{2}+uv+v^{2})e^{-u^{2}-uv-v^{2}}}{\tanh(\Phi u)\tanh(\Phi v)}\;.
\end{equation}
Contrary to the case of the diffusion constant, the three regimes EW, I and KPZ give different expressions for the third cumulant of the current. In particular, the third cumulant is positive in the EW regime and becomes negative in the I and KPZ regimes. The presence of three regimes is also observed on the higher cumulants of the current. In \cite{ADLvW08.1,PM09.1}, all the cumulants were obtained in the EW regime, as well as in the weakly asymmetric scaling $1-x\sim1/L$. Previously \cite{DL98.1,DA99.1,LK99.1}, all the cumulants were obtained in the KPZ regime. From the exact expression (\ref{E(gamma) forests}), we will obtain in section \ref{Section cumulants(Phi)} the expression of all the cumulants in the strongly asymmetric scaling $1-x\sim1/\sqrt{L}$. The only regime left is the intermediate regime.\\\indent
We observe that in this regime, the third cumulant of the current is simply given by the limit $\nu\to\infty$ of the expression in the weakly asymmetric scaling, which agrees with the limit $\Phi\to0$ of the expression in the strongly asymmetric scaling \cite{P08.1}. We checked that the limits $\nu\to\infty$ and $\Phi\to0$ give also the same expression in the case of the fourth cumulant. We conjecture that this is true for all the cumulants of the current, which allows to obtain the expression in the intermediate regime by taking the limit $\nu\to\infty$ of the expression in the weakly asymmetric scaling (as it is much more difficult to take the limit $\Phi\to0$ of the expression in the strongly asymmetric scaling).\\\indent
We can finally write the following expressions in the various scaling limits for the higher cumulants of the current $E_{k}$ ($k\geq3$):
\begin{equation}
\label{Ek regimes}
\left\{\begin{array}{ll}
\frac{E_{k}}{p}\sim(1-x)h_{k}(\infty)\rho^{(k+1)/2}(1-\rho)^{(k+1)/2}L^{3(k-1)/2}&\text{if $\frac{1}{\sqrt{L}}\ll1-x$ (KPZ)}\\
&\\
\frac{E_{k}}{p}\sim2\Phi h_{k}(\Phi)\rho^{k/2}(1-\rho)^{k/2}L^{(3k-4)/2}&\text{if $1-x\sim\frac{2\Phi}{\sqrt{\rho(1-\rho)L}}$}\\
&\\
\frac{E_{k}}{p}\sim\frac{B_{2k-2}}{(k-1)!}(1-x)^{k}\rho^{k}(1-\rho)^{k}L^{2k-2}&\text{if $\frac{1}{L}\ll1-x\ll\frac{1}{\sqrt{L}}$ (I)}\\
&\\
\frac{E_{k}}{p}\sim\sum\limits_{j=\lceil k/2\rceil}^{k}\C{j}{k-j}\frac{k!B_{2j-2}}{j!(j-1)!}\nu^{2j-k}\rho^{j}(1-\rho)^{j}L^{k-2}&\text{if $1-x\sim\frac{\nu}{L}$}\\
&\\
\frac{E_{k}}{p}\sim\left|\begin{array}{ll}\frac{k!B_{k-2}}{\left(\frac{k}{2}\right)!\left(\frac{k-2}{2}\right)!}(\rho(1-\rho))^{k/2}L^{k-2}&(\text{$k$ even})\\&\\\frac{1-x}{2}\,\frac{(k+1)!B_{k-1}}{\left(\frac{k+1}{2}\right)!\left(\frac{k-1}{2}\right)!}(\rho(1-\rho))^{(k+1)/2}L^{k-1}&(\text{$k$ odd})\end{array}\right.&\text{if $1-x\ll\frac{1}{L}$ (EW)}
\end{array}\right.\;.
\end{equation}
The $B_{j}$ are the Bernoulli numbers, and the functions $h_{k}$ are given in equation (\ref{hr(Phi)}) by $k-1$ nested integrals. There are two distinct expressions for $E_{k}$ in the EW regime depending on the parity of $k$. This is related to the fact that for the symmetric model, the symmetry between the forward and the backward directions on the ring forces the odd cumulants of the current to be equal to zero, unlike the even cumulants.
\end{subsection}

\end{section}

\begin{section}{Bethe Ansatz for the asymmetric exclusion process}
\label{Section Bethe Ansatz}
In this section, we write the functional Bethe equation whose polynomial solution $Q$ gives access to the cumulants of the stationary state current in the exclusion process. Then, we recall the construction due to Pronko and Stroganov of a polynomial $P$ solution of a ``dual'' functional Bethe equation. Combining the polynomials $P$ and $Q$ into a function $w$, we finally obtain a closed equation for the unknown quantity $w$. This last equation will be solved perturbatively to all order in the fugacity $\gamma$ in the next sections.

\begin{subsection}{Reminder of the functional formulation of the Bethe equations}
The generating function of the cumulants of the current $E(\gamma)$ is the eigenvalue with largest real part of a deformation $M(\gamma)$ of the Markov matrix of the system \cite{DL98.1,PM08.1}. The matrix $M(\gamma)$ is related to the Hamiltonian of the XXZ spin chain defined in terms of the usual spin $1/2$ operators by
\begin{equation}
H_{\text{XXZ}}=-\frac{1}{2}\sum_{i=1}^{L}\left(S_{i}^{(x)}S_{i+1}^{(x)}+S_{i}^{(y)}S_{i+1}^{(y)}+\Delta S_{i}^{(z)}S_{i+1}^{(z)}\right)\;,
\end{equation}
with the ``twisted'' boundary condition ($2S^{(\pm)}\equiv S^{(x)}\pm iS^{(y)}$)
\begin{equation}
S_{L+1}^{(+)}=\left(\frac{e^{\gamma}}{\sqrt{x}}\right)^{L}S_{1}^{(+)}\qquad\qquad S_{L+1}^{(-)}=\left(\frac{\sqrt{x}}{e^{\gamma}}\right)^{L}S_{1}^{(-)}\qquad\qquad S_{L+1}^{(z)}=S_{1}^{(z)}\;,
\end{equation}
and with a parameter $\Delta$ given by
\begin{equation}
\Delta=\frac{\sqrt{x}+\sqrt{x^{-1}}}{2}\geq1\;.
\end{equation}
More precisely, $H_{\text{XXZ}}$ is related to $M(\gamma)$ through the similarity transformation (see \textit{e.g.} \cite{GM06.1} for the case $\gamma=0$; the generalization to nonzero $\gamma$ is straightforward)
\begin{equation}
\label{HXXZ <-> M(gamma)}
H_{\text{XXZ}}\sim-\frac{1}{p\sqrt{x}}M(\gamma)-L\frac{\Delta}{2}\openone\;.
\end{equation}
Since the Hamiltonian $H_{\text{XXZ}}$ is known to be integrable, the matrix $M(\gamma)$ is diagonalizable using the Bethe Ansatz. We note that $E(\gamma)$, which is the largest eigenvalue of $M(\gamma)$, also gives the ground state of the (non-hermitian) quantum Hamiltonian $H_{\text{XXZ}}$.\\\indent
Using the Bethe Ansatz, each eigenvalue $E$ of $M(\gamma)$ can be expressed in terms of a solution of the \textit{Bethe equations} of the system
\begin{equation}
\label{EB[y]}
e^{L\gamma}\left(\frac{1-y_{i}}{1-xy_{i}}\right)^{L}=-\prod_{j=1}^{n}\frac{y_{i}-xy_{j}}{xy_{i}-y_{j}}\;
\end{equation}
as
\begin{equation}
\label{E[y]}
\frac{E}{p}=(1-x)\sum_{i=1}^{n}\left(\frac{1}{1-y_{i}}-\frac{1}{1-xy_{i}}\right)\;.
\end{equation}
The Bethe equations (\ref{EB[y]}) form a set of $n$ coupled polynomial equations for the quantities $y_{1}$, \ldots, $y_{n}$. Different solutions of this set of equations give through (\ref{E[y]}) different eigenvalues of $M(\gamma)$. Introducing the polynomial
\begin{equation}
Q(t)=\prod_{j=1}^{n}(t-y_{j})\;,
\end{equation}
the Bethe equations can be rewritten \cite{PM08.1} as the functional equation
\begin{equation}
\label{EB[Q,R]}
Q(t)R(t)=e^{L\gamma}(1-t)^{L}Q(xt)+x^{n}(1-xt)^{L}Q(t/x)\;,
\end{equation}
where the two unknown polynomials $Q$ and $R$ are of respective degrees $n$ and $L$. This functional equation is the scalar version of Baxter's equation \cite{B82.1}. As the Bethe equations (\ref{EB[y]}), it has several solutions corresponding to different eigenstates of the matrix $M(\gamma)$. The solution corresponding to the largest eigenvalue of $M(\gamma)$ is the only one such that
\begin{equation}
\label{Q(gamma=0)}
Q(t)=t^{n}+\O{\gamma}\;.
\end{equation}
This solution also verifies \cite{PM08.1}
\begin{equation}
\label{Q(1)}
\frac{x^{n}Q(1/x)}{Q(1)}=e^{n\gamma}\;.
\end{equation}
The corresponding eigenvalue $E(\gamma)$ is given by
\begin{equation}
\label{E[Q]}
\frac{E(\gamma)}{p}=-(1-x)\frac{d}{dt}\log\left(\frac{x^{n}Q(t/x)}{Q(t)}\right)_{|t=1}\;.
\end{equation}
The functional equation (\ref{EB[Q,R]}) has been used in \cite{P08.1} to calculate the three first cumulants of the current. In the present article, we will show that another functional equation, equivalent to (\ref{EB[Q,R]}), is more suitable to calculate systematically all the cumulants of the current.
\end{subsection}

\begin{subsection}{Functional equation ``beyond the equator''}
Given a polynomial $Q$ of degree $n\leq L/2$ solution of the functional Bethe equation (\ref{EB[Q,R]}), it is possible to construct a polynomial $P$ of degree $L-n$ such that
\begin{equation}
\label{P(0)}
P(0)=1\;,
\end{equation}
and
\begin{equation}
\label{EB[P,Q]}
\boxed{\boxed{(1-x^{n}e^{-L\gamma})Q(0)(1-t)^{L}=Q(t)P(t/x)-x^{n}e^{-L\gamma}Q(t/x)P(t)}}\;.
\end{equation}
The construction of the polynomial $P$, due to Pronko and Stroganov \cite{PS99.1}, is explained in appendix \ref{Appendix proof EB[P,Q]}. As the functional Bethe equation (\ref{EB[Q,R]}), equation (\ref{EB[P,Q]}) has several solutions corresponding to several eigenvalues of the deformed Markov matrix $M(\gamma)$. In the following, we will calculate the solution $Q(t)$ corresponding to the largest eigenvalue $E(\gamma)$ by solving (\ref{EB[P,Q]}) instead of (\ref{EB[Q,R]}). From equations (\ref{Q(gamma=0)}), (\ref{P(0)}) and (\ref{EB[P,Q]}), this solution is characterized by $Q(t)=t^{n}+\O{\gamma}$ and
\begin{equation}
\label{P(gamma=0)}
P(t)=1+\O{\gamma}\;.
\end{equation}
We note that using (\ref{EB[P,Q]}) to replace $Q(xt)$ and $Q(t/x)$ in equation (\ref{EB[Q,R]}) by expressions in terms of both polynomials $P$ and $Q$ gives an equation in which all the $Q(t)$ cancel:
\begin{equation}
\label{EB[P,R]}
P(t)R(t)=x^{n}(1-t)^{L}P(xt)+e^{L\gamma}(1-xt)^{L}P(t/x)\;.
\end{equation}
From this last equation, we observe that $(P,x^{L-n}e^{-L\gamma}R)$ is in fact solution of the initial functional Bethe equation (\ref{EB[Q,R]}) with $n$ replaced by $L-n$ and $\gamma$ replaced by $\log x-\gamma$.
\end{subsection}

\begin{subsection}{Resolution of the functional Bethe equation}
\label{Section solution functional Bethe equation}
The functional equation (\ref{EB[P,Q]}) depends on two unknown polynomials $P$ and $Q$, which makes it difficult to solve directly. In this section, we will gather the polynomials $P$ and $Q$ corresponding to the stationary state into a unique function $w$ which preserves all the information from both $P$ and $Q$. We will then rewrite the functional equation (\ref{EB[P,Q]}) as a closed equation (\ref{w[X[w]]}) for the quantity $w$. This equation, which was not known before, is the key to a systematic calculation of all the cumulants of the stationary state current in the exclusion process on a ring.\\\indent
The characterization (\ref{Q(gamma=0)}) and (\ref{P(gamma=0)}) of the polynomials $P$ and $Q$ corresponding to the stationary state makes it natural to solve the functional equation (\ref{EB[P,Q]}) perturbatively near $\gamma=0$. We write the perturbative expansion of $Q(t)$ and $P(t)$ near $\gamma=0$ as
\begin{align}
&Q(t)=t^{n}+Q_{1}(t)\gamma+Q_{2}(t)\gamma^{2}+\ldots\\
&P(t)=1+P_{1}(t)\gamma+P_{2}(t)\gamma^{2}+\ldots
\end{align}
Since the term of higher degree of $Q(t)$ is equal to $t^{n}$, all the $Q_{k}(t)$ are polynomials in $t$ of degree $n-1$. Similarly, since $P(0)=1$, all the $P_{k}(0)$ are equal to zero. In the perturbative expansion near $\gamma=0$, the quantity
\begin{equation}
\log\left(\frac{Q(t)}{t^{n}}\right)=\frac{Q_{1}(t)}{t^{n}}\gamma+\left(\frac{Q_{2}(t)}{t^{n}}-\frac{Q_{1}(t)^{2}}{2t^{2n}}\right)\gamma^{2}+\ldots\;
\end{equation}
has thus only strictly negative powers in $t$, while the quantity
\begin{equation}
\log(P(t))=P_{1}(t)\gamma+\left(P_{2}(t)-\frac{P_{1}(t)^{2}}{2}\right)\gamma^{2}+\ldots\;
\end{equation}
has only strictly positive powers in $t$. The fact that both $\log(Q(t)/t^{n})$ and $\log(P(t))$ can be expressed as a sum of powers of $t$ in such a way that no power of $t$ appears in both quantities, which is a consequence of the perturbative expansion near $\gamma=0$, will be crucial in the following. We write
\begin{align}
\label{Q[alphaj]}
\log\left(\frac{Q(t)}{t^{n}}\right)&=\sum_{j<0}[\alpha]_{j}\frac{x^{j}t^{j}}{1-x^{j}}\\
\label{P[betaj]}
\log\left(P(t)\right)&=\sum_{j>0}[\beta]_{j}\frac{x^{j}t^{j}}{1-x^{j}}\;,
\end{align}
where the $[\alpha]_{j}$ and the $[\beta]_{j}$ are formal series in $\gamma$. The sums over $j$ cover respectively the strictly negative integers and the strictly positive integers. At each order in $\gamma$, only a finite number of $j$ give a nonzero contribution to the sums. From the coefficients $[\alpha]_{j}$ and $[\beta]_{j}$ we also define
\begin{align}
\label{alpha[alphaj]}
\alpha(t)&\equiv\sum_{j<0}[\alpha]_{j}t^{j}\\
\label{beta[betaj]}
\beta(t)&\equiv\sum_{j>0}[\beta]_{j}t^{j}\;.
\end{align}
In the previous expression, $\alpha(t)$ has only strictly negative powers in $t$: at each order in $\gamma$, it is a polynomial in $1/t$ without constant term. Similarly, $\beta(t)$ has only strictly positive powers in $t$: at each order in $\gamma$, it is a polynomial in $t$ without constant term. We now introduce the operator $X$ acting on any series with both positive and negative powers in $t$
\begin{equation}
u(t)=\sum_{j}[u]_{j}t^{j}\;
\end{equation}
as
\begin{equation}
\boxed{X[u(t)]=\sum_{j}[u]_{j}t^{j}\frac{1+x^{|j|}}{1-x^{|j|}}}\;,
\end{equation}
with the convention
\begin{equation}
\frac{1+x^{|0|}}{1-x^{|0|}}=\lambda\;
\end{equation}
for an arbitrary constant $\lambda$. Because of the absolute value in its definition, the operator $X$ acts differently on quantities with only negative powers in $t$ and on quantities with only positive powers in $t$. Using (\ref{Q[alphaj]}) and the definition (\ref{alpha[alphaj]}) of $\alpha(t)$, we find in particular the following relations:
\begin{align}
\log\left(\frac{Q(t)}{t^{n}}\right)&=-\frac{\alpha(t)}{2}-\frac{X[\alpha(t)]}{2}\\
\log\left(\frac{x^{n}Q(t/x)}{t^{n}}\right)&=\frac{\alpha(t)}{2}-\frac{X[\alpha(t)]}{2}\;.
\end{align}
Similarly, using (\ref{P[betaj]}), and the definition (\ref{beta[betaj]}) of $\beta(t)$, we obtain
\begin{align}
\log\left(P(t)\right)&=-\frac{\beta(t)}{2}+\frac{X[\beta(t)]}{2}\\
\log\left(P(t/x)\right)&=\frac{\beta(t)}{2}+\frac{X[\beta(t)]}{2}\;.
\end{align}
The four previous equations allow us to express the rhs of the functional equation (\ref{EB[P,Q]}) only in terms of $\alpha(t)$, $\beta(t)$ and the operator $X$: all the dependency on the asymmetry $x$ has been absorbed in these three quantities. The functional equation (\ref{EB[P,Q]}) now becomes
\begin{equation}
\label{F[alpha-beta]=0}
\left(1-x^{n}e^{-L\gamma}\right)Q(0)\frac{(1-t)^{L}}{t^{n}}=e^{-\frac{\alpha(t)-\beta(t)}{2}-\frac{X[\alpha(t)-\beta(t)]}{2}}-e^{-L\gamma+\frac{\alpha(t)-\beta(t)}{2}-\frac{X[\alpha(t)-\beta(t)]}{2}}\;.
\end{equation}
We observe that the previous equation depends on $\alpha$ and $\beta$ only through the difference $\alpha(t)-\beta(t)$. Thus, we define
\begin{equation}
\label{w[alpha,beta]}
w(t)=\frac{\alpha(t)}{2}-\frac{L\gamma}{2}-\frac{\beta(t)}{2}\;.
\end{equation}
Using the notations
\begin{equation}
h(t)=\frac{(1-t)^{L}}{t^{n}}\;,
\end{equation}
and
\begin{equation}
\label{C[gamma,Q(0)]}
\boxed{C=-\frac{e^{\frac{\lambda L\gamma}{2}}(e^{\frac{L\gamma}{2}}-x^{n}e^{-\frac{L\gamma}{2}})}{2}\,Q(0)}\;,
\end{equation}
and recalling that $X[1]=\lambda$, equation (\ref{F[alpha-beta]=0}) finally becomes
\begin{equation}
\label{w[X[w]]}
\boxed{\boxed{w(t)=\arcsinh\left(Ch(t)e^{X[w(t)]}\right)}}\;.
\end{equation}
This closed functional equation depends only on one unknown quantity $w(t)$, unlike the functional equation (\ref{EB[P,Q]}) which involved two unknown polynomials $P$ and $Q$. This is the key to the systematic calculation of all the cumulants of the current that will be performed in the following.\\\indent
We note that in the initial functional equation (\ref{EB[P,Q]}), the fact that $P$ and $Q$ are polynomials was crucial to constrain the set of the solutions: for a given polynomial $P$ of degree $L-n$, there is in general no polynomial $Q$ of degree $n$ such that $P$ and $Q$ form a solution of (\ref{EB[P,Q]}). In equation (\ref{w[X[w]]}) however, we do not need to use the fact that $\alpha$ and $\beta$ are constructed using polynomials anymore: we already used it to prove that $\alpha(t)$ has only negative powers in $t$ and $\beta(t)$ has only positive powers in $t$.
\end{subsection}

\begin{subsection}{Remaining steps for the calculation of the generating function \texorpdfstring{$E(\gamma)$}{E(gamma)}}
Expanding equation (\ref{w[X[w]]}) as a formal series in the parameter $C$, we note that it admits a unique power series $w(t)=\O{C}$ as a solution. At first orders in $C$, this solution can be obtained directly from the expansion in $C$ of (\ref{w[X[w]]}). Up to order $5$ in $C$, we have
\begin{align}
\label{w[h,C] order 5}
w(t)=&hC+hX[h]C^{2}+\left(\frac{1}{2}hX[h]^{2}+hX[hX[h]]-\frac{1}{6}h^{3}\right)C^{3}\nonumber\\
&+\left(\frac{1}{6}hX[h]^{3}+hX[h]X[hX[h]]+\frac{1}{2}hX[hX[h]^{2}]+hX[hX[hX[h]]]-\frac{1}{2}h^{3}X[h]-\frac{1}{6}hX[h^{3}]\right)C^{4}\nonumber\\
&+\left(\frac{1}{24}hX[h]^{4}+\frac{1}{2}hX[hX[h]]^{2}+\frac{1}{2}hX[h]^{2}X[hX[h]]+\frac{1}{2}hX[h]X[hX[h]^{2}]+\frac{1}{6}hX[hX[h]^{3}]\right.\nonumber\\
&\qquad\left.+hX[h]X[hX[hX[h]]]+hX[hX[h]X[hX[h]]]+\frac{1}{2}hX[hX[hX[h]^{2}]]+hX[hX[hX[hX[h]]]]\right.\nonumber\\
&\qquad\left.-\frac{3}{4}h^{3}X[h]^{2}-\frac{1}{2}h^{3}X[hX[h]]-\frac{1}{6}hX[h]X[h^{3}]-\frac{1}{2}hX[h^{3}X[h]]-\frac{1}{6}hX[hX[h^{3}]]+\frac{3}{40}h^{5}\right)C^{5}\nonumber\\
&+\O{C^{6}}\;,
\end{align}
where we wrote $h$ for $h(t)$ to lighten the notations. In section \ref{Section tree structures for w(t)}, we will perform a systematic calculation of $w(t)$ to all order in the parameter $C$. This calculation will involve tree structures. Indeed, we note on the previous equation that each term is of the form a rational number times a function depending on $t$ through $h(t)$ and some nesting of $X$ operators. It is convenient to write such a term as a tree representing the nesting of the $X$ operators. The edges of the tree are labeled by $X$ and the nodes are labeled by $h$ to the power an odd integer. For example, we have the correspondence
\begin{equation}
\label{trees w[h,C] order 5}
\begin{array}{ccc}
-\frac{1}{6}hX[h]X[h^{3}] & \leftrightarrow & -\frac{1}{6} \begin{array}{c}\begin{picture}(20,20)\cadre{20}{20}\edge{(10,15)}{(-1,-2)}{5}\edge{(10,15)}{(1,-2)}{5}\put(9,16){$h$}\put(4,1.5){$h$}\put(14,1.5){$h^{3}$}\put(4,9){$X$}\put(13.5,9){$X$}\end{picture}\end{array}\\
\frac{1}{2}hX[h]^{2}X[hX[h]] & \leftrightarrow & \frac{1}{2} \begin{array}{c}\begin{picture}(30,30)\cadre{30}{30}\edge{(15,25)}{(-1,-1)}{10}\edge{(15,25)}{(0,-1)}{10}\edge{(15,25)}{(1,-1)}{10}\edge{(25,15)}{(0,-1)}{10}\put(14,26){$h$}\put(4,11.5){$h$}\put(14,11.5){$h$}\put(26,14){$h$}\put(24,1.5){$h$}\put(6,19){$X$}\put(15.5,18){$X$}\put(21.5,19){$X$}\put(25.5,8){$X$}\end{picture}\end{array}
\end{array}\;.
\end{equation}
Then, once $w(t)$ is known, $\alpha(t)$ can be obtained by extracting its negative powers in $t$ using (\ref{w[alpha,beta]}). This is done in section \ref{Section expansion t}. But, from (\ref{Q[alphaj]}) and (\ref{alpha[alphaj]}), we have
\begin{equation}
\label{alpha[Q]}
\alpha(t)=\log\left(\frac{x^{n}Q(t/x)}{Q(t)}\right)\;.
\end{equation}
Using (\ref{Q(1)}) and (\ref{E[Q]}), we will thus obtain a parametric expression for $E(\gamma)$ in section \ref{Section E(gamma) parametric}: equation (\ref{E[Q]}) will give $E(\gamma)$ in terms of the parameter $C$, while equation (\ref{Q(1)}) will give $\gamma$ in terms of $C$. The parameter $C$ is of order $\gamma$ from (\ref{C[gamma,Q(0)]}) and (\ref{Q(gamma=0)}). It means that in the parametric expression for $E(\gamma)$, the parameter $C$ can be eliminated order by order in $\gamma$. This elimination is done systematically in section \ref{Section E(gamma) explicit} using the Lagrange inversion formula. It will finally give an explicit expression for all the cumulants of the current in terms of forest structures.
\end{subsection}

\end{section}

\begin{section}{Solution of the closed functional equation in terms of tree structures}
\label{Section tree structures for w(t)}
In this section, we will show that the closed functional equation (\ref{w[X[w]]}) for the quantity $w(t)$ defined in (\ref{w[alpha,beta]}) can be used to obtain an explicit expression of $w(t)$ involving tree structures. We will first define these tree structures and then show how they appear from the functional equation (\ref{w[X[w]]}).

\begin{subsection}{Rooted trees with composite nodes}
\label{Section rooted trees}
We call ``\textit{elementary node}'' an object that will be represented as a dot $\elementaryNode$. The elementary node is the brick with which we will build the tree structures necessary to express the fluctuations of the current in the asymmetric exclusion process. We call generically ``\textit{composite node}'' an object containing an odd number of elementary nodes. In this section, we will consider only composite nodes without internal structure on the elementary nodes they contain. The size $|c|$ of a composite node $c$ is the number of elementary nodes it contains.\\\indent
We define the set $\mathcal{C}$ of composite nodes without internal structure. A composite node $c\in\mathcal{C}$ will be represented by a set of dots corresponding to the elementary nodes it contains, surrounded by a closed line. We have
\begin{equation}
\mathcal{C}=\left\{\Ca,\;\CCCa,\;\CCCCCa,\;\CCCCCCCa,\;\ldots\right\}\;,
\end{equation}
where we have only drawn the four composite nodes of $\mathcal{C}$ up to size $7$.\\\indent
We now build \textit{trees} whose nodes are composite nodes. The size of a tree $g$, which will be noted $|g|$, is defined as the sum of the sizes of the composite nodes of $g$. It is also the total number of elementary nodes contained in the composite nodes of $g$. We define the set $\mathcal{G}^{\circ}$ of the rooted trees with nodes belonging to $\mathcal{C}$. For a tree $g\in\mathcal{G}^{\circ}$, one elementary node is called the root (or elementary root) and will be represented by a small circle $\elementaryRoot$ instead of a black dot. The composite node containing the elementary root is called the composite root. For a tree $g$ element of $\mathcal{G}^{\circ}$, we call $e(g)$ the set of the elementary nodes of $g$ and $c(g)$ the set of the composite nodes of $g$. For a composite node $c\in c(g)$, we define the number $v_{c}$ of the composite nodes of $g$ which are neighbors of $c$ (that is, the number of composite nodes linked to $c$ by an edge).\\\indent
For each positive integer $r$, we also define the set $\mathcal{G}_{r}^{\circ}$ of the trees in $\mathcal{G}^{\circ}$ of size $r$. The five first sets $\mathcal{G}_{r}^{\circ}$ are
\begin{align}
&\mathcal{G}_{1}^{\circ}=\left\{\rootedGa\right\}\qquad\mathcal{G}_{2}^{\circ}=\left\{\rootedGGa\right\}\qquad\mathcal{G}_{3}^{\circ}=\left\{\rootedGGGa,\;\rootedGGGb,\;\rootedGGGc\right\}\nonumber\\
&\mathcal{G}_{4}^{\circ}=\left\{\rootedGGGGa,\;\rootedGGGGb,\;\rootedGGGGc,\;\rootedGGGGd,\;\rootedGGGGe,\;\rootedGGGGf\right\}\nonumber\\
&\mathcal{G}_{5}^{\circ}=\left\{\rootedGGGGGa,\;\rootedGGGGGb,\;\rootedGGGGGc,\;\rootedGGGGGd,\;\rootedGGGGGe,\;\rootedGGGGGf,\;\rootedGGGGGg,\;\right.\nonumber\\
&\quad\left.\begin{picture}(0,19)\end{picture}\rootedGGGGGh,\;\rootedGGGGGi,\;\rootedGGGGGj,\;\rootedGGGGGk,\;\rootedGGGGGl,\;\rootedGGGGGm,\;\rootedGGGGGn,\;\rootedGGGGGo\right\}\;.
\end{align}
The first sets $\mathcal{G}_{r}^{\circ}$ can be explicitly constructed using a computer. For $r$ between $1$ and $16$, we find that the number of trees in $\mathcal{G}_{r}^{\circ}$ is given by
\begin{equation}
\begin{array}{ccccccccccccccccc}
r & 1 & 2 & 3 & 4 & 5 & 6 & 7 & 8 & 9 & 10 & 11 & 12 & 13 & 14 & 15 & 16\\
\operatorname{card}\mathcal{G}_{r}^{\circ} & 1 & 1 & 3 & 6 & 15 & 36 & 94 & 245 & 663 & 1815 & 5062 & 14269 & 40706 & 117103 & 339673 & 991834
\end{array}\;.
\end{equation}
In the following, we will see that the trees of $\mathcal{G}^{\circ}$ correspond to the good tree structures to express $w(t)$.
\end{subsection}

\begin{subsection}{\texorpdfstring{$w(t)$}{w(t)} as a sum over rooted trees}
The expansion of the $\arcsinh$ in equation (\ref{w[X[w]]}) gives
\begin{equation}
\label{w[X[w]] expansion sinh}
w(t)=\sum_{\substack{r=1\\(r\;\text{odd})}}^{\infty}\frac{(-1)^{\frac{r-1}{2}}(r!!)^{2}}{r^{2}(r!)}C^{r}h(t)^{r}e^{rX[w(t)]}\;,
\end{equation}
with $r!!=r\times(r-2)\times(r-4)\times\ldots\times3\times1$ for any positive odd integer $r$. We write the expansion of $w(t)$ as a formal series in $C$
\begin{equation}
\label{w[wk,C]}
w(t)=\sum_{k=1}^{\infty}w_{k}(t)C^{k}\;,
\end{equation}
and we expand $e^{rX[w(t)]}$ in powers of $C$. This expansion generates tree structures, as we have seen in (\ref{w[h,C] order 5}) and (\ref{trees w[h,C] order 5}) up to order $5$ in $C$. In appendix \ref{Appendix w(t) sum rooted trees}, we prove from equation (\ref{w[X[w]] expansion sinh}) that
\begin{equation}
\label{wk[chi,S] rooted}
w_{k}(t)=\sum_{g\in\mathcal{G}_{k}^{\circ}}\frac{\chi(g)}{S_{\circ}(g)}\;.
\end{equation}
This expression involves the set of trees $\mathcal{G}_{k}^{\circ}$ defined previously. The factor $\chi(g)$ can be computed by induction on a tree $g$ using
\begin{equation}
\label{chi(g)}
\boxed{\chi(g)=h(t)^{r}\prod_{\substack{\text{$g'$ subtree of $g$}\\\text{connected to the}\\\text{composite root}}}X[\chi(g')]}\;,
\end{equation}
if the tree $g$ is made of a composite root of size $r$ to which the trees $g'$ are attached (the product is taken to be equal to $1$ if the tree $g$ is made of a single composite node). The symmetry factor $S_{\circ}(g)$ is given by
\begin{equation}
\label{Srooted(g)}
\frac{1}{S_{\circ}(g)}=\left(\prod_{c\in c(g)}\frac{(-1)^{\frac{|c|-1}{2}}(|c|!!)^{2}|c|^{v_{c}}}{|c|^{3}|c|!}\right)\times\frac{\left(\substack{\text{number of equivalent}\\\text{choices of the}\\\text{elementary root of $g$}}\right)}{P(g)}\;,
\end{equation}
where $P(g)$ is the number of permutations of the composite nodes of $g$ leaving $g$ invariant (except for the choice of the composite root). We recall that $c(g)$ is the set of the composite nodes of $g$, and $v_{c}$ the number of composite nodes neighbors of $c$.\\\indent
We note that $S_{\circ}(g)$ depends on the position of the root of $g$ only from the number of equivalent choices of the elementary root. This will allow us to replace the sum over the set $\mathcal{G}_{k}^{\circ}$ of rooted trees in $w_{k}(t)$ by a sum over a set of unrooted trees, whose number is smaller than the number of rooted trees.
\end{subsection}

\begin{subsection}{Unrooted trees with composite nodes}
We call $\mathcal{G}$ the set of unrooted trees whose nodes are composite. The set $\mathcal{G}$ can be obtained from the set $\mathcal{G}^{\circ}$ of rooted trees by replacing the elementary roots of the trees by ordinary elementary nodes. Several rooted trees from $\mathcal{G}^{\circ}$ will then correspond to the same unrooted tree from $\mathcal{G}$.\\\indent
For a tree $g$ element of $\mathcal{G}$, we call again $e(g)$ the set of the elementary nodes of $g$ and $c(g)$ the set of the composite nodes of $g$. For a composite node $c\in c(g)$, we also define the number $v_{c}$ of the composite nodes of $g$ which are neighbors of $c$. We call $\mathcal{G}_{r}$ the set of unrooted trees of size $r$, that is, the set of unrooted trees containing $r$ elementary nodes. The five first sets $\mathcal{G}_{r}$ are
\begin{align}
&\mathcal{G}_{1}=\left\{\Ga\right\}\qquad\mathcal{G}_{2}=\left\{\GGa\right\}\qquad\mathcal{G}_{3}=\left\{\GGGa,\;\GGGb\right\}\nonumber\\
&\mathcal{G}_{4}=\left\{\GGGGa,\;\GGGGb,\;\GGGGc\right\}\nonumber\\
&\mathcal{G}_{5}=\left\{\GGGGGa,\;\GGGGGb,\;\GGGGGc,\;\GGGGGd,\;\GGGGGe,\;\GGGGGf\right\}\;.
\end{align}
In the following, we will use a simplified notation for the trees of $\mathcal{G}$. All the composite nodes which contain only one elementary node will simply be represented as a dot $\elementaryNode$, while the other composite nodes will be represented by their size surrounded by a circle (\textit{e.g.} $\compositeNode{3}$, $\compositeNode{5}$, \ldots). We emphasize that with these simplified notations, a dot $\elementaryNode$ will represent a composite node of size $1$ and not an elementary node. Using the new notations, the six first sets $\mathcal{G}_{r}$ are
\begin{align}
\label{sGr}
&\mathcal{G}_{1}=\left\{\sGa\right\}\qquad\mathcal{G}_{2}=\left\{\sGGa\right\}\qquad\mathcal{G}_{3}=\left\{\sGGGa,\;\sGGGb\right\}\qquad\mathcal{G}_{4}=\left\{\sGGGGa,\;\sGGGGb,\;\sGGGGc\right\}\nonumber\\
&\mathcal{G}_{5}=\left\{\sGGGGGa,\;\sGGGGGb,\;\sGGGGGc,\;\sGGGGGd,\;\sGGGGGe,\;\sGGGGGf\right\}\nonumber\\
&\mathcal{G}_{6}=\left\{\sGGGGGGa\;,\sGGGGGGb\;,\sGGGGGGc\;,\sGGGGGGd\;,\sGGGGGGe\;,\sGGGGGGf\right.\nonumber\\
&\qquad\qquad\qquad\qquad\left.\begin{picture}(0,8)\end{picture}\sGGGGGGg\;,\sGGGGGGh\;,\sGGGGGGi\;,\sGGGGGGj\;,\sGGGGGGk\;,\sGGGGGGl\right\}\;.
\end{align}
The first sets $\mathcal{G}_{r}$ can be explicitly constructed using a computer. For $r$ between $1$ and $16$, we find that the number of trees in $\mathcal{G}_{r}$ is given by
\begin{equation}
\begin{array}{ccccccccccccccccc}
r & 1 & 2 & 3 & 4 & 5 & 6 & 7 & 8 & 9 & 10 & 11 & 12 & 13 & 14 & 15 & 16\\
\operatorname{card}\mathcal{G}_{r} & 1 & 1 & 2 & 3 & 6 & 12 & 25 & 55 & 126 & 304 & 745 & 1893 & 4893 & 12916 & 34562 & 93844
\end{array}\;.
\end{equation}
\end{subsection}

\begin{subsection}{\texorpdfstring{$w(t)$}{w(t)} as a sum over unrooted trees}
We will now rewrite the expression (\ref{wk[chi,S] rooted}) of $w_{k}(t)$ using a sum over $g\in\mathcal{G}_{k}$ instead of the sum over $g\in\mathcal{G}_{k}^{\circ}$. We have
\begin{equation}
w_{k}(t)=\sum_{g\in\mathcal{G}_{k}}\sum_{\substack{g'\in\mathcal{G}_{k}^{\circ}\\\text{$g'$ has the same}\\\text{tree structure as $g$}}}\frac{\chi(g')}{S_{\circ}(g')}\;,
\end{equation}
where the trees $g'$ are all the trees of $\mathcal{G}_{k}^{\circ}$ that can be obtained from $g$ by choosing an elementary root. Calling $g_{j}$ the tree of $\mathcal{G}_{k}^{\circ}$ obtained from $g$ by choosing for elementary root the $j$-th elementary node of $g$ (for some arbitrary order on the elementary nodes of $g$), we can also write
\begin{equation}
w_{k}(t)=\sum_{g\in\mathcal{G}_{k}}\sum_{j=1}^{k}\frac{1}{\left(\substack{\text{number of equivalent}\\\text{choices of the}\\\text{elementary root of $g_{j}$}}\right)}\frac{\chi(g_{j})}{S_{\circ}(g_{j})}\;.
\end{equation}
From the expression (\ref{Srooted(g)}) of the rooted symmetry factor $S_{\circ}(g_{j})$, we define an unrooted symmetry factor $S(g)$ associated to the unrooted tree $g$ by
\begin{equation}
\label{S(g)}
\boxed{\frac{1}{S(g)}=\left(\prod_{c\in c(g)}\frac{(-1)^{\frac{|c|-1}{2}}(|c|!!)^{2}|c|^{v_{c}}}{|c|^{3}|c|!}\right)\times\frac{1}{P(g)}}\;.
\end{equation}
We recall that $c(g)$ is the set of the composite nodes of $g$, $v_{c}$ the number of composite nodes of $g$ neighbors of $c$, and $P(g)$ the number of permutations of the composite nodes of $g$ that leave $g$ invariant. With this unrooted symmetry factor, we finally obtain
\begin{equation}
\label{wk[chi,S] unrooted}
\boxed{\boxed{w_{k}(t)=\sum_{g\in\mathcal{G}_{k}}\frac{1}{S(g)}\sum_{j=1}^{k}\chi(g_{j})}}\;.
\end{equation}
The symmetry factors of the unrooted trees up to size $6$ are given in table \ref{Table symmetry factors}.
\begin{table}
\begin{flushleft}
\begin{tabular}{|c||c|c|cc|ccc|}\hline
&&&&&&&\\
$g\in\mathcal{G}$ & \;\sGa & \;\sGGa & \;\sGGGa & \;\sGGGb & \;\sGGGGa & \;\sGGGGb & \;\sGGGGc\\
&&&&&&&\\
&&&&&&&\\
$S(g)$ & 1 & 2 & 2 & -18 & 2 & 6 & -6\\
&&&&&&&\\
$P(g)$ & 1 & 2 & 2 & 1 & 2 & 6 & 1\\\hline
\end{tabular}\\
\begin{tabular}{|c||cccccc|}\hline
&&&&&&\\
$g\in\mathcal{G}$ & \;\sGGGGGa & \;\sGGGGGb & \;\sGGGGGc & \;\sGGGGGd & \;\sGGGGGe & \;\sGGGGGf\\
&&&&&&\\
&&&&&&\\
$S(g)$ & 2 & 2 & 24 & -6 & -4 & 200/3\\
&&&&&&\\
$P(g)$ & 2 & 2 & 24 & 1 & 2 & 1\\\hline
\end{tabular}\\
\begin{tabular}{|c||cccccc|}\hline
&&&&&&\\
$g\in\mathcal{G}$ & \;\sGGGGGGa & \;\sGGGGGGb & \;\sGGGGGGc & \;\sGGGGGGd & \;\sGGGGGGe & \;\sGGGGGGf\\
&&&&&&\\
&&&&&&\\
$S(g)$ & 2 & 2 & 8 & 2 & 6 & 120\\
&&&&&&\\
$P(g)$ & 2 & 2 & 8 & 2 & 6 & 120\\\hline
\end{tabular}\\
\begin{tabular}{|c||cccccc|}\hline
&&&&&&\\
$g\in\mathcal{G}$ & \;\sGGGGGGg & \;\sGGGGGGh & \;\sGGGGGGi & \;\sGGGGGGj & \;\sGGGGGGk & \;\sGGGGGGl\\
&&&&&&\\
&&&&&&\\
$S(g)$ & -6 & -2 & -12 & -4 & 40/3 & 72\\
&&&&&&\\
$P(g)$ & 1 & 1 & 2 & 6 & 1 & 2\\\hline
\end{tabular}
\caption{Symmetry factors of the trees of $\mathcal{G}$ up to size $6$, as given by equation (\ref{S(g)}).}
\label{Table symmetry factors}
\end{flushleft}
\end{table}
\end{subsection}

\end{section}

\begin{section}{Exact parametric solution of the functional Bethe equation}
\label{Section expansion t}
In this section, we write explicitly the expansion of $w(t)$ in powers of $C$ and $t$, and give a parametric expression of the polynomial $Q(t)$. In order to do that, it is useful to add an internal tree structure to the composite nodes of the trees.

\begin{subsection}{Composite nodes with internal tree structure}
In section \ref{Section rooted trees}, we defined the set $\mathcal{C}$ of composite nodes without internal structure. We now define the set $\widetilde{\mathcal{C}}$ of the composite nodes with an internal unrooted tree structure linking all the elementary nodes contained in the composite node. A composite node $c\in\widetilde{\mathcal{C}}$ will be represented by a set of dots $\elementaryNode$ representing the elementary nodes it contains, linked by edges representing the tree structure on the elementary nodes, and surrounded by a closed line. For the five composite nodes of $\widetilde{\mathcal{C}}$ up to size $5$, we have
\begin{equation}
\widetilde{\mathcal{C}}=\left\{\tildeCa,\;\tildeCCCa,\;\tildeCCCCCa,\;\tildeCCCCCb,\;\tildeCCCCCc,\;\ldots\right\}\;.
\end{equation}
We recall that the size of a composite node (\textit{i.e.} the number of elementary nodes it contains) is required to be an odd integer.\\\indent
We now define the set $\widetilde{\mathcal{G}}$ of the rooted trees with elementary nodes belonging to $\widetilde{\mathcal{C}}$, oriented edges labeled by $i_{1}$, $i_{2}$, \ldots, the edges between composite nodes being identified to edges between elementary nodes belonging to different composite nodes. The labels of the edges of $g\in\widetilde{\mathcal{G}}$ are chosen all different. For $g\in\widetilde{\mathcal{G}}$, one node is called the root (or elementary root) of $g$ and is represented by a small circle $\elementaryRoot$ instead of a dot $\elementaryNode$. Again, the composite node containing the elementary root will be called the composite root. We have
\begin{align}
\label{tildeG}
\widetilde{\mathcal{G}}=&\left\{\tildeGa,\;\tildeGGa,\;\tildeGGaa,\;\tildeGGGa,\;\tildeGGGaa,\;\tildeGGGaaa,\;\tildeGGGaaaa,\right.\nonumber\\
&\nonumber\\
&\qquad\tildeGGGb,\;\tildeGGGbb,\;\tildeGGGbbb,\;\tildeGGGbbbb,\;\tildeGGGbbbbb,\;\tildeGGGbbbbbb,\;\tildeGGGbbbbbbb,\;\tildeGGGbbbbbbbb,\;\tildeGGGc,\;\tildeGGGcc,\nonumber\\
&\nonumber\\
&\nonumber\\
&\left.\begin{picture}(0,10)\end{picture}\tildeGGGccc,\;\tildeGGGcccc,\;\tildeGGGd,\;\tildeGGGdd,\;\tildeGGGddd,\;\tildeGGGdddd,\;\tildeGGGddddd,\;\tildeGGGdddddd,\;\tildeGGGddddddd,\;\tildeGGGdddddddd,\;\ldots\right\}\;,
\end{align}
where we represented all the trees of $\widetilde{\mathcal{G}}$ up to size $3$. We recall that the size of a tree is equal to the number of elementary nodes it contains. As before, we introduce the set $\widetilde{\mathcal{G}}_{r}$ made of the trees of $\widetilde{\mathcal{G}}$ of size $r$. Equivalently, the set $\widetilde{\mathcal{G}}_{r}$ is the set of the trees of $\widetilde{\mathcal{G}}$ with $r-1$ edges. The four first sets $\widetilde{\mathcal{G}}_{r}$ are (again, we do not draw circles around the composite nodes of size $1$ to simplify the notations)
\begin{align}
\label{sTildeGr}
&\widetilde{\mathcal{G}}_{1}=\left\{\sTildeGa\right\}\qquad\qquad\qquad\widetilde{\mathcal{G}}_{2}=\left\{\sTildeGGa,\;\ldots\right\}\nonumber\\
&\widetilde{\mathcal{G}}_{3}=\left\{\sTildeGGGa,\;\sTildeGGGb,\;\sTildeGGGc,\;\sTildeGGGd,\;\ldots\right\}\nonumber\\
&\widetilde{\mathcal{G}}_{4}=\left\{\sTildeGGGGa,\;\sTildeGGGGb,\;\sTildeGGGGc,\;\sTildeGGGGd,\;\sTildeGGGGe,\begin{picture}(0,12)\end{picture}\right.\nonumber\\
&\left.\qquad\sTildeGGGGf,\;\sTildeGGGGg,\;\sTildeGGGGh,\;\sTildeGGGGi,\;\sTildeGGGGj,\;\sTildeGGGGk,\;\ldots\right\}\;.
\end{align}
The \ldots\ represent the trees of $\widetilde{\mathcal{G}}_{r}$ which can be obtained from the ones drawn by permuting the labels of the edges or reversing the directions of the edges.\\\indent
As in the case of the trees of $\mathcal{G}$, we define for a tree $g\in\widetilde{\mathcal{G}}$ the set $e(g)$ of the elementary nodes of $g$ and the set $c(g)$ of the composite nodes of $g$. For example, the four first trees of $\widetilde{\mathcal{G}}_{4}$ drawn in (\ref{sTildeGr}) have four composite nodes (all of size $1$), while the seven following trees have one composite node of size $1$ and one composite node of size $3$.\\\indent
In the following, we will make a distinction between two types of edges in the trees: ``inner edges'', which link elementary nodes belonging to the same composite node, and ``outer edges'' which link elementary nodes belonging to distinct composite nodes. We will also call ``inner labels'' the labels of the inner edges and ``outer labels'' the labels of the outer edges. The set of the outer edges of a tree $g$ will be noted $o(g)$. For example, the four first trees of $\widetilde{\mathcal{G}}_{4}$ drawn in (\ref{sTildeGr}) have three outer edges and no inner edge, while the seven following trees have one outer edge and two inner edges.
\end{subsection}

\begin{subsection}{Function \texorpdfstring{$W_{\varphi}^{\eta,\xi}$}{W} defined on the trees}
\label{Section functions trees}
We will now define functions acting on the sets $\mathcal{G}$ or $\widetilde{\mathcal{G}}$, from which the cumulants of the current will be expressed in the following.\\\indent
Let $g$ be a tree of $\widetilde{\mathcal{G}}$. For an outer edge $o\in o(g)$, the subtree of $g$ beginning at $o$ is the tree made of the edge $o$ and of all the edges which can be reached from $o$ by moving on the edges of $g$ (independently of the direction they are pointing to) away from the elementary root of $g$. We then call $m(o)$ the linear combination (with coefficients $+1$ and $-1$) of the indices $i_{j}$ which label the edges (inner and outer) of the subtree of $g$ beginning at $o$, with coefficients chosen in the following way: the labels of the edges pointing in the direction of the root have a coefficient $-1$ while the labels of the edges pointing away from the root have a coefficient $+1$. We illustrate this definitions on the example of the tree
\begin{equation}
\label{example g element of Gtilde9}
g=\begin{array}{c}\sTildeGGGGGGGGGy\\\\\\\\\end{array}\;.
\end{equation}
This tree $g$ has four outer edges $o_{3}$, $o_{4}$, $o_{7}$ and $o_{8}$, labeled respectively by $i_{3}$, $i_{4}$, $i_{7}$ and $i_{8}$. They are such that $m(o_{3})=i_{3}$, $m(o_{4})=i_{4}-i_{5}+i_{6}+i_{7}-i_{8}$, $m(o_{7})=i_{7}$ and $m(o_{8})=-i_{8}$.\\\indent
We also want to associate a linear combination of the indices $i_{j}$ to each elementary node. For an elementary node $e\in e(g)$ of a tree $g\in\widetilde{\mathcal{G}}$ ($e$ may be the elementary root of $g$), we call $\ell(e)$ the sum of the labels of the edges (inner or outer, linking $e$ to another elementary node) pointing to $e$, minus the sum of the labels of the edges pointing away from $e$. For the nine elementary nodes $e\in e(g)$ of the tree $g$ drawn in (\ref{example g element of Gtilde9}), the $\ell(e)$ are $i_{3}$, $i_{1}-i_{3}$, $-i_{1}-i_{2}$, $i_{2}-i_{4}$, $i_{4}+i_{5}-i_{6}$, $-i_{5}-i_{7}$, $i_{7}$, $i_{6}+i_{8}$ and $-i_{8}$.\\\indent
For $g\in\widetilde{\mathcal{G}}$, we define a tree $g^{*}\in\widetilde{\mathcal{G}}$ by attaching all the edges (inner and outer) of $g$ to the elementary root. The edges keep their labels, their direction toward the elementary root (an edge pointing in the direction of the root in $g$ will still point in the direction of the root in $g^{*}$, and conversely), and their inner or outer nature. All the composite nodes of $g^{*}$ are thus of size $1$ except possibly the composite root. For example, if we consider the tree $g$ drawn in (\ref{example g element of Gtilde9}), we have
\begin{equation}
\label{example g* element of Gtilde9}
g^{*}=\begin{array}{c}\sTildeGGGGGGGGGz\\\\\end{array}\;.
\end{equation}
For this tree $g^{*}$, if we call $o_{3}$, $o_{4}$, $o_{7}$ and $o_{8}$ the outer edges labeled by $i_{3}$, $i_{4}$, $i_{7}$ and $i_{8}$, we have $m(o_{3})=i_{3}$, $m(o_{4})=i_{4}$, $m(o_{7})=i_{7}$ and $m(o_{8})=-i_{8}$. For the nine elementary nodes $e\in e(g^{*})$, we also have the following values for the $\ell(e)$: $i_{1}$, $i_{2}$, $i_{3}$, $i_{4}$, $-i_{5}$, $i_{6}$, $i_{7}$, $-i_{8}$ ($e$ elementary node different from the elementary root of $g^{*}$) and $-i_{1}-i_{2}-i_{3}-i_{4}+i_{5}-i_{6}-i_{7}+i_{8}$ ($e$ elementary root of $g^{*}$).\\\indent
We will now define functions mapping a tree of $\mathcal{G}$ or $\widetilde{\mathcal{G}}$ to a number. For $g\in\widetilde{\mathcal{G}}$, and $\varphi$ and $\eta$ two arbitrary numerical functions, we define
\begin{equation}
\label{U(g)}
\boxed{U_{\varphi,\eta}(g)=\left(\sum_{e\in e(g)}\varphi(\ell(e))\right)\left(\prod_{e\in e(g)}\eta(\ell(e))\right)}\;.
\end{equation}
For the trees $g$ and $g^{*}$ represented respectively in (\ref{example g element of Gtilde9}) and (\ref{example g* element of Gtilde9}), we have for example
\begin{align}
&U_{\varphi,\eta}(g)=[\varphi(i_{3})+\varphi(i_{1}-i_{3})+\varphi(-i_{1}-i_{2})+\varphi(i_{2}-i_{4})+\varphi(i_{4}+i_{5}-i_{6})\nonumber\\
&\qquad\qquad\qquad\qquad\qquad\qquad\qquad\qquad+\varphi(-i_{5}-i_{7})+\varphi(i_{7})+\varphi(i_{6}+i_{8})+\varphi(-i_{8})]\nonumber\\
&\qquad\qquad\times\eta(i_{3})\eta(i_{1}-i_{3})\eta(-i_{1}-i_{2})\eta(i_{2}-i_{4})\eta(i_{4}+i_{5}-i_{6})\eta(-i_{5}-i_{7})\eta(i_{7})\eta(i_{6}+i_{8})\eta(-i_{8})\\
&U_{\varphi,\eta}(g^{*})=[\varphi(i_{1})+\varphi(i_{2})+\varphi(i_{3})+\varphi(i_{4})+\varphi(-i_{5})+\varphi(i_{6})+\varphi(i_{7})\nonumber\\
&\qquad\qquad\qquad\qquad\qquad\qquad+\varphi(-i_{8})+\varphi(-i_{1}-i_{2}-i_{3}-i_{4}+i_{5}-i_{6}-i_{7}+i_{8})]\nonumber\\
&\qquad\qquad\times\eta(i_{1})\eta(i_{2})\eta(i_{3})\eta(i_{4})\eta(-i_{5})\eta(i_{6})\eta(i_{7})\eta(-i_{8})\eta(-i_{1}-i_{2}-i_{3}-i_{4}+i_{5}-i_{6}-i_{7}+i_{8})\;.
\end{align}
By definition, $U_{\varphi,\eta}(g)$ depends only on the tree structure (with directed edges) of the elementary nodes of $g$, and is independent of the position of the elementary root and of the composite nodes in $g$. For an arbitrary numerical function $\xi$, we also define
\begin{equation}
\label{V(g)}
\boxed{V_{\xi}(g)=\prod_{o\in o(g)}\xi(m(o))}\;.
\end{equation}
For a tree $g$ without any outer edge (\textit{i.e.} a tree formed of only one composite node), we take the product in $V_{\xi}(g)$ to be equal to $1$. For the trees $g$ and $g^{*}$ represented in (\ref{example g element of Gtilde9}) and (\ref{example g* element of Gtilde9}), we have
\begin{align}
&V_{\xi}(g)=\xi(i_{3})\xi(i_{4}-i_{5}+i_{6}+i_{7}-i_{8})\xi(i_{7})\xi(-i_{8})\\
&V_{\xi}(g^{*})=\xi(i_{3})\xi(i_{4})\xi(i_{7})\xi(-i_{8})\;.
\end{align}
Unlike $U_{\varphi,\eta}(g)$, $V_{\xi}(g)$ depends on the position of the composite root, of the composite nodes of $g$, and on the orientation of the edges, but not on the internal tree structure of the elementary nodes inside the composite nodes (except for the direction of the edges).\\\indent
We now choose a map $\theta$ from $\mathcal{G}$ to $\widetilde{\mathcal{G}}$ preserving the tree structure of the composite nodes; the map $\theta$ applied on $g\in\mathcal{G}$ roots $g$, adds an internal tree structure on the elementary nodes of each composite node, transforms the edges between composite nodes into edges between elementary nodes, and adds a direction and a label to each edge. Thus, the map $\theta$ changes a tree from $\mathcal{G}$ into a tree from $\widetilde{\mathcal{G}}$.\\\indent
For example, the first tree of $\mathcal{G}_{4}$ drawn in (\ref{sGr}) can be mapped by $\theta$ to the first or the second tree of $\widetilde{\mathcal{G}}_{4}$ drawn in (\ref{sTildeGr}), and also to every tree that can be obtained from it by changing the orientation of the edges and any permutation of the labels. In the same way, the second tree of $\mathcal{G}_{4}$ can be sent by $\theta$ to the third or fourth tree of $\widetilde{\mathcal{G}}_{4}$. Finally, the third tree of $\mathcal{G}_{4}$ can be mapped by $\theta$ to the fifth up to eleventh tree of $\widetilde{\mathcal{G}}_{4}$ drawn in (\ref{sTildeGr}).\\\indent
We now define for $g\in\mathcal{G}_{r}$
\begin{equation}
\label{W(g) tree 1}
\boxed{W_{\varphi}^{\eta,\xi}(g)=\sum_{i_{1}\in\mathbb{Z}}\ldots\!\sum_{i_{r-1}\in\mathbb{Z}}U_{\varphi,\eta}(\theta(g))V_{\xi}(\theta(g)^{*})}\;,
\end{equation}
with the convention $W_{\varphi}^{\eta,\xi}(\elementaryNode)=\varphi(0)\eta(0)$. This definition of $W_{\varphi}^{\eta,\xi}(g)$ is equivalent (see appendix \ref{Appendix definitions W equivalent}) to the definition
\begin{equation}
\label{W(g) tree 2}
\boxed{W_{\varphi}^{\eta,\xi}(g)=\sum_{i_{1}\in\mathbb{Z}}\ldots\!\sum_{i_{r-1}\in\mathbb{Z}}U_{\varphi,\eta}(\theta(g)^{*})V_{\xi}(\theta(g))}\;.
\end{equation}
It can be proved (see appendix \ref{Appendix proof W independent of theta}) that the function $W_{\varphi}^{\eta,\xi}$ does not depend on the choice of the map $\theta$ if the function $\xi$ is even, which will be the case in all the situations we will consider in the next sections. We emphasize that if the function $\eta$ has a finite support, the sums over the indices $i_{1}$, \ldots, $i_{r-1}$ in the definition of $W_{\varphi}^{\eta,\xi}(g)$ are finite sums. In the following, the generating function of the cumulants of the current will be expressed in terms of the function $W_{\varphi}^{\eta,\xi}$.
\end{subsection}

\begin{subsection}{Parametric expression of the polynomial \texorpdfstring{$Q$}{Q}}
In appendix \ref{Appendix expansion w(t) powers t}, we expand the expression (\ref{wk[chi,S] unrooted}) of $w(t)$ in powers of $t$. Then, keeping only the negative powers in $t$, we obtain an expression for the polynomial $Q(t)$ in terms of the function $W_{\varphi}^{\eta,\xi}$ defined previously. We define the functions $\varphi_{l}$, $\eta$ and $\xi_{\lambda}$ by
\begin{align}
\label{phi(z)}
&\varphi_{l}(z)=\frac{(n+z)}{(L-l)}\,\frac{\C{L-n-z}{l}}{\C{L}{l}}\\
\label{eta(z)}
&\eta(z)=\frac{\C{L}{n+z}}{\C{L}{n}}\\
\label{xi(z)}
&\xi_{\lambda}(z)=\left\{\begin{array}{cl}\lambda & \text{if $z=0$}\\\frac{1+x^{|z|}}{1-x^{|z|}} & \text{if $z\neq 0$}\end{array}\right.\;.
\end{align}
Then, the polynomial $Q$ solution of the functional Bethe equation (\ref{EB[Q,R]}) corresponding to the stationary state (\ref{Q(gamma=0)}) is given by
\begin{equation}
\label{A(t) trees}
\boxed{\boxed{\ln\left[\frac{x^{n}Q(t/x)}{Q(t)}\right]=-2\sum_{k=1}^{\infty}\sum_{l=0}^{\infty}\left(\frac{B}{2}\right)^{k}(1-t)^{l}\sum_{g\in\mathcal{G}_{k}}\frac{W_{\varphi_{l}}^{\eta,\xi_{\lambda}}(g)}{S(g)}}}\;,
\end{equation}
where $B$, which depends on the arbitrary parameter $\lambda$ unlike $Q(t)$, is equal to
\begin{equation}
\label{B[Q(0)]}
\boxed{\boxed{B=(-1)^{n-1}\C{L}{n}e^{\frac{\lambda L\gamma}{2}}\left(e^{\frac{L\gamma}{2}}-x^{n}e^{-\frac{L\gamma}{2}}\right)Q(0)}}\;.
\end{equation}
This expression for $Q(t)$ is precisely the one that we conjectured in \cite{P09.1}, where it was checked up to order $5$ in $\gamma$ and $1-t$ for all the systems up to size $L=12$ by using the systematic perturbative solution of the functional Bethe equation (\ref{EB[Q,R]}) presented in \cite{P08.1}.\\\indent
We note that divergences appear in equation (\ref{A(t) trees}) in the terms such that $L\leq l$, because of the denominator of $\varphi_{l}(z)$. An analytical continuation for complex $L$ shows that these divergences cancel in fact in the limit where $L$ becomes an integer as soon as $\varphi_{l}(z)$ is multiplied by $\eta(z)$, which is systematically the case in $W_{\varphi_{l}}^{\eta,\xi_{\lambda}}(g)$. The terms such that $L\leq l$ in (\ref{A(t) trees}) must then be understood by taking the analytic continuation for complex $L$ and by taking the limit $L$ integer.\\\indent
We emphasize that the function $\xi_{\lambda}$ is even. This ensures that the previous expressions do not depend on the choice of the map $\theta$ which enters the definition of $W_{\varphi_{l}}^{\eta,\xi_{\lambda}}(g)$.
\end{subsection}

\end{section}

\begin{section}{Parametric expression of the current fluctuations}
\label{Section E(gamma) parametric}
In this section, we use the solution of the functional Bethe equation derived previously to give a parametric expression of the generating function of the cumulants of the current $E(\gamma)$. Then, we show that this expression allows to recover the known exact result in the totally asymmetric model \cite{DL98.1}, and the result in the partially asymmetric model with non-vanishing asymmetry in the thermodynamic limit \cite{LK99.1}.

\begin{subsection}{Parametric form of the generating function \texorpdfstring{$E(\gamma)$}{E(gamma)}}
\label{Section E(B) gamma(B)}
Taking $t=1$ in equation (\ref{A(t) trees}) gives, using (\ref{Q(1)}), an expression for $\gamma$ in terms of $B$:
\begin{equation}
\gamma=-\frac{2}{n}\sum_{k=1}^{\infty}\left(\frac{B}{2}\right)^{k}\sum_{g\in\mathcal{G}_{k}}\frac{W_{\varphi_{0}}^{\eta,\xi_{\lambda}}(g)}{S(g)}\;.
\end{equation}
The function $\varphi_{0}$ is the affine function
\begin{equation}
\varphi_{0}(z)=\frac{n}{L}+\frac{z}{L}\;.
\end{equation}
In the definition of $U_{\varphi,\eta}(g)$ for a tree $g\in\widetilde{\mathcal{G}}$, a summation over all the elementary nodes $e$ of $g$ of the $\varphi(\ell(e))$ is performed. But, by definition of $\ell(e)$, the summation over $e$ of the $\ell(e)$ is equal to zero. Indeed, for each edge of $g$, there is only one node to which the edge is pointing toward and one node from which it is pointing away, which implies that each index $i_{j}$ appears in one of the $\ell(e)$ with a coefficient $+1$ and in another one with a coefficient $-1$. The linear term of $\varphi$ then does not contribute to $U_{\varphi,\eta}$, nor to $W_{\varphi}^{\eta,\xi_{\lambda}}$. We obtain
\begin{equation}
\label{gamma(B) trees}
\boxed{\boxed{\gamma=-\frac{2}{L}\sum_{k=1}^{\infty}\left(\frac{B}{2}\right)^{k}\sum_{g\in\mathcal{G}_{k}}\frac{W_{z\mapsto 1}^{\eta,\xi_{\lambda}}(g)}{S(g)}}}\;.
\end{equation}
We have written $z\mapsto 1$ for the function equal to $1$ everywhere. Taking now the derivative in $t=1$ of equation (\ref{A(t) trees}), the expression (\ref{E[Q]}) for the generating function of the cumulants of the current implies that
\begin{equation}
\frac{E(\gamma)}{p}=-2(1-x)\sum_{k=1}^{\infty}\left(\frac{B}{2}\right)^{k}\sum_{g\in\mathcal{G}_{k}}\frac{W_{\varphi_{1}}^{\eta,\xi_{\lambda}}(g)}{S(g)}\;.
\end{equation}
The function $\varphi_{1}$ is a polynomial of degree $2$:
\begin{equation}
\varphi_{1}(z)=\frac{n(L-n)}{L(L-1)}+\frac{(L-2n)z}{L(L-1)}-\frac{z^{2}}{L(L-1)}\;.
\end{equation}
As previously for $\varphi_{0}$, the linear term in $\varphi_{1}$ does not contribute to $W_{\varphi_{1}}^{\eta,\xi_{\lambda}}(g)$, whereas the constant term gives again (\ref{gamma(B) trees}). We finally find
\begin{equation}
\label{E(gamma) trees}
\boxed{\boxed{\frac{E(\gamma)-J\gamma}{p}=\frac{2(1-x)}{L(L-1)}\sum_{k=2}^{\infty}\left(\frac{B}{2}\right)^{k}\sum_{g\in\mathcal{G}_{k}}\frac{W_{z\mapsto z^{2}}^{\eta,\xi_{\lambda}}(g)}{S(g)}}}\;.
\end{equation}
Here, $z\mapsto z^{2}$ is the square function, and the mean value of the current $J$ is given by (\ref{J(x)}). Because of the definition of $W$ applied to the only tree of $\mathcal{G}_{1}$, the term $k=1$ does not contribute to the sum in the previous equation.\\\indent
Equations (\ref{gamma(B) trees}) and (\ref{E(gamma) trees}) give a parametric expression of the generating function of the cumulants of the current: the parameter $B$ can be eliminated between both equations to give an expansion of $E(\gamma)$ in powers of $\gamma$ which does not involve $B$. We will see in section \ref{Section E(gamma) explicit} that the elimination of the parameter $B$ can in fact be performed in a systematic way to all orders in $\gamma$.
\end{subsection}

\begin{subsection}{Totally asymmetric limit}
\label{Section TASEP limit}
We note that the parametric expression (\ref{gamma(B) trees}), (\ref{E(gamma) trees}) of the generating function of the cumulants of the current $E(\gamma)$ looks very much like the one obtained by Derrida and Lebowitz \cite{DL98.1} for the totally asymmetric model. We will now prove that the result of \cite{DL98.1} can be recovered from (\ref{gamma(B) trees}) and (\ref{E(gamma) trees}). In the limit $x\to 0$, the function $\xi_{\lambda}$ becomes
\begin{equation}
\left(\xi_{\lambda}\right)_{|x\to 0}(z)=\left|\begin{array}{ll}\lambda&\text{if $z=0$}\\1&\text{if $z\neq 0$}\end{array}\right.\;.
\end{equation}
We choose the arbitrary parameter $\lambda$ to be equal to $1$ so that $\xi_{\lambda}(z)=1$ for each value of $z$. From (\ref{V(g)}), it implies that $V_{\xi_{\lambda}}(\theta(g))$ is equal to $1$ for any tree $g\in\mathcal{G}_{k}$. We also note that if the map $\theta$ defined in section \ref{Section functions trees} is chosen so that all the edges in $\theta(g)$ point away from the elementary root, then $\theta(g)^{*}$ is the same tree for all $g\in\mathcal{G}_{k}$ (we recall that the map $\theta$ transforms a tree of $\mathcal{G}$ into a tree of $\widetilde{\mathcal{G}}$). From (\ref{U(g)}), and with the function $\eta$ defined in (\ref{eta(z)}), we find
\begin{align}
U_{\varphi,\eta}(\theta(g)^{*})=(\varphi(i_{1})+&\ldots+\varphi(i_{k-1})+\varphi(-i_{1}-\ldots-i_{k-1}))\nonumber\\
&\times\frac{\C{L}{n+i_{1}}\ldots\C{L}{n+i_{k-1}}\times\C{L}{n-i_{1}-\ldots-i_{k-1}}}{\C{L}{n}^{k}}\;.
\end{align}
Thus, neither $W_{z\mapsto 1}^{\eta,z\mapsto 1}(g)$ nor $W_{z\mapsto z^{2}}^{\eta,z\mapsto 1}(g)$ depend on the tree $g$ anymore. They can be calculated by considering the generating function $(1-t_{1})^{L}\ldots(1-t_{k})^{L}$. We obtain
\begin{align}
\label{W1eta1}
W_{z\mapsto 1}^{\eta,z\mapsto 1}(g)&=k\frac{\C{kL}{kn}}{\C{L}{n}^{k}}\\
\label{Wz2eta1}
W_{z\mapsto z^{2}}^{\eta,z\mapsto 1}(g)&=\frac{k(k-1)n(L-n)}{kL-1}\,\frac{\C{kL}{kn}}{\C{L}{n}^{k}}\;.
\end{align}
The parametric expression (\ref{gamma(B) trees}), (\ref{E(gamma) trees}) of $E(\gamma)$ for the totally asymmetric model then rewrites
\begin{align}
\gamma=&-\frac{1}{L}\sum_{k=1}^{\infty}\frac{kB^{k}}{2^{k-1}}\,\frac{\C{kL}{kn}}{\C{L}{n}^{k}}\times\left(\sum_{g\in\mathcal{G}_{k}}\frac{1}{S(g)}\right)\\
\frac{E(\gamma)-J\gamma}{p}=&\frac{n(L-n)}{L(L-1)}\sum_{k=2}^{\infty}\frac{k(k-1)B^{k}}{(kL-1)2^{k-1}}\,\frac{\C{kL}{kn}}{\C{L}{n}^{k}}\times\left(\sum_{g\in\mathcal{G}_{k}}\frac{1}{S(g)}\right)\;.
\end{align}
In appendix \ref{Appendix GF trees}, we calculated a generating function for the trees of $\mathcal{G}_{k}$ with the symmetry factor $S$. In $z=1$, equations (\ref{GF trees def}) and (\ref{GF trees}) give the sum over the trees in the previous equation:
\begin{equation}
\sum_{g\in\mathcal{G}_{k}}\frac{1}{S(g)}=\frac{2^{k-1}}{k^{2}}\;.
\end{equation}
Expressing the term $J\gamma$ of $E(\gamma)$ using the expression of $\gamma$ in terms of $B$, we finally obtain
\begin{align}
\gamma=&-\frac{1}{L}\sum_{k=1}^{\infty}\frac{B^{k}}{k}\,\frac{\C{kL}{kn}}{\C{L}{n}^{k}}\\
\frac{E(\gamma)}{p}=&-\frac{n(L-n)}{L}\sum_{k=1}^{\infty}\frac{B^{k}}{kL-1}\,\frac{\C{kL}{kn}}{\C{L}{n}^{k}}\;.
\end{align}
By absorbing into $B$ the binomial coefficient to the power $k$, we recover the result of \cite{DL98.1}.
\end{subsection}

\begin{subsection}{Thermodynamic limit with finite asymmetry}
For a system with particle density $\rho=n/L$ fixed, it is possible to take the thermodynamic limit of the equations (\ref{gamma(B) trees}) and (\ref{E(gamma) trees}) giving $E(\gamma)$ under a parametric form if the asymmetry $1-x$ stays finite when $L\to\infty$. This allows to recover the result obtained by Lee and Kim \cite{LK99.1} in this particular limit. The steps of the derivation are similar to the case of the totally asymmetric limit studied in the previous section. From the large $L$ expansion
\begin{equation}
\eta(z)\sim\left(\frac{1-\rho}{\rho}\right)^{z}e^{-\frac{z^{2}+(1-2\rho)z}{2\rho(1-\rho)L}}\;,
\end{equation}
and from the expression (\ref{W(g) tree 2}) of $W_{\varphi}^{\eta,\xi}(g)$, the parametric expression (\ref{gamma(B) trees}) and (\ref{E(gamma) trees}) for $E(\gamma)$ becomes for large $L$
\begin{align}
\gamma&\sim-\frac{2}{L}\sum_{k=1}^{\infty}\left(\frac{B}{2}\right)^{k}\sum_{g\in\mathcal{G}_{k}}\frac{W_{z\mapsto1,z\mapsto e^{-\frac{z^{2}}{2\rho(1-\rho)L}}}^{\;z\mapsto\xi_{\lambda}(z)}(g)}{S(g)}\\
\frac{E(\gamma)-J\gamma}{p-q}&\sim\frac{2}{L^{2}}\sum_{k=2}^{\infty}\left(\frac{B}{2}\right)^{k}\sum_{g\in\mathcal{G}_{k}}\frac{W_{z\mapsto z^{2},z\mapsto e^{-\frac{z^{2}}{2\rho(1-\rho)L}}}^{\;z\mapsto\xi_{\lambda}(z)}(g)}{S(g)}\;.
\end{align}
We used the fact that the terms in $e^{z}$ of $\eta(z)$ cancel in $W_{\varphi}^{\eta,\xi}(g)$, which is similar to the cancellation of the linear term of $\varphi$ that we used in section \ref{Section E(B) gamma(B)}.\\\indent
Because of the $-z^{2}/L$ in the exponentials, the indices $i_{j}$ in $W$ contributing the most to the sums are of order $\sqrt{L}$. Thus, we set $i_{j}=u_{j}\sqrt{L\rho(1-\rho)}$. The Riemann sums over the indices $i_{j}$ in $W$ can now be rewritten as integrals. We note from equation (\ref{xi(z)}) that if $\lambda$ is equal to $1$ and $1-x$ remains strictly between $0$ and $1$ in the limit $L\to\infty$, we can replace $\xi_{\lambda}(z)$ by $1$. It implies that we can replace $V_{\xi_{\lambda}}(\theta(g))$ by $1$ in the Riemann integrals, using the definition (\ref{W(g) tree 2}) of $W$. We also note that if the map $\theta$ is chosen so that all the edges of $\theta(g)$ point away from the elementary root, then $U_{\varphi,\eta}(\theta(g)^{*})$ does not depend on the tree $g$ anymore. It means that we can use the generating function of the trees (\ref{GF trees}) as in the case of the totally asymmetric limit. We obtain
\begin{align}
\gamma&\sim-\sum_{k=1}^{\infty}\frac{B^{k}(\rho(1-\rho))^{\frac{k-1}{2}}L^{\frac{k-3}{2}}}{k}\int_{-\infty}^{\infty}du_{1}\ldots\int_{-\infty}^{\infty}du_{k-1}\nonumber\\
&\qquad\qquad\qquad\qquad\qquad\qquad\qquad\qquad\qquad\qquad e^{-(u_{1}^{2}+\ldots+u_{k-1}^{2}+u_{1}u_{2}+u_{1}u_{3}+\ldots+u_{k-2}u_{k-1})}\\
\frac{E(\gamma)-J\gamma}{p-q}&\sim\sum_{k=2}^{\infty}\frac{B^{k}(\rho(1-\rho))^{\frac{k+1}{2}}L^{\frac{k-3}{2}}}{k^{2}}\int_{-\infty}^{\infty}du_{1}\ldots\int_{-\infty}^{\infty}du_{k-1}\nonumber\\
2(u_{1}^{2}&+\ldots+u_{k-1}^{2}+u_{1}u_{2}+u_{1}u_{3}+\ldots+u_{k-2}u_{k-1})e^{-(u_{1}^{2}+\ldots+u_{k-1}^{2}+u_{1}u_{2}+u_{1}u_{3}+\ldots+u_{k-2}u_{k-1})}\;.
\end{align}
We can now perform the integration over the indices $u_{j}$. Setting finally $D=-B\sqrt{2\pi\rho(1-\rho)L}$, we find the result of Lee and Kim \cite{LK99.1} at first order in $L$:
\begin{align}
\gamma&\sim-\sqrt{\frac{1}{2\pi\rho(1-\rho)L^{3}}}\sum_{k=1}^{\infty}\frac{(-D)^{k}}{k^{3/2}}\\
\frac{E(\gamma)-J\gamma}{p-q}&\sim\sqrt{\frac{\rho(1-\rho)}{2\pi L^{3}}}\sum_{k=1}^{\infty}\left(\frac{(-D)^{k}}{k^{3/2}}-\frac{(-D)^{k}}{k^{5/2}}\right)\;.
\end{align}
In this limit, the dependency in the asymmetry in $E(\gamma)$ reduces to the global factor $p-q=p(1-x)$.
\end{subsection}

\end{section}

\begin{section}{Explicit expression for the cumulants of the current}
\label{Section E(gamma) explicit}
We wrote in the previous section a parametric expression for the generating function of the cumulants of the current $E(\gamma)$ of the partially asymmetric model. This expression involved sums over tree sets. In this section, we will give another expression for $E(\gamma)$, equivalent to the previous one, as an explicit formal series in $\gamma$. It will involve sums over forest sets. This new expression for $E(\gamma)$ will give explicit expressions for the cumulants of the current. In particular, we will recover the known results for the three first cumulants.

\begin{subsection}{Forests with composite nodes}
\label{Section def forests}
We call ``\textit{forest}'' an acyclic graph whose connected components are trees. For a forest $h$, we will note $\overline{h}$ the number of trees in $h$ and $|h|$ the size of $h$, defined to be the sum of the sizes of all the trees it contains. The size of a forest is thus equal to the sum of the sizes of all its composite nodes. It is also equal to the total number of elementary nodes it contains. A forest will be represented as a list of trees surrounded by square brackets $[\quad]$. We will consider in the following two sets of forests.\\\indent
We first define the set $\mathcal{H}$ of the forests made of trees elements of $\mathcal{G}$ of size larger or equal to $2$. We call $\mathcal{H}_{r}$ the subset of $\mathcal{H}$ made of the forests $h$ such that $|h|-\overline{h}=r$. The four first sets $\mathcal{H}_{r}$ are
\begin{align}
&\mathcal{H}_{1}=\left\{\left[\sGGa\right]\right\}\\
&\mathcal{H}_{2}=\left\{\left[\sGGGa\right],\;\left[\sGGGb\right],\;\left[\sGGa,\sGGa\right]\right\}\nonumber\\
&\mathcal{H}_{3}=\left\{\left[\sGGGGa\right],\;\left[\sGGGGb\right],\;\left[\sGGGGc\right],\;\left[\sGGa,\sGGGa\right],\;\left[\sGGa,\sGGGb\right],\;\left[\sGGa,\sGGa,\sGGa\right]\right\}\nonumber\\
&\mathcal{H}_{4}=\left\{\left[\sGGGGGa\right],\;\left[\sGGGGGb\right],\;\left[\sGGGGGc\right],\;\left[\sGGGGGd\right],\;\left[\sGGGGGe\right],\;\left[\sGGGGGf\right],\right.\nonumber\\
&\qquad\qquad\left.\left[\sGGa,\sGGGGa\right],\;\left[\sGGa,\sGGGGb\right],\;\left[\sGGa,\sGGGGc\right],\;\left[\sGGGa,\sGGGa\right],\;\left[\sGGGa,\sGGGb\right],\right.\nonumber\\
&\qquad\qquad\left.\begin{picture}(0,7)\end{picture}\left[\sGGGb,\sGGGb\right],\;\left[\sGGa,\sGGa,\sGGGa\right],\;\left[\sGGa,\sGGa,\sGGGb\right],\;\left[\sGGa,\sGGa,\sGGa,\sGGa\right]\right\}\;.\nonumber
\end{align}
The first sets $\mathcal{H}_{r}$ can be generated using a computer. The number of forests of the thirteen first sets $\mathcal{H}_{r}$ is given in the following table:
\begin{equation}
\begin{array}{cccccccccccccc}
r & 1 & 2 & 3 & 4 & 5 & 6 & 7 & 8 & 9 & 10 & 11 & 12 & 13\\
\operatorname{card}\mathcal{H}_{r} & 1 & 3 & 6 & 15 & 33 & 80 & 186 & 454 & 1109 & 2772 & 7012 & 18053 & 47066
\end{array}\;.
\end{equation}
We also define the set $\widetilde{\mathcal{H}}$ of the forests made of trees belonging to $\widetilde{\mathcal{G}}$ of size larger or equal to $2$, with edges relabeled by $i_{1}$, $i_{2}$, \ldots\ so that all the labels are different. We call $\widetilde{\mathcal{H}}_{r}$ the subset of $\widetilde{\mathcal{H}}$ made of the forests $h$ such that $|h|-\overline{h}=r$. We note that for $h\in\widetilde{\mathcal{H}}_{r}$, the number of edges in $h$ is precisely equal to $r$, as in each tree of $\widetilde{\mathcal{G}}$ the number of edges is one less than the number of elementary nodes. On the contrary, for $h\in\mathcal{H}_{r}$, the number $r$ is not equal to the number of edges of $h$ since there is no edge inside the composite nodes. The two first sets $\widetilde{\mathcal{H}}_{r}$ are
\begin{align}
&\widetilde{\mathcal{H}}_{1}=\left\{\left[\sTildeGGa\right],\;\ldots\right\}\qquad\text{and}\qquad\widetilde{\mathcal{H}}_{2}=\left\{\left[\sTildeGGGa\right],\;\left[\sTildeGGGb\right],\;\left[\sTildeGGGc\right],\;\left[\sTildeGGGd\right],\;\left[\sTildeGGa,\sTildeGGaa\right],\;\ldots\right\}\;,
\end{align}
where the \ldots\ represent the forests obtained from the ones drawn by permuting the indices $i_{j}$ and by changing the directions of the edges.\\\indent
As in the case of the trees, we define for a forest $h$ of $\mathcal{H}$ or $\widetilde{\mathcal{H}}$ the set $c(h)$ of the composite nodes of $h$, the set $e(h)$ of the elementary nodes of $h$, and the set $o(h)$ of the outer edges of $h$ (\textit{i.e.} the set of the edges of $h$ linking two elementary nodes from different composite nodes, and not the edges linking two elementary nodes of the same composite node).
\end{subsection}

\begin{subsection}{Function \texorpdfstring{$W_{\varphi}^{\eta,\xi}$}{W} defined on the forests}
Most of the functions which were defined previously on the trees of $\mathcal{G}$ and of $\widetilde{\mathcal{G}}$ in section \ref{Section functions trees} naturally extend to forests. We choose a map $\theta$ from $\mathcal{H}$ to $\widetilde{\mathcal{H}}$ which preserves the forest structure of the composite nodes. The map $\theta$ transforms a forest $h\in\mathcal{H}_{r}$ into a forest $h'\in\widetilde{\mathcal{H}}_{r}$. It roots each tree of $h$, adds an internal tree structure to the elementary nodes inside each composite node, transforms the edges between composite nodes into edges between elementary nodes, and adds a direction and a label to all of the edges.\\\indent
The operator $*$ applied on a forest $h\in\widetilde{\mathcal{H}}$ is defined as the forest constituted of the trees of $h$ to which the operator $*$ has been applied to: the forest $h^{*}$ is thus made of trees having all their elementary nodes linked to the elementary root.\\\indent
The functions $U_{\varphi,\eta}$ and $V_{\xi}$ can now be extended naturally to the forests of $\widetilde{\mathcal{H}}$. For arbitrary numerical functions $\varphi$, $\eta$ and $\xi$, we define
\begin{align}
\label{U(h)}
&U_{\varphi,\eta}(h)=\left(\sum_{e\in e(h)}\varphi(\ell(e))\right)\left(\prod_{e\in e(h)}\eta(\ell(e))\right)\\
\label{V(h)}
&V_{\xi}(h)=\prod_{o\in o(h)}\xi(m(o))\;.
\end{align}
The function $W_{\varphi}^{\eta,\xi}$ also extends naturally to the forests. For $h\in\mathcal{H}_{r}$, we define
\begin{equation}
\label{W(h) forest 1}
\boxed{W_{\varphi}^{\eta,\xi}(h)=\sum_{i_{1}\in\mathbb{Z}}\ldots\!\sum_{i_{r}\in\mathbb{Z}}U_{\varphi,\eta}(\theta(h))V_{\xi}(\theta(h)^{*})}\;.
\end{equation}
For the same reason as in the case of the trees, we have another equivalent definition for $W_{\varphi}^{\eta,\xi}$ applied to a forest:
\begin{equation}
\label{W(h) forest 2}
\boxed{W_{\varphi}^{\eta,\xi}(h)=\sum_{i_{1}\in\mathbb{Z}}\ldots\!\sum_{i_{r}\in\mathbb{Z}}U_{\varphi,\eta}(\theta(h)^{*})V_{\xi}(\theta(h))}\;.
\end{equation}
As in the case of the trees, $W_{\varphi}^{\eta,\xi}(h)$ does not depend on the choice of the map $\theta$ if the function $\xi$ is even.
\end{subsection}

\begin{subsection}{Explicit formula for the cumulants of the current}
Equation (\ref{E(gamma) trees}) gives the generating function of the cumulants of the current $E(\gamma)$ in terms of a parameter $B$, which can be determined from the equation (\ref{gamma(B) trees}) giving $\gamma$ in terms of $B$. This parametric representation of $E(\gamma)$ does not give a direct access to the cumulants of the current. To obtain them, the parameter $B$ must be eliminated between equations (\ref{E(gamma) trees}) and (\ref{gamma(B) trees}). This elimination can be done perturbatively in $\gamma$. At the first orders in $\gamma$, it gives explicit expressions for the first cumulants of the current.\\\indent
The use of the Lagrange inversion formula (see \textit{e.g.} \cite{W05.1} or \cite{FS09.1}) allows to go further, by inverting equation (\ref{gamma(B) trees}) explicitly to all orders in $\gamma$. This gives an expression of $B$ as a formal series in $\gamma$. Inserting this expression of $B$ in equation (\ref{E(gamma) trees}) finally gives $E(\gamma)$ as an explicit formal series in $\gamma$. The calculation is explained in appendix \ref{Appendix E(gamma) explicit Lagrange}, where we obtain the following result:
\begin{equation}
\label{E(gamma) forests}
\boxed{\boxed{\frac{E(\gamma)-J\gamma}{p}=\frac{1-x}{L-1}\sum_{r=2}^{\infty}\frac{\gamma^{r}}{r!}\left(-\frac{L}{2}\right)^{r-1}\!\!\sum_{h\in\mathcal{H}_{r-1}}\frac{W_{z\mapsto z^{2}}^{\eta,\xi_{\lambda}}(h)}{S_{f}(h)}}}\;.
\end{equation}
The functions $\eta$ and $\xi_{\lambda}$ were defined previously in equations (\ref{eta(z)}) and (\ref{xi(z)}). We observe that the sums over the trees in the parametric equations (\ref{E(gamma) trees}) and (\ref{gamma(B) trees}) for $E(\gamma)$ have became sums over forests. The forest symmetry factor $S_{f}(h)$ is defined by
\begin{equation}
\label{Sf(h)}
\boxed{\frac{1}{S_{f}(h)}=(-1)^{\overline{h}}\frac{(|h|-1)!}{P_{f}(h)}\prod_{g\in h}\frac{|g|}{S(g)}}\;.
\end{equation}
The tree symmetry factor $S(g)$ was defined in equation (\ref{S(g)}). The factor $P_{f}(h)$ is equal to the cardinal of the permutation group of the identical trees of $h$. If the forest $h$ is made of $m_{1}$ times a first tree, $m_{2}$ times a second tree, \ldots, then $P_{f}(h)$ is simply equal to the product of the factorials of the $m_{i}$.\\\indent
Up to order $5$ in $\gamma$, the expression (\ref{E(gamma) forests}) of $E(\gamma)$ gives (using the notation $W$ for $W_{z\mapsto z^{2}}^{\eta,\xi_{\lambda}}$)
\begin{align}
&\frac{(L-1)E(\gamma)}{p(1-x)}=n(L-n)\gamma+\frac{L\gamma^{2}}{4}W\!\left[\sGGa\right]-\frac{L^{2}\gamma^{3}}{72}\left(9W\!\left[\sGGGa\right]-W\!\left[\sGGGb\right]-9W\!\left[\sGGa,\sGGa\right]\right)\nonumber\\
&+\frac{L^{3}\gamma^{4}}{48}\left(W\!\left[\sGGGGb\right]+3W\!\left[\sGGGGa\right]-W\!\left[\sGGGGc\right]-9W\!\left[\sGGGa,\sGGa\right]+W\!\left[\sGGGb,\sGGa\right]\right.\nonumber\\
&\qquad\qquad\qquad\qquad\qquad\qquad\qquad\qquad\qquad\qquad\qquad\qquad\qquad\left.\begin{picture}(0,5)\end{picture}+5W\!\left[\sGGa,\sGGa,\sGGa\right]\right)\nonumber\\
&-\frac{L^{4}\gamma^{5}}{28800}\left(900W\!\left[\sGGGGGa\right]+900W\!\left[\sGGGGGb\right]+75W\!\left[\sGGGGGc\right]-300W\!\left[\sGGGGGd\right]\right.\nonumber\\
&\qquad\qquad\begin{picture}(0,7)\end{picture}-450W\!\left[\sGGGGGe\right]+27W\!\left[\sGGGGGf\right]-3600W\!\left[\sGGGGa,\sGGa\right]-1200W\!\left[\sGGGGb,\sGGa\right]\nonumber\\
&\qquad\qquad\begin{picture}(0,7)\end{picture}+1200W\!\left[\sGGGGc,\sGGa\right]-2025W\!\left[\sGGGa,\sGGGa\right]+450W\!\left[\sGGGa,\sGGGb\right]-25W\!\left[\sGGGb,\sGGGb\right]\nonumber\\
&\qquad\qquad\left.\begin{picture}(0,7)\end{picture}+8100W\!\left[\sGGGa,\sGGa,\sGGa\right]-900W\!\left[\sGGGb,\sGGa,\sGGa\right]-3150W\!\left[\sGGa,\sGGa,\sGGa,\sGGa\right]\right)\nonumber\\
&+\O{\gamma^{6}}\;.
\end{align}
We observe that this expression of $E(\gamma)$ allows to recover the known formulas (\ref{D(x)}) and (\ref{E3(x)}) of the diffusion constant and of the third cumulant of the current. We also recover the expression (\ref{E4(x)}) announced for the fourth cumulant $E_{4}(x)$.
\end{subsection}

\begin{subsection}{Thermodynamic limit with vanishing asymmetry}
\label{Section cumulants(Phi)}
For a system with particle density $\rho=n/L$ fixed, it is possible to take the limit of the expression (\ref{E(gamma) forests}) of $E(\gamma)$ if the asymmetry $1-x$ is of order $1/\sqrt{L}$. We set
\begin{equation}
\Phi=-\frac{\log x\sqrt{L\rho(1-\rho)}}{2}\;.
\end{equation}
If $1-x$ goes to zero when $L$ goes to infinity, we have
\begin{equation}
\label{1-x[Phi]}
1-x\sim\frac{2\Phi}{\sqrt{L\rho(1-\rho)}}\;.
\end{equation}
We want to replace $x$ by its expression in terms of $\Phi$ in the function $\xi_{\lambda}$ used in (\ref{E(gamma) forests}). From the definition of $\Phi$, we can write
\begin{equation}
\xi_{\lambda}(z)=\left|\begin{array}{lll}\lambda&&\text{if $z=0$}\\\frac{1}{\tanh\left(\frac{|z|\Phi}{\sqrt{L\rho(1-\rho)}}\right)}&&\text{if $z\neq 0$}\end{array}\right.\;.
\end{equation}
We also want to take the limit $L\to\infty$ in the function $\eta$ used in (\ref{E(gamma) forests}). From Stirling's formula, we have
\begin{equation}
\eta(z)\sim\left(\frac{1-\rho}{\rho}\right)^{z}e^{-\frac{z^{2}+(1-2\rho)z}{2\rho(1-\rho)L}}\;.
\end{equation}
From the two previous equations, using the definition (\ref{W(g) tree 2}) of $W_{\varphi}^{\eta,\xi}(g)$, the expression (\ref{E(gamma) forests}) of $E(\gamma)$ becomes in the limit $L\to\infty$
\begin{align}
&\frac{E(\gamma)-J\gamma}{p}\sim(1-x)\sum_{r=2}^{\infty}\frac{(-1)^{r-1}L^{r-2}\gamma^{r}}{2^{r-1}r!}\nonumber\\
&\qquad\qquad\qquad\times\sum_{h\in\mathcal{H}_{r-1}}\sum_{i_{1}\in\mathbb{Z}}\ldots\sum_{i_{r-1}\in\mathbb{Z}}\frac{U_{z\mapsto z^{2},z\mapsto e^{-\frac{z^{2}}{2\rho(1-\rho)L}}}(\theta(h)^{*})\times V_{z\mapsto\tanh\left(\frac{|z|\Phi}{\sqrt{L\rho(1-\rho)}}\right)}(\theta(h))}{S_{f}(h)}\;.
\end{align}
We set $u_{j}=i_{j}/\sqrt{L\rho(1-\rho)}$. When $L$ goes to $\infty$, the Riemann sums over the indices $i_{j}$ become integrals. We finally obtain
\begin{equation}
\label{E(gamma) forests integral}
\frac{E(\gamma)-J\gamma}{p}\sim(1-x)\sum_{r=2}^{\infty}h_{r}(\Phi)(\rho(1-\rho))^{(r+1)/2}L^{3(r-1)/2}\frac{\gamma^{r}}{r!}\;,
\end{equation}
with
\begin{equation}
\label{hr(Phi)}
h_{r}(\Phi)=\frac{(-1)^{r-1}}{2^{r-1}}\int_{-\infty}^{\infty}du_{1}\ldots\int_{-\infty}^{\infty}du_{r-1}\sum_{h\in\mathcal{H}_{r-1}}\frac{U_{z\mapsto z^{2},z\mapsto e^{-\frac{z^{2}}{2}}}(\theta(h)^{*})\times V_{z\mapsto\tanh(|z|\Phi)}(\theta(h))}{S_{f}(h)}\;.
\end{equation}
For $r=2$ and $r=3$, we recover the functions $h_{r}$ given in section \ref{Section 3 regimes} after a few simplifications. The exponential $e^{-z^{2}/2}$ in the definition of $h_{r}(\Phi)$ ensures the convergence at infinity of the integrals. The convergence at $u_{j}=0$ is more complicated to prove. We checked it explicitly for $r$ between $2$ and $5$. For general $r$, it should be related to the cancellation of the arbitrary constant $\lambda$ from the finite size expressions.\\\indent
We note that the expression (\ref{E(gamma) forests integral}) of the cumulants of the current is in fact also true if $1/\sqrt{L}\ll1-x$, as all the expressions remain bounded and nonzero in the limit $\Phi\to\infty$. It gives the cumulants of the current in the Kardar-Parisi-Zhang regime (see section \ref{Section 3 regimes}) as
\begin{equation}
\frac{E_{r}}{p}\sim(1-x)h_{r}(\infty)(\rho(1-\rho))^{(r+1)/2}L^{3(r-1)/2}\;.
\end{equation}
The case $1-x\ll1/\sqrt{L}$ corresponding to the limit $\Phi\to0$ is more complicated. From an exact calculation for $r=3$ \cite{P08.1}, and from a numerical evaluation of the integrals in $h_{r}(\Phi)$ for $r=4$, we conjecture that for $r\geq3$, the limit $\Phi\to0$ of $h_{r}(\Phi)$ gives
\begin{equation}
h_{r}(\Phi)\sim\frac{2^{r-1}B_{2r-2}}{(r-1)!}\Phi^{r-1}\;,
\end{equation}
where the $B_{k}$ are the Bernoulli numbers. This gives for the $r$-th cumulant of the current (replacing $\Phi$ by its expression in terms of $1-x$)
\begin{equation}
\label{cumulants regime I}
E_{r}\sim\frac{B_{2r-2}}{(r-1)!}(1-x)^{r}(\rho(1-\rho))^{r}L^{2r-2}\;.
\end{equation}
We observe that this expression matches exactly the one obtained by taking the limit $\nu\to\infty$ in the expression of the cumulant $E_{r}$ in the weakly asymmetric scaling $1-x\sim\nu/L$ \cite{PM09.1}. We conjecture that the expression (\ref{cumulants regime I}) gives the correct value of the $r$-th cumulant of the current in all the intermediate regime $1/L\ll1-x\ll1/\sqrt{L}$, which lies between the Edwards-Wilkinson regime $1-x\ll1/L$ and the Kardar-Parisi-Zhang regime $1/\sqrt{L}\ll1-x$. Proving that (\ref{cumulants regime I}) does hold in this whole regime would require the calculation of finite size corrections to the Riemann integrals. It was done in \cite{P08.1} in the case of the third cumulant.
\end{subsection}

\end{section}

\begin{section}{Conclusion}
The fluctuations of the stationary state current in the periodic asymmetric exclusion process have already been studied much in the past. The diffusion constant was first obtained in the totally asymmetric model \cite{DEM93.1}, and the result was then generalized to the case of the partially asymmetric model \cite{DM97.1,PM08.1}. Higher cumulants of the current have also been calculated, both in the totally asymmetric model \cite{DL98.1,DA99.1} and in the partially asymmetric model \cite{LK99.1,P08.1,PM09.1}.\\\indent
In this paper, we generalized these previous results for the cumulants of the current. Solving the functional Bethe equation of the model with an arbitrary asymmetry between the hopping rates of the particles, we obtained exact finite size expressions for all the cumulants of the current. These expressions are given in terms of sums over tree sets, and have a nice combinatorial structure. It would be interesting to obtain them in a more direct way, starting from a graph theoretic formulation of the problem such as the one used in \cite{ZS06.1,ZS07.1} to describe the stationary measure. In particular, having a combinatorial interpretation of the binomial coefficients, of the trees, and of the indices $i_{j}$ that appear in the expressions for the cumulants would be nice. Another interesting question is whether such combinatorial structures also appear in the case of the open ASEP and of the ASEP with several classes of particles.\\\indent
Taking the thermodynamic limit in our exact formulas for the cumulants of the current in various scalings for the asymmetry, we observe three different regimes for the fluctuations of the current. In the regime of weakest asymmetry, which contains the model with symmetric hopping rates, it is known that the dynamics of the system is described at large scales by the Edwards-Wilkinson equation. Similarly, in the regime of strongest asymmetry which contains the totally asymmetric model, it is also known that the system is described at large scales by the Kardar-Parisi-Zhang equation. These two regimes are separated by an intermediate regime. It would be interesting to know how the system can be described at large scales in this regime. It would also be nice to know whether the presence of this intermediate regime between the Edwards-Wilkinson and the Kardar-Parisi-Zhang regimes is a general feature or is particular to the periodic exclusion process.\\\indent
The fluctuations of the current have also been studied for the exclusion process on the infinite line $\mathbb{Z}$, using random matrix techniques. We emphasize that these results are not given by simply taking the infinite size limit in our expressions for the cumulants of the current, as the long time limit giving the stationary state does not commute with the large system size limit. Finding the crossover between the current fluctuations on the ring and on the infinite line is still an open question.\\\indent
We note that our calculation of the cumulants of the current involves finding the ``ground state'' of the non hermitian Hamiltonian of the XXZ spin chain with $\Delta\geq1$ and twisted boundary conditions. Our solution relies on a perturbative expansion of the twist parameter near a point where all the Bethe roots are equal. A generalization of this method to the calculation of other eigenvalues would be useful.

\subsection*{Acknowledgments}
It is a pleasure to thank Olivier Golinelli and Kirone Mallick for many useful discussions.
\end{section}

\appendix
\begin{section}{Proof of the functional equation ``beyond the equator''}
\label{Appendix proof EB[P,Q]}
In this appendix, we construct a polynomial $P$ of degree $L-n$ related by the functional equation (\ref{EB[P,Q]}) to the polynomial $Q$ of degree $n$ solution of the functional Bethe equation (\ref{EB[Q,R]}). This construction is a straightforward generalization to the case of nonzero twist of the construction in the article \cite{PS99.1} by Pronko and Stroganov.\\\indent
For a system with $n\leq L/2$, we first divide the functional Bethe equation (\ref{EB[Q,R]}) by $Q(t/x)Q(t)Q(xt)$. We obtain
\begin{equation}
\label{R/Q(t/x)/Q(xt)}
\frac{R(t)}{Q(t/x)Q(xt)}=\frac{e^{L\gamma}(1-t)^{L}}{Q(t/x)Q(t)}+\frac{x^{n}(1-xt)^{L}}{Q(t)Q(xt)}\;.
\end{equation}
If all the zeros of the polynomials $Q(t/x)$ and $Q(xt)$ in the variable $t$ are different, which is generically the case, the rational function $(1-t)^{L}/(Q(t/x)Q(t))$ can be written in the form
\begin{equation}
\label{1/QQ[U,V,W]}
\frac{(1-t)^{L}}{Q(t/x)Q(t)}=\frac{U(t)}{Q(t/x)}+\frac{V(t)}{Q(t)}+W(t)\;,
\end{equation}
where the polynomial $W$ is of degree $L-2n$, and the polynomials $U$ and $V$ are of degree at most $n-1$. In terms of the polynomials $U$, $V$ and $W$, the functional equation (\ref{R/Q(t/x)/Q(xt)}) reads
\begin{equation}
\frac{R(t)}{Q(t/x)Q(xt)}=\frac{e^{L\gamma}V(t)+x^{n}U(xt)}{Q(t)}+\frac{e^{L\gamma}U(t)}{Q(t/x)}+\frac{x^{n}V(xt)}{Q(xt)}+e^{L\gamma}W(t)+x^{n}W(xt)\;.
\end{equation}
If all the zeros of $Q(t)$ are different from the zeros of $Q(t/x)Q(xt)$, which is again generically the case, the lhs of the previous equation has no pole in the zeros of $Q(t)$, unlike the rhs. The previous equation then implies that the coefficients of these poles vanish. Since the polynomials $U$ and $V$ are of degree strictly lower than the degree of $Q$, we must necessarily have
\begin{equation}
V(t)=-x^{n}e^{-L\gamma}U(xt)\;.
\end{equation}
With the help of the previous equation between the polynomials $U$ and $V$, we can now eliminate $V$ in equation (\ref{1/QQ[U,V,W]}). We obtain
\begin{equation}
\label{1/QQ[U,W]}
\frac{(1-t)^{L}}{Q(t/x)Q(t)}=\frac{U(t)}{Q(t/x)}-\frac{x^{n}}{e^{L\gamma}}\frac{U(xt)}{Q(t)}+W(t)\;.
\end{equation}
It is now natural to write the polynomial $W$ in the form
\begin{equation}
W(t)=X(t)-\frac{x^{n}}{e^{L\gamma}}X(xt)\;,
\end{equation}
with a polynomial $X$ of degree $L-2n$. If we call $w_{k}$ the coefficient of $t^{k}$ in $W(t)$, then the polynomial $X$ is given by
\begin{equation}
X(t)=\sum_{k=0}^{L-2n}\frac{w_{k}t^{k}}{1-x^{n+k}e^{-L\gamma}}\;.
\end{equation}
In terms of the polynomial $X$, equation (\ref{1/QQ[U,W]}) becomes
\begin{equation}
\label{1/QQ[U,X]}
\frac{(1-t)^{L}}{Q(t/x)Q(t)}=\left(\frac{U(t)}{Q(t/x)}+X(t)\right)-\frac{x^{n}}{e^{L\gamma}}\left(\frac{U(xt)}{Q(t)}+X(xt)\right)\;.
\end{equation}
We now introduce the polynomial $P$, of degree $L-n$, defined by
\begin{equation}
P(t)=(1-x^{n}e^{-L\gamma})Q(0)(U(xt)+Q(t)X(xt))\;.
\end{equation}
In terms of the polynomial $P$, equation (\ref{1/QQ[U,X]}) finally becomes
\begin{equation}
(1-x^{n}e^{-L\gamma})Q(0)(1-t)^{L}=Q(t)P(t/x)-x^{n}e^{-L\gamma}Q(t/x)P(t)\;.
\end{equation}
We have recovered equation (\ref{EB[P,Q]}). In particular, in $t=0$, this equation implies
\begin{equation}
P(0)=1\;.
\end{equation}
\end{section}

\begin{section}{Proof of the expression of \texorpdfstring{$w(t)$}{w(t)} as a sum over rooted trees}
\label{Appendix w(t) sum rooted trees}
In this appendix, we prove the expression (\ref{wk[chi,S] rooted}) of $w(t)$ as a sum over rooted trees. We start from the functional equation (\ref{w[X[w]] expansion sinh}) and from the expansion in powers of $C$ of $w(t)$ (\ref{w[wk,C]}). The expansion of $e^{rX[w(t)]}$ in powers of $C$ gives
\begin{align}
e^{rX[w(t)]}&=\exp\left(r\sum_{k=1}^{\infty}X[w_{k}(t)]C^{k}\right)=\sum_{s=0}^{\infty}\frac{r^{s}}{s!}\left(\sum_{k=1}^{\infty}X[w_{k}(t)]C^{k}\right)^{s}\nonumber\\
&=\sum_{s=0}^{\infty}\frac{r^{s}}{s!}\sum_{k_{1}=1}^{\infty}\ldots\sum_{k_{s}=1}^{\infty}X[w_{k_{1}}(t)]\ldots X[w_{k_{s}}(t)]C^{k_{1}+\ldots+k_{s}}\nonumber\\
&=\sum_{s=0}^{\infty}\frac{r^{s}}{s!}\sum_{\substack{l_{1},l_{2},l_{3},\ldots\in\mathbb{N},\\l_{1}+l_{2}+l_{3}+\ldots=s}}\C{s}{l_{1},l_{2},l_{3},\ldots}\prod_{i=1}^{\infty}(X[w_{i}(t)]C^{i})^{l_{i}}\nonumber\\
&=\sum_{s=0}^{\infty}r^{s}\sum_{\substack{l_{1},l_{2},l_{3},\ldots\in\mathbb{N},\\l_{1}+l_{2}+l_{3}+\ldots=s}}\prod_{i=1}^{\infty}\frac{(X[w_{i}(t)]C^{i})^{l_{i}}}{l_{i}!}\;.
\end{align}
Between the second and the third line in the previous expression, we called $l_{i}$ the number of indices $k_{j}$ such that $k_{j}=i$. The multinomial coefficient $\C{s}{l_{1},l_{2},l_{3},\ldots}$ counts the number of $s$-tuples of indices $(k_{1},\ldots,k_{s})$ corresponding to the same sequence of indices $(l_{1},l_{2},l_{3},\ldots)$.\\\indent
The equation (\ref{w[X[w]] expansion sinh}) for $w(t)$ now reads
\begin{equation}
\label{w[X[w]] expansion exp}
w(t)=\sum_{\substack{r=1\\(r\;\text{odd})}}^{\infty}\sum_{s=0}^{\infty}\sum_{\substack{l_{1},l_{2},l_{3},\ldots\in\mathbb{N},\\l_{1}+l_{2}+l_{3}+\ldots=s}}\frac{(-1)^{\frac{r-1}{2}}(r!!)^{2}r^{s}}{r^{2}r!}C^{r}h(t)^{r}\prod_{i=1}^{\infty}\frac{(X[w_{i}(t)]C^{i})^{l_{i}}}{l_{i}!}\;.
\end{equation}
At order $k$ in $C$, this expression gives $w_{k}(t)$ in terms of the $w_{i}(t)$ for $i<k$. It means that the previous equation allows to solve recursively $w(t)$ in powers of $C$. At first order in $C$, we find $w(t)=h(t)C+\O{C^{2}}$. We associate to the term at first order in $C$ the tree $\rootedGa$ of $\mathcal{G}^{\circ}$. More precisely, defining $\chi(\rootedGa)=h(t)$ and $S_{\circ}(\rootedGa)=1$, we have
\begin{equation}
w_{1}(t)=\frac{\chi(\rootedGa)}{S_{\circ}(\rootedGa)}=\sum_{g\in\mathcal{G}_{1}^{\circ}}\frac{\chi(g)}{S_{\circ}(g)}\;.
\end{equation}
We will now show by recurrence on the integer $k$ that it is possible to define suitable $\chi(g)$ and $S_{\circ}(g)$ for all the trees in $\mathcal{G}_{k}^{\circ}$ so that
\begin{equation}
\label{wk[chi,S] rooted 2}
w_{k}(t)=\sum_{g\in\mathcal{G}_{k}^{\circ}}\frac{\chi(g)}{S_{\circ}(g)}\;.
\end{equation}
The factor $S_{\circ}(g)$ is a symmetry factor associated to the tree $g$, while the factor $\chi(g)$ contain all the dependency on the asymmetry $x$ of $w_{k}(t)$. If we assume that for any $i<k$, $w_{i}(t)$ is given by (\ref{wk[chi,S] rooted 2}), then
\begin{align}
\frac{X[w_{i}(t)]^{l_{i}}}{l_{i}!}&=\frac{1}{l_{i}!}\sum_{\substack{m_{1}^{(i)},\ldots,m_{|\mathcal{G}_{i}^{\circ}|}^{(i)}\in\mathbb{N}\\m_{1}^{(i)}+\ldots+m_{|\mathcal{G}_{i}^{\circ}|}^{(i)}=l_{i}}}\C{l_{i}}{m_{1}^{(i)},\ldots,m_{|\mathcal{G}_{i}^{\circ}|}^{(i)}}\prod_{j=1}^{|\mathcal{G}_{i}^{\circ}|}\frac{X[\chi(g_{j}^{(i)})]^{m_{j}^{(i)}}}{S_{\circ}(g_{j}^{(i)})^{m_{j}^{(i)}}}\nonumber\\
&=\sum_{\substack{m_{1}^{(i)},\ldots,m_{|\mathcal{G}_{i}^{\circ}|}^{(i)}\in\mathbb{N}\\m_{1}^{(i)}+\ldots+m_{|\mathcal{G}_{i}^{\circ}|}^{(i)}=l_{i}}}\prod_{j=1}^{|\mathcal{G}_{i}^{\circ}|}\frac{X[\chi(g_{j}^{(i)})]^{m_{j}^{(i)}}}{m_{j}^{(i)}!\;S_{\circ}(g_{j}^{(i)})^{m_{j}^{(i)}}}\;.
\end{align}
We called $g_{1}^{(i)},\ldots,g_{|\mathcal{G}_{i}^{\circ}|}^{(i)}$ the trees of the set $\mathcal{G}_{i}^{\circ}$. The multinomial coefficient $\C{l_{i}}{m_{1}^{(i)},\ldots,m_{|\mathcal{G}_{i}^{\circ}|}^{(i)}}$ counts the number of terms in the expansion of $X[w_{i}(t)]^{l_{i}}$ such that the trees $g_{j}^{(i)}$ appear $m_{j}^{(i)}$ times. We obtain from (\ref{w[X[w]] expansion exp})
\begin{align}
\label{wk[chi,S] induction}
w_{k}(t)=&\sum_{\substack{r=1\\(r\;\text{odd})}}^{\infty}\sum_{s=0}^{\infty}\sum_{\substack{l_{1},l_{2},l_{3},\ldots\in\mathbb{N},\\l_{1}+l_{2}+l_{3}+\ldots=s\\r+l_{1}+2l_{2}+3l_{3}+\ldots=k}}\sum_{\substack{m_{1}^{(1)},\ldots,m_{|\mathcal{G}_{1}^{\circ}|}^{(1)}\in\mathbb{N}\\m_{1}^{(2)},\ldots,m_{|\mathcal{G}_{2}^{\circ}|}^{(2)}\in\mathbb{N}\\\ldots\\m_{1}^{(1)}+\ldots+m_{|\mathcal{G}_{1}^{\circ}|}^{(1)}=l_{1}\\m_{1}^{(2)}+\ldots+m_{|\mathcal{G}_{2}^{\circ}|}^{(2)}=l_{2}\\\ldots}}\nonumber\\
&\left(\frac{(-1)^{\frac{r-1}{2}}(r!!)^{2}r^{s}}{r^{2}r!}\prod_{i=1}^{\infty}\prod_{j=1}^{|\mathcal{G}_{i}^{\circ}|}\frac{1}{m_{j}^{(i)}!\;S_{\circ}(g_{j}^{(i)})^{m_{j}^{(i)}}}\right)\left(h(t)^{r}\prod_{i=1}^{\infty}\prod_{j=1}^{|\mathcal{G}_{i}^{\circ}|}X[\chi(g_{j}^{(i)})]^{m_{j}^{(i)}}\right)\;.
\end{align}
We will now associate a tree $g\in\mathcal{G}_{k}$ to each value of the tuple $(r,s,l_{1},l_{2},l_{3},\ldots,m_{1}^{(1)},\ldots,m_{|\mathcal{G}_{1}^{\circ}|}^{(1)},m_{1}^{(2)},\ldots,m_{|\mathcal{G}_{2}^{\circ}|}^{(2)},\ldots)$ in the previous expression for $w_{k}(t)$. The (odd) integer $r$ will be the size of the composite root of $g$, that is, the number of elementary nodes contained in the same composite nodes as the elementary root (including the elementary root). The integer $s$ will be the number of composite nodes directly linked by an edge to the composite root of $g$. The tree $g$ is then made of a composite root of size $r$ to which the trees $g_{j}^{(i)}$ are attached $m_{j}^{(i)}$ times for all $i$ and $j$. With the notations of the previous equation, we observe that (\ref{wk[chi,S] rooted}) holds if we define $\chi(g)$ and $S_{\circ}(g)$ by the recursive formulas
\begin{equation}
\chi(g)=h(t)^{r}\prod_{\substack{\text{$g'$ subtree of $g$}\\\text{connected to the}\\\text{composite root}}}X[\chi(g')]\;,
\end{equation}
and
\begin{equation}
\frac{1}{S_{\circ}(g)}=\frac{(-1)^{\frac{r-1}{2}}(r!!)^{2}r^{s+1}}{r^{3}r!}\prod_{i=1}^{\infty}\prod_{j=1}^{|\mathcal{G}_{i}^{\circ}|}\frac{1}{m_{j}^{(i)}!\;S_{\circ}(g_{j}^{(i)})^{m_{j}^{(i)}}}\;.
\end{equation}
As we observed on the expansion (\ref{trees w[h,C] order 5}) of $w(t)$ up to order $5$ in $C$, $\chi(g)$ is obtained by labeling the composite nodes $c$ of $g$ by $h(t)^{|c|}$, by labeling the edges of $g$ by the $X$ operator, and by reading the tree from the composite root, applying $X$ on the term coming from the subtree attached to the corresponding edge, and multiplying the terms coming from different subtrees attached to the same composite node.\\\indent
The recursive expression for $S_{\circ}(g)$ can be written explicitly as
\begin{align}
\frac{1}{S_{\circ}(g)}=&\left(\prod_{c\in c(g)}\frac{(-1)^{\frac{|c|-1}{2}}(|c|!!)^{2}|c|^{1+\text{number of subtrees attached to $c$ in $g$}}}{|c|^{3}|c|!}\right)\nonumber\\
&\times\prod_{\substack{\text{$g_{1}$ distinct subtrees}\\\text{attached to the}\\\text{composite root of $g$}\\\text{(with multiplicity $m_{1}$)}}}\left[\frac{1}{m_{1}!}\times\prod_{\substack{\text{$g_{2}$ distinct subtrees}\\\text{attached to the}\\\text{composite root of $g_{1}$}\\\text{(with multiplicity $m_{2}$)}}}\left[\frac{1}{m_{2}!}\times\ldots\right]^{m_{1}}\right]\;.
\end{align}
We see that we can rewrite it as
\begin{equation}
\frac{1}{S_{\circ}(g)}=\left(\prod_{c\in c(g)}\frac{(-1)^{\frac{|c|-1}{2}}(|c|!!)^{2}|c|^{v_{c}}}{|c|^{3}|c|!}\times\left(\substack{\text{size of the}\\\text{composite root}\\\text{of $g$}}\right)\right)\times\frac{\left(\substack{\text{number of equivalent}\\\text{choices of the}\\\text{composite root of $g$}}\right)}{P(g)}\;,
\end{equation}
where $P(g)$ is the number of permutations of the composite nodes of $g$ leaving it invariant (except for the position of the root). The size of the composite root of $g$ comes from the fact that the number of neighbors of the composite root is equal to the number of the subtrees attached to it in $g$, whereas the number of neighbors of any other composite node is equal to $1$ plus the number of the subtrees attached to it in $g$. But, the number of equivalent choices of the composite root is equal to the number of equivalent choices of the elementary root of $g$ divided by the size of the composite root of $g$. Thus, we can write
\begin{equation}
\frac{1}{S_{\circ}(g)}=\left(\prod_{c\in c(g)}\frac{(-1)^{\frac{|c|-1}{2}}(|c|!!)^{2}|c|^{v_{c}}}{|c|^{3}|c|!}\right)\times\frac{\left(\substack{\text{number of equivalent}\\\text{choices of the}\\\text{elementary root of $g$}}\right)}{P(g)}\;.
\end{equation}
\end{section}

\begin{section}{Generating function of the trees}
\label{Appendix GF trees}
In section \ref{Section tree structures for w(t)}, it was shown that $w_{r}(t)$ is given by a sum over the trees of $\mathcal{G}_{r}$ with the symmetry factor $S$ defined in equation (\ref{S(g)}). In this appendix, we study another interesting example of such a sum over $\mathcal{G}_{r}$ with the symmetry factor $S$: the ``generating function of $\mathcal{G}_{r}$'', which will be noted $Z_{r}(z)$. It is used in section \ref{Section TASEP limit} to recover the known expression for the current fluctuations in the totally asymmetric limit. It is defined by
\begin{equation}
\label{GF trees def}
Z_{r}(z)=\sum_{g\in\mathcal{G}_{r}}\frac{z^{\operatorname{card}c(g)}}{S(g)}\;,
\end{equation}
where $\operatorname{card}c(g)$ is the number of composite nodes of $g$. We will show that this generating function is given by
\begin{equation}
\label{GF trees}
Z_{r}(z)=\frac{z}{r\times r!}\prod_{j=1}^{r-1}\left[r(z+1)-2j\right]\;.
\end{equation}
In order to prove this expression, we first replace the sum over unrooted trees by a sum over rooted trees, by doing the reverse of the reasoning we used to go from the expression (\ref{wk[chi,S] rooted}) to the expression (\ref{wk[chi,S] unrooted}) of $w_{k}(t)$. We obtain
\begin{equation}
Z_{r}(z)=\frac{1}{r}\sum_{g'\in\mathcal{G}_{r}^{\circ}}\frac{z^{\operatorname{card}c(g')}}{S_{\circ}(g')}\;.
\end{equation}
We now consider the generating function
\begin{equation}
U(C,z)=\sum_{r=1}^{\infty}rZ_{r}(z)C^{r}=\sum_{g'\in\mathcal{G}_{r}^{\circ}}\frac{z^{\operatorname{card}c(g')}C^{\operatorname{card}e(g')}}{S_{\circ}(g')}\;,
\end{equation}
where $\operatorname{card}e(g')$ is the number of elementary nodes of $g'$. From a calculation similar to the one performed in appendix \ref{Appendix w(t) sum rooted trees} to extract from equation (\ref{w[X[w]]}) the expression (\ref{wk[chi,S] rooted}) of $w(t)$, the generating function $U\equiv U(C,z)$ is the solution of
\begin{equation}
U=z\arcsinh(Ce^{U})\;.
\end{equation}
The parameter $C$ can be written in terms of $U$
\begin{equation}
C=e^{-U}\sinh\left(\frac{U}{z}\right)\;.
\end{equation}
The Lagrange inversion formula (see \textit{e.g.} \cite{W05.1,FS09.1}) then gives $U$ as
\begin{equation}
U(C,z)=\sum_{r=1}^{\infty}\frac{1}{r}\left[\frac{e^{ru}}{\sinh^{r}(u/z)}\right]_{(u^{-1})}C^{r}=\frac{1}{2i\pi}\sum_{r=1}^{\infty}\frac{C^{r}}{r}\oint_{\mathcal{C}_{0}}du\,\frac{2^{r}e^{ru(z+1)/z}}{(e^{2u/z}-1)^{r}}\;,
\end{equation}
where the contour $\mathcal{C}_{0}$ surrounds $0$ but not the other poles of $1/(e^{2u/z}-1)$. We perform the change of variables $w=e^{2u/z}$ in the contour integral. The contour of integration $\mathcal{C}_{1}$ now surrounds $1$. We find
\begin{equation}
\left[\frac{e^{ru}}{\sinh^{r}(u/z)}\right]_{(u^{-1})}=\frac{1}{2i\pi}\oint_{\mathcal{C}_{1}}\frac{z\,dw}{2w}\,\frac{2^{r}w^{\frac{r(z+1)}{2}}}{(w-1)^{r}}=2^{r-1}z\C{\frac{r(z+1)}{2}-1}{r-1}\;.
\end{equation}
We finally find
\begin{equation}
U(C,z)=z\sum_{r=1}^{\infty}\frac{C^{r}}{r!}\prod_{j=1}^{r-1}(r(z+1)-2j)\;.
\end{equation}
We recover the expression (\ref{GF trees}) for $Z_{r}(z)$. We checked this expression for $r$ between $1$ and $16$ by constructing $149388$ trees of $\mathcal{G}$ and by computing their symmetry factors.
\end{section}

\begin{section}{Proof of the equivalence between the two definitions of \texorpdfstring{$W_{\varphi}^{\eta,\xi}$}{W}}
\label{Appendix definitions W equivalent}
In this appendix, we show that both definitions (\ref{W(g) tree 1}) and (\ref{W(g) tree 2}) of $W_{\varphi}^{\eta,\xi}(g)$ are equivalent. This will be illustrated on several examples of trees  $g\in\mathcal{G}$.\\\indent
If $g$ is a tree from $\mathcal{G}_{r}$, $W_{\varphi}^{\eta,\xi}(g)$ can be expressed as a sum over all integer values (both negative and nonnegative) of the indices $i_{1}$, \ldots, $i_{r-1}$ of the product of the functions $U_{\varphi,\eta}$ and $V_{\xi}$ applied to the trees $\theta(g)$ and $\theta(g)^{*}$ elements of $\widetilde{\mathcal{G}}_{r}$. Taking $\theta(g)$ as the argument of $U_{\varphi,\eta}$ and $\theta(g)^{*}$ as the argument of $V_{\xi}$ corresponds to the definition (\ref{W(g) tree 1}) of $W_{\varphi}^{\eta,\xi}(g)$, while the opposite choice corresponds to the definition (\ref{W(g) tree 2}) of $W_{\varphi}^{\eta,\xi}(g)$. Starting from (\ref{W(g) tree 2}), changes of variables can be performed on the indices $i_{1}$, \ldots, $i_{r-1}$. In general, these changes of variables are such that $U_{\varphi,\eta}(\theta(g)^{*})$ is transformed into an expression which can not be expressed as $U_{\varphi,\eta}(g')$ for some tree $g'\in\widetilde{\mathcal{G}}_{r}$. The same is true for $V_{\xi}(\theta(g))$. However, there exists changes of variables corresponding to transforming the tree $\theta(g)^{*}$ into a tree $g_{1}\in\widetilde{\mathcal{G}}_{r}$ and the tree $\theta(g)$ into a tree $g_{1}'\in\widetilde{\mathcal{G}}_{r}$ as we will show in the following. We will see that it is in fact possible to find a sequence of such changes of variables of the indices $i_{j}$ corresponding to a sequence of doublets of trees $(g_{j},g_{j}')$ beginning with $(\theta(g)^{*},\theta(g))$ and ending with $(\theta(g),\theta(g)^{*})$:
\begin{align}
W_{\varphi}^{\eta,\xi}(g)&=\sum_{i_{1}\in\mathbb{Z}}\ldots\!\sum_{i_{r-1}\in\mathbb{Z}}U_{\varphi,\eta}(\theta(g)^{*})V_{\xi}(\theta(g))\qquad\qquad\text{(definition (\ref{W(g) tree 2}) of $W_{\varphi}^{\eta,\xi}(g)$)}\\
&=\sum_{i_{1}\in\mathbb{Z}}\ldots\!\sum_{i_{r-1}\in\mathbb{Z}}U_{\varphi,\eta}(g_{1})V_{\xi}(g_{1}')=\ldots=\sum_{i_{1}\in\mathbb{Z}}\ldots\!\sum_{i_{r-1}\in\mathbb{Z}}U_{\varphi,\eta}(g_{m})V_{\xi}(g_{m}')\\
&=\sum_{i_{1}\in\mathbb{Z}}\ldots\!\sum_{i_{r-1}\in\mathbb{Z}}U_{\varphi,\eta}(\theta(g))V_{\xi}(\theta(g)^{*})\qquad\qquad\text{(definition (\ref{W(g) tree 1}) of $W_{\varphi}^{\eta,\xi}(g)$)}\;.
\end{align}
This will prove the equivalence between the two definitions (\ref{W(g) tree 1}) and (\ref{W(g) tree 2}) of $W_{\varphi}^{\eta,\xi}(g)$.

\begin{subsection}{Changes of variables corresponding to changes of trees}
We consider the expression
\begin{equation}
\label{U(g1)V(g2) change variables}
\sum_{i_{1}\in\mathbb{Z}}\ldots\!\sum_{i_{r-1}\in\mathbb{Z}}U_{\varphi,\eta}(g_{1})V_{\xi}(g_{1}')\;,
\end{equation}
where $g_{1}$ and $g_{1}'$ are trees from $\widetilde{\mathcal{G}}_{r}$. The changes of variables $i_{j}\to-i_{j}$ simply correspond to transforming the trees $g_{1}$ and $g_{1}'$ into $g_{2}$ and $g_{2}'$ obtained from $g_{1}$ and $g_{1}'$ by reversing the direction of the edge labeled by $i_{j}$. On the contrary, other changes of variables of the $i_{j}$, only correspond to changing $U_{\varphi,\eta}(g_{1})$ into $U_{\varphi,\eta}(g_{2})$ in the previous expression, but do not necessarily correspond to transforming $V_{\xi}(g_{1}')$ into an expression which can be put in the form $V_{\xi}(g_{2}')$. Other changes of variables can correspond to changing $V_{\xi}(g_{1}')$ into $V_{\xi}(g_{2}')$ but do not transform $U_{\varphi,\eta}(g_{1})$ into an expression of the form $U_{\varphi,\eta}(g_{2})$. In this subsection, we will study separately the changes of variables changing $g_{1}$ to $g_{2}$ (while transforming $V_{\xi}(g_{1}')$ in an arbitrary way), and the ones changing $g_{1}'$ to $g_{2}'$ (while transforming $U_{\varphi,\eta}(g_{1})$ arbitrarily).\\\indent
We begin with the changes of variables modifying the tree $g_{1}$. If an elementary node $e$ of $g_{1}$ is surrounded by $k$ edges pointing to $e$ (labeled by $a_{1}$, \ldots, $a_{k}$) and by $l$ edges pointing away from $e$ (labeled by $b_{1}$, \ldots, $b_{l}$), then $e$ contributes to $U_{\varphi,\eta}(g_{1})$ only through $\ell(e)$, which is equal to the sum of the $a_{j}$ minus the sum of the $b_{j}$:
\begin{equation}
\begin{array}{ccc}
\begin{picture}(20,8)\arrow{(3,-6)}{(1,1)}{7}{4}\arrow{(0,1)}{(1,0)}{10}{6}\arrow{(3,8)}{(1,-1)}{7}{4}\node{(10,1)}\arrow{(10,1)}{(1,-1)}{7}{4}\arrow{(10,1)}{(1,0)}{10}{6}\arrow{(10,1)}{(1,1)}{7}{4}\put(9,-2){\small$e$}\put(6,6){\small$a_{1}$}\put(0.5,2){\small$\ldots$}\put(1,-4){\small$a_{k}$}\put(11,6){\small$b_{1}$}\put(16,2){\small$\ldots$}\put(16,-4){\small$b_{l}$}\end{picture}&\quad\Rightarrow\quad&\ell(e)=(a_{1}+\ldots+a_{k})-(b_{1}+\ldots+b_{l})\\
&&
\end{array}\;.
\end{equation}
Changing one of the $a_{i}$ into $a_{i}$ minus the sum of some of the $a_{j}$ (different from $a_{i}$) plus the sum of some of the $b_{j}$ is equivalent to removing these $a_{j}$ and these $b_{j}$ from $\ell(e)$ and to add them to $\ell(e')$, where $e'$ is the elementary node which is linked by the edge labeled $a_{i}$ to the elementary node $e$. Thus, it is equivalent to detach from $e$ the edges labeled by the relevant $a_{j}$ and $b_{j}$ and to attach them back to $e'$. This change of variables corresponds then to a transfer in the tree $g_{1}$ of some edges from $e$ to $e'$. This transfer is local in the sense that the nodes $e$ and $e'$ are neighbors. Similarly, the change of variables transforming one of the $b_{i}$ into $b_{i}$ minus the sum of some of the $b_{j}$ (different from $b_{i}$) plus the sum of some of the $a_{j}$ corresponds to a transfer of edges from $e$ to the node $e''$ such that the edge between $e$ and $e''$ is labeled $b_{i}$ in $g_{1}$.\\\indent
We now continue with the changes of variables modifying the tree $g_{1}'$. We call $a_{1}$, \ldots, $a_{k}$ the indices labeling the edges which point away from the elementary root of $g_{1}'$ and $b_{1}$, \ldots, $b_{l}$ the other indices, which point to the elementary root. Let $p$ be an edge (inner or outer) of $g_{1}'$ labeled by one of the $a_{i}$. The contribution of $a_{i}$ to $V_{\xi}(g)$ comes from all the $m(o)$ for $o$ an outer edge of $g_{1}'$ located between the composite root of $g_{1}'$ and the edge $p$ (including $m(p)$ if $p$ is an outer edge):
\begin{equation}
\begin{array}{ccc}
\begin{array}{c}
g_{1}'=\begin{picture}(15,14)\cadre{15}{14}\put(5.5,8.8){\circle{10.6}}\rnode{(5.5,12.5)}\arrow{(5.05279,11.60557)}{(-1,-2)}{3}{1.7}\arrow{(5.94721,11.60557)}{(1,-2)}{3}{1.7}\put(0.5,8.5){\small$a_{1}$}\put(7.8,8.5){\small$a_{2}$}\node{(2.5,6.5)}\node{(8.5,6.5)}\arrow{(2.5,6.5)}{(0,-1)}{6}{4}\arrow{(8.5,6.5)}{(0,-1)}{6}{4}\put(-0.8,2.5){\small$a_{3}$}\put(9.3,2.6){\small$a_{4}$}\node{(2.5,0.5)}\put(8.5,-3.2){\circle{10.6}}\node{(8.5,0.5)}\arrow{(5.5,-5.5)}{(+1,+2)}{3}{2}\arrow{(8.5,0.5)}{(1,-2)}{3}{2}\put(3.5,-3.3){\small$b_{1}$}\put(10.6,-3.3){\small$a_{5}$}\node{(5.5,-5.5)}\node{(11.5,-5.5)}\arrow{(5.5,-5.5)}{(0,-1)}{6}{4}\arrow{(11.5,-11.5)}{(0,+1)}{6}{4}\put(2.4,-9.5){\small$a_{6}$}\put(12.1,-9.6){\small$b_{2}$}\node{(5.5,-11.5)}\node{(11.5,-11.5)}\put(8.5,-4.6){\small$p$}\end{picture}\\\\\\
\end{array}
&\Rightarrow&
\begin{array}{c}
V_{\xi}(g_{1}')=\xi(a_{4}+a_{5}+a_{6}-b_{1}-b_{2})\\
\qquad\qquad\qquad\times\xi(a_{2}+a_{4}+a_{5}+a_{6}-b_{1}-b_{2})\\
\qquad\qquad\qquad\times\text{(terms that do not contain $a_{5}$)}
\end{array}
\end{array}\;.
\end{equation}
We consider the change of variables changing $a_{i}$ to $a_{i}$ minus the the sum of the $a_{j}$ (different from $a_{i}$) labeling the edges of the subtree of $g_{1}'$ beginning at the edge $p$ plus the sum of the $b_{j}$ labeling the edges of the same subtree. This change of variables consists in detaching from $g_{1}'$ all the subtrees beginning at the neighboring edges of $p$ farther from the elementary root than $p$, and attaching them to the elementary root of $g_{1}'$. Contrary to the changes of variables considered previously for $g_{1}$, this change of variables is non-local: we attach to the elementary root of $g_{1}'$ some subtrees initially arbitrarily far from it. Similarly, we can consider an edge $p$ of $g_{1}'$ labeled by one of the $b_{i}$. The corresponding change of variables consists in changing $b_{i}$ into $b_{i}$ minus the sum of the $b_{j}$ (different from $b_{i}$) labeling the edges of the subtree of $g_{1}'$ beginning with the edge $p$ plus the sum of the $a_{j}$ labeling the edges of the same subtree. The meaning of this change of variables for the tree $g_{1}'$ is the same as for the previous change of variables involving $b_{i}$ instead of $a_{i}$.\\\indent
Depending on the respective structure of the trees $g_{1}$ and $g_{1}'$, it is possible that the two sets of changes of variables that we introduced in the two previous paragraphs have a non-empty intersection. These changes of variables then preserve the structure of the expression (\ref{U(g1)V(g2) change variables}), modifying both $g_{1}$ and $g_{1}'$. These are the changes of variables that we will use in the following to show that both definitions of $W_{\varphi}^{\eta,\xi}$ are equivalent.
\end{subsection}

\begin{subsection}{Example 1: linear tree with composite nodes of size \texorpdfstring{$1$}{1}}
We begin by considering the case where $g$ is the linear tree $g=\sGGGGGGa$, and by making for $\theta(g)$ the following choice:
\begin{equation}
\begin{array}{ccccccc}
\theta(g)=&\sTildeGGGGGGa&&\text{and}&&\theta(g)^{*}=&\sTildeGGGGGGb\\
&&&&&&\\
&&&&&&\\
&&&&&&
\end{array}\;.
\end{equation}
The definition (\ref{W(g) tree 2}) of $W_{\varphi}^{\eta,\xi}(g)$ gives
\begin{align}
W_{\varphi}^{\eta,\xi}(g)&=\sum_{i_{1}\in\mathbb{Z}}\ldots\!\sum_{i_{5}\in\mathbb{Z}}U_{\varphi,\eta}\left(\sTildeGGGGGGb\right)\times V_{\xi}\left(\sTildeGGGGGGa\right)\\
&=\sum_{i_{1}\in\mathbb{Z}}\ldots\!\sum_{i_{5}\in\mathbb{Z}}\varphi\eta(i_{1},i_{2},i_{3},i_{4},i_{5},-i_{5}-i_{4}-i_{3}-i_{2}-i_{1})\nonumber\\
&\qquad\qquad\qquad\times\xi(i_{1},i_{1}+i_{2},i_{1}+i_{2}+i_{3},i_{1}+i_{2}+i_{3}+i_{4},i_{1}+i_{2}+i_{3}+i_{4}+i_{5})\nonumber\;,
\end{align}
where we used the notations
\begin{align}
\varphi\eta(a_{1},\ldots,a_{l})&\equiv(\varphi(a_{1})+\ldots+\varphi(a_{l}))\times\eta(a_{1})\times\ldots\times\eta(a_{l})\\
\xi(b_{1},\ldots,b_{m})&\equiv\xi(b_{1})\times\ldots\times\xi(b_{m})\;,
\end{align}
in order to lighten the expressions. In the previous expression of $W_{\varphi}^{\eta,\xi}(g)$, we will make the following sequence of changes of variables:
\begin{equation}
\begin{array}{ccc}
i_{5}\to -i_{1}-i_{2}-i_{3}-i_{4}+i_{5}\;, &\qquad& i_{4}\to -i_{1}-i_{2}-i_{3}+i_{4}\;,\\
&&\\
i_{3}\to -i_{1}-i_{2}+i_{3}\;,&\text{and}& i_{2}\to -i_{1}+i_{2}\;.
\end{array}\;
\end{equation}
The first change of variables $i_{5}\to -i_{1}-i_{2}-i_{3}-i_{4}+i_{5}$ gives
\begin{align}
W_{\varphi}^{\eta,\xi}(g)&=\sum_{i_{1}\in\mathbb{Z}}\ldots\!\sum_{i_{5}\in\mathbb{Z}}\varphi\eta(i_{1},i_{2},i_{3},i_{4},i_{5}-i_{4}-i_{3}-i_{2}-i_{1},-i_{5})\nonumber\\
&\qquad\qquad\qquad\qquad\qquad\qquad\qquad\qquad\times\xi(i_{1},i_{1}+i_{2},i_{1}+i_{2}+i_{3},i_{1}+i_{2}+i_{3}+i_{4},i_{5})\nonumber\\
&=\sum_{i_{1}\in\mathbb{Z}}\ldots\!\sum_{i_{5}\in\mathbb{Z}}U_{\varphi,\eta}\left(\sTildeGGGGGGaaaa\right)V_{\xi}\left(\sTildeGGGGGGbbbb\right)\;.
\end{align}
The second change of variables $i_{4}\to -i_{1}-i_{2}-i_{3}+i_{4}$ leads to
\begin{align}
W_{\varphi}^{\eta,\xi}(g)&=\sum_{i_{1}\in\mathbb{Z}}\ldots\!\sum_{i_{5}\in\mathbb{Z}}\varphi\eta(i_{1},i_{2},i_{3},i_{4}-i_{3}-i_{2}-i_{1},i_{5}-i_{4},-i_{5})\xi(i_{1},i_{1}+i_{2},i_{1}+i_{2}+i_{3},i_{4},i_{5})\nonumber\\
&=\sum_{i_{1}\in\mathbb{Z}}\ldots\!\sum_{i_{5}\in\mathbb{Z}}U_{\varphi,\eta}\left(\sTildeGGGGGGaaa\right)V_{\xi}\left(\sTildeGGGGGGbbb\right)\;.
\end{align}
By the third change of variables $i_{3}\to -i_{1}-i_{2}+i_{3}$, we obtain
\begin{align}
W_{\varphi}^{\eta,\xi}(g)&=\sum_{i_{1}\in\mathbb{Z}}\ldots\!\sum_{i_{5}\in\mathbb{Z}}\varphi\eta(i_{1},i_{2},i_{3}-i_{2}-i_{1},i_{4}-i_{3},i_{5}-i_{4},-i_{5})\times\xi(i_{1},i_{1}+i_{2},i_{3},i_{4},i_{5})\nonumber\\
&=\sum_{i_{1}\in\mathbb{Z}}\ldots\!\sum_{i_{5}\in\mathbb{Z}}U_{\varphi,\eta}\left(\sTildeGGGGGGaa\right)V_{\xi}\left(\sTildeGGGGGGbb\right)\;.
\end{align}
Finally, the fourth change of variables $i_{2}\to -i_{1}+i_{2}$ gives
\begin{align}
W_{\varphi}^{\eta,\xi}(g)&=\sum_{i_{1}\in\mathbb{Z}}\ldots\!\sum_{i_{5}\in\mathbb{Z}}\varphi\eta(i_{1},i_{2}-i_{1},i_{3}-i_{2},i_{4}-i_{3},i_{5}-i_{4},-i_{5})\times\xi(i_{1},i_{2},i_{3},i_{4},i_{5})\nonumber\\
&=\sum_{i_{1}\in\mathbb{Z}}\ldots\!\sum_{i_{5}\in\mathbb{Z}}U_{\varphi,\eta}\left(\sTildeGGGGGGa\right)V_{\xi}\left(\sTildeGGGGGGb\right)\;,
\end{align}
which is the expression corresponding to the definition (\ref{W(g) tree 1}) of $W_{\varphi}^{\eta,\xi}(g)$.\\\indent
We observe that at each step, we recover the ``upper part'' of the tree $g$ (starting at $i_{5}$) in the tree on the left, and the ``lower part'' of the tree $g$ (starting from $i_{1}$) in the tree on the right. At each step, we destroy the uppermost part of the tree on the right and we rebuild it in the tree on the left: the complete structure of $g$ is thus preserved between the two trees.\\\indent
The previous proof for the tree $g$ can be easily extended to a linear tree of arbitrary size. The introduction of composite nodes of size strictly larger than $1$ does not change much if $\theta(g)$ is chosen such that the tree structure of the elementary nodes is linear.
\end{subsection}

\begin{subsection}{Example 2: branched tree with composite nodes of size \texorpdfstring{$>1$}{>1}}
We now consider the example of the tree
\begin{align}
&g=\sGGGGGGGGGd\;,\\
&\nonumber
\end{align}
and we choose $\theta(g)$ such that
\begin{align}
&\theta(g)=\sTildeGGGGGGGGGy\qquad\text{and}\qquad \theta(g)^{*}=\sTildeGGGGGGGGGz\;.\\
&\nonumber\\
&\nonumber
\end{align}
Starting from
\begin{align}
&W_{\varphi}^{\eta,\xi}(g)=\sum_{i_{1}\in\mathbb{Z}}\ldots\!\sum_{i_{8}\in\mathbb{Z}}U_{\varphi,\eta}\left(\sTildeGGGGGGGGGz\right)V_{\xi}\left(\sTildeGGGGGGGGGy\right)\\
&=\varphi\eta(i_{1},i_{2},i_{3},i_{4},-i_{5},i_{6},i_{7},-i_{8},-i_{1}-i_{2}-i_{3}-i_{4}+i_{5}-i_{6}-i_{7}+i_{8})\nonumber\\
&\qquad\qquad\qquad\qquad\qquad\qquad\qquad\qquad\qquad\qquad\xi(i_{3},i_{7},-i_{8},i_{4}-i_{5}+i_{6}+i_{7}-i_{8})\;,\nonumber
\end{align}
we will step by step destroy the upper part of the tree on the right while rebuilding it in the tree on the left, similarly to what we did in the previous section for the case of the linear tree. We begin with the change of variables $i_{1}\to i_{1}-i_{3}$, $i_{2}\to i_{2}-i_{4}+i_{5}-i_{6}-i_{7}+i_{8}$. We obtain
\begin{align}
W_{\varphi}^{\eta,\xi}(g)&=\varphi\eta(i_{1}-i_{3},i_{2}-i_{4}+i_{5}-i_{6}-i_{7}+i_{8},i_{3},i_{4},-i_{5},i_{6},i_{7},-i_{8},-i_{1}-i_{2})\nonumber\\
&\qquad\qquad\qquad\qquad\qquad\qquad\qquad\qquad\qquad\qquad\times\xi(i_{3},i_{7},-i_{8},i_{4}-i_{5}+i_{6}+i_{7}-i_{8})\nonumber\\
&=\sum_{i_{1}\in\mathbb{Z}}\ldots\!\sum_{i_{8}\in\mathbb{Z}}U_{\varphi,\eta}\left(\sTildeGGGGGGGGGzz\right)V_{\xi}\left(\sTildeGGGGGGGGGyy\right)\;.
\end{align}
The change of variables $i_{4}\to i_{4}+i_{5}-i_{6}-i_{7}+i_{8}$ then leads to
\begin{align}
W_{\varphi}^{\eta,\xi}(g)&=\varphi\eta(i_{1}-i_{3},i_{2}-i_{4},i_{3},i_{4}+i_{5}-i_{6}-i_{7}+i_{8},-i_{5},i_{6},i_{7},-i_{8},-i_{1}-i_{2})\xi(i_{3},i_{7},-i_{8},i_{4})\nonumber\\
&=\sum_{i_{1}\in\mathbb{Z}}\ldots\!\sum_{i_{8}\in\mathbb{Z}}U_{\varphi,\eta}\left(\sTildeGGGGGGGGGzzz\right)V_{\xi}\left(\sTildeGGGGGGGGGyyy\right)\;.
\end{align}
Finally, we perform the change of variables $i_{5}\to i_{5}+i_{7}$, $i_{6}\to i_{6}+i_{8}$ and we obtain
\begin{align}
W_{\varphi}^{\eta,\xi}(g)&=\varphi\eta(i_{3},i_{1}-i_{3},-i_{1}-i_{2},i_{2}-i_{4},i_{4}+i_{5}-i_{6},-i_{5}-i_{7},i_{6}+i_{8},i_{7},-i_{8})\xi(i_{3},i_{4},i_{7},-i_{8})\nonumber\\
&=\sum_{i_{1}\in\mathbb{Z}}\ldots\!\sum_{i_{8}\in\mathbb{Z}}U_{\varphi,\eta}\left(\sTildeGGGGGGGGGy\right)V_{\xi}\left(\sTildeGGGGGGGGGz\right)\;.
\end{align}
We observe that for any tree, the procedure that we have presented to transform the expression (\ref{W(g) tree 2}) into the expression (\ref{W(g) tree 1}) of $W_{\varphi}^{\eta,\xi}(g)$ works, which proves the equivalence between these two definitions. We note that we do not need at this stage to have a function $\xi$ even. This is only necessary to ensure that $W_{\varphi}^{\eta,\xi}(g)$ does not depend of the map $\theta$ that has been chosen (\textit{c.f.} appendix \ref{Appendix proof W independent of theta}).
\end{subsection}

\end{section}

\begin{section}{Independence of \texorpdfstring{$W_{\varphi}^{\eta,\xi}$}{W} with respect to the choice of the map \texorpdfstring{$\theta$}{theta}}
\label{Appendix proof W independent of theta}
We prove in this appendix that the function $W_{\varphi}^{\eta,\xi}$ defined in (\ref{W(g) tree 1}) and (\ref{W(g) tree 2}) is independent of the choice of the map $\theta$ provided that the function $\xi$ is even. We recall that the map $\theta$ transforms the trees of $\mathcal{G}$ into trees of $\widetilde{\mathcal{G}}$. We must show that for any tree $g\in\mathcal{G}$, $W_{\varphi}^{\eta,\xi}(g)$ is independent of the choice of the labeling and of the directions of the edges of $\theta(g)$, of the choice of the elementary root in $\theta(g)$, and of the choice of the internal tree structure of the composite nodes of $\theta(g)$. We will see that each of these properties can be easily seen from either one of the two definition (\ref{W(g) tree 1}) and (\ref{W(g) tree 2}) of $W_{\varphi}^{\eta,\xi}(g)$, the equivalence of which is proved in appendix \ref{Appendix definitions W equivalent}.\\\indent
We begin with the independence of $W_{\varphi}^{\eta,\xi}(g)$ with respect of the choice of the labeling of $\theta(g)$. Permuting the indices which label the edges of $\theta(g)$ in (\ref{W(g) tree 1}) or (\ref{W(g) tree 2}) is equivalent to permuting the indices $i_{j}$ in $U_{\varphi,\eta}(\theta(g))V_{\xi}(\theta(g)^{*})$ or $U_{\varphi,\eta}(\theta(g)^{*})V_{\xi}(\theta(g))$. The summation over the $i_{j}$ implies then that $W_{\varphi}^{\eta,\xi}(g)$ stays unchanged by the permutation of the labels of the edges.\\\indent
We continue with the independence with respect to the choice of the directions of the edges. Reversing the directions of the edge labeled by $i_{j}$ is equivalent to replacing $i_{j}$ by $-i_{j}$ in $U_{\varphi,\eta}(\theta(g))V_{\xi}(\theta(g)^{*})$ or $U_{\varphi,\eta}(\theta(g)^{*})V_{\xi}(\theta(g))$. Again, the summation over the $i_{j}$ in the definition of $W_{\varphi}^{\eta,\xi}(g)$ ensures the independence with respect to the choice of the directions of the edges in $\theta(g)$.\\\indent
We now move on to the independence with respect to the position of the elementary root of $\theta(g)$. It is now useful to consider the definition (\ref{W(g) tree 1}) of $W_{\varphi}^{\eta,\xi}(g)$ instead of the other one. By definition, $U_{\varphi,\eta}(\theta(g))$ does not depend on the position of the elementary root in $\theta(g)$. On the contrary, $V_{\xi}(\theta(g)^{*})$ depends on it: depending on the position of the root of $\theta(g)$, the orientation of some of the edges of $\theta(g)^{*}$ can be modified. This corresponds to changing some of the $i_{j}$ which appear in $V_{\xi}(\theta(g)^{*})$ into $-i_{j}$. But this transformation of some $i_{j}$ into $-i_{j}$ only takes place in $V_{\xi}(\theta(g)^{*})$ and not in $U_{\varphi,\eta}(\theta(g))$. Thus, we can not use the summation over the $i_{j}$ to conclude. We need to use the following remark: the factor $V_{\xi}(\theta(g)^{*})$ depends on the indices $i_{j}$ only through $\xi(i_{j})$ (there is no linear combination of the indices $i_{j}$ when $V_{\xi}$ is applied on the tree $\theta(g)^{*}$). The transformations of $V_{\xi}(\theta(g)^{*})$ resulting from the modification of the position of the root of $\theta(g)$ are thus equivalent to modifying some factors $\xi(i_{j})$ into $\xi(-i_{j})$. It is thus crucial that the function $\xi$ is even so that the independence of $W_{\varphi}^{\eta,\xi}(g)$ with respect to the position of the root in $\theta(g)$ is verified.\\\indent
We still must show the independence of $W_{\varphi}^{\eta,\xi}(g)$ with respect to the choice of the internal tree structure of the composite nodes by the map $\theta$. We will use here the definition (\ref{W(g) tree 2}) of $W_{\varphi}^{\eta,\xi}(g)$. By definition, $V_{\xi}(\theta(g))$ does not depend on the internal tree structure on the composite nodes of $g$ (but depends however on the orientation of the edges in this internal tree structure). The tree $\theta(g)^{*}$ does not depend either on the internal tree structure of the composite nodes of $g$. We thus conclude that $W_{\varphi}^{\eta,\xi}$ does not depend on the choice of the composite nodes of $\theta(g)$.\\\indent
We have finally proved that $W_{\varphi}^{\eta,\xi}(g)$ does not depend on the choice of the map $\theta$. The fact that the function $\xi$ is even is only needed to prove the independence with respect to the choice of the root of $\theta(g)$.
\end{section}

\begin{section}{Proof of the parametric expression of the polynomial \texorpdfstring{$Q$}{Q}}
\label{Appendix expansion w(t) powers t}
In this appendix, we perform the expansion of the expression (\ref{wk[chi,S] unrooted}) of $w(t)$ in powers of $t$, which leads to the expression (\ref{A(t) trees}) of the polynomial $Q(t)$ in terms of the function $W_{\varphi}^{\eta,\xi}$ defined in (\ref{W(g) tree 1}) and (\ref{W(g) tree 2}).

\begin{subsection}{Explicit expression for \texorpdfstring{$w(t)$}{w(t)}}
We start with the expression (\ref{wk[chi,S] unrooted}) of $w(t)$:
\begin{equation}
w_{k}(t)=\sum_{g\in\mathcal{G}_{k}}\frac{1}{S(g)}\sum_{j=1}^{k}\chi(g_{j})\;,
\end{equation}
where the function $\chi$ is defined in equation (\ref{chi(g)}). We recall that the trees $g_{j}\in\mathcal{G}_{k}^{\circ}$ are the trees obtained from $g$ by choosing for elementary root the $j$-th elementary node of $g$ (for some ordering of the elementary nodes of $g$). We will write $w(t)$ in a more explicit form by expanding all the functions $h(t)=(1-t)^{L}/t^{n}$ in powers of $t$, except the one corresponding to the elementary root.\\\indent
We first label the $k$ elementary nodes of $g\in\mathcal{G}_{k}$ by $i_{1}$, \ldots, $i_{k}$ so that in the tree $g_{j}$, the index $i_{j}$ labels the elementary root. Then, for all $m$ different from $j$, we expand the $h(t)$ corresponding to the $m$-th elementary node as
\begin{equation}
h(t)=\sum_{i_{m}\in\mathbb{Z}}h_{i_{m}}t^{i_{m}}\;.
\end{equation}
For $g\in\mathcal{G}_{k}$, we have
\begin{equation}
\label{expansion chi(gj) t 1}
\sum_{j=1}^{k}\chi(g_{j})=h(t)\sum_{j=1}^{k}\sum_{i_{1},\ldots,i_{j-1},i_{j+1},\ldots,i_{k}\in\mathbb{Z}}h_{i_{1}}\ldots h_{i_{j-1}}h_{i_{j+1}}\ldots h_{i_{k}}t^{i_{1}+\ldots+i_{j-1}+i_{j+1}\ldots+i_{k}}V(g_{j})\;,
\end{equation}
where $V(g_{j})$, which corresponds to the action of the operators $X$ in $\chi(g_{j})$, is equal to a product of factors of the form $\xi_{\lambda}(z)$ with $z$ a sum of indices from $i_{1}$, \ldots, $i_{k}$ and
\begin{equation}
\xi_{\lambda}(z)=\left\{\begin{array}{cl}\lambda & \text{if $z=0$}\\\frac{1+x^{|z|}}{1-x^{|z|}} & \text{if $z\neq 0$}\end{array}\right.\;.
\end{equation}
More precisely, $V(g_{j})$ is equal to the product over all the composite nodes $c$ of $g_{j}$ (except the composite root) of $\xi_{\lambda}$ applied to the sum of the indices labeling the subtree of $g_{j}$ made of the composite nodes that can be attained from $c$ by moving away from the composite root. For example, for the labeled tree
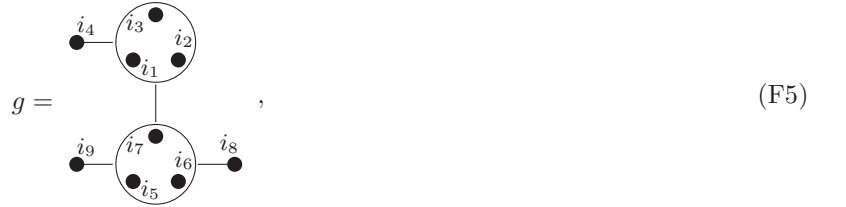
\begin{equation}
\begin{array}{cc}
\label{example g labeled}
g = & \begin{picture}(23,14)\cadre{23}{14}\put(11.5,8.8){\circle{10.6}}\node{(11.5,12.5)}\put(7.5,10.5){\small$i_{3}$}\node{(8.5,6.5)}\put(9.5,4.5){\small$i_{1}$}\node{(14.5,6.5)}\put(14,8.5){\small$i_{2}$}\edge{(5.8,8.8)}{(-1,0)}{5}\edge{(11.5,3.2)}{(0,-1)}{5}\node{(1,8.8)}\put(1,10.2){\small$i_{4}$}\put(11.5,-7.4){\circle{10.6}}\node{(11.5,-3.7)}\put(7.5,-5.7){\small$i_{7}$}\node{(8.5,-9.7)}\put(9.5,-11.7){\small$i_{5}$}\node{(14.5,-9.7)}\put(14,-7.7){\small$i_{6}$}\edge{(5.8,-7.4)}{(-1,0)}{5}\edge{(17.1,-7.4)}{(1,0)}{5}\node{(1,-7.4)}\put(1,-6){\small$i_{9}$}\node{(22,-7.4)}\put(20,-5.5){\small$i_{8}$}\end{picture}\\\\\\
\end{array}\;,
\end{equation}
we have
\begin{align}
&V(g_{1})=V(g_{2})=V(g_{3})=\xi_{\lambda}(i_{4})\xi_{\lambda}(i_{5}+i_{6}+i_{7}+i_{8}+i_{9})\xi_{\lambda}(i_{8})\xi_{\lambda}(i_{9})\\
&V(g_{4})=\xi_{\lambda}(i_{1}+i_{2}+i_{3}+i_{5}+i_{6}+i_{7}+i_{8}+i_{9})\xi_{\lambda}(i_{5}+i_{6}+i_{7}+i_{8}+i_{9})\xi_{\lambda}(i_{8})\xi_{\lambda}(i_{9})\\
&V(g_{5})=V(g_{6})=V(g_{7})=\xi_{\lambda}(i_{1}+i_{2}+i_{3}+i_{4})\xi_{\lambda}(i_{4})\xi_{\lambda}(i_{8})\xi_{\lambda}(i_{9})\\
&V(g_{8})=\xi_{\lambda}(i_{1}+i_{2}+i_{3}+i_{4}+i_{5}+i_{6}+i_{7}+i_{9})\xi_{\lambda}(i_{9})\xi_{\lambda}(i_{1}+i_{2}+i_{3}+i_{4})\xi_{\lambda}(i_{4})\\
&V(g_{9})=\xi_{\lambda}(i_{1}+i_{2}+i_{3}+i_{4}+i_{5}+i_{6}+i_{7}+i_{8})\xi_{\lambda}(i_{8})\xi_{\lambda}(i_{1}+i_{2}+i_{3}+i_{4})\xi_{\lambda}(i_{4})\;.
\end{align}
We now rename $i_{k}$ by $i_{j}$ in equation (\ref{expansion chi(gj) t 1}). We obtain
\begin{equation}
\sum_{j=1}^{k}\chi(g_{j})=h(t)\sum_{j=1}^{k}\sum_{i_{1},\ldots,i_{k-1}\in\mathbb{Z}}h_{i_{1}}\ldots h_{i_{k-1}}t^{i_{1}+\ldots+i_{k-1}}\left(V(g_{j})_{i_{k}\to i_{j}}\right)\;.
\end{equation}
We will now use the following property (which can be seen easily on the example of the labeled tree $g$ defined in (\ref{example g labeled}), using the fact that the function $\xi_{\lambda}$ is even):
\begin{equation}
\label{V(gj)<->V(gk)}
V(g_{j})_{i_{k}\to i_{j}}=V(g_{k})_{i_{j}\to-i_{1}-\ldots-i_{k-1}}\;.
\end{equation}
Then, for $j\neq k$, we perform the change of variables $i_{j}\to-i_{1}-\ldots-i_{k-1}$. We find
\begin{align}
\sum_{j=1}^{k}\chi(g_{j})=&h(t)\sum_{j=1}^{k-1}\sum_{i_{1},\ldots,i_{k-1}\in\mathbb{Z}}h_{i_{1}}\ldots h_{i_{j-1}}h_{-i_{1}-\ldots-i_{k-1}}h_{i_{j+1}}\ldots h_{i_{k-1}}t^{-i_{j}}V(g_{k})\nonumber\\
&+h(t)\sum_{i_{1},\ldots,i_{k-1}\in\mathbb{Z}}h_{i_{1}}\ldots h_{i_{k-1}}\frac{h_{-i_{1}-\ldots-i_{k-1}}}{h_{-i_{1}-\ldots-i_{k-1}}}t^{i_{1}+\ldots+i_{k-1}}V(g_{k})\;.
\end{align}
Factoring the product of the $h_{m}$, we can write
\begin{align}
\sum_{j=1}^{k}\chi(g_{j})=h(t)\sum_{i_{1},\ldots,i_{k-1}\in\mathbb{Z}}&\left(\frac{1}{h_{i_{1}}t^{i_{1}}}+\ldots+\frac{1}{h_{i_{k-1}}t^{i_{k-1}}}+\frac{1}{h_{-i_{1}-\ldots-i_{k-1}}t^{-i_{1}-\ldots-i_{k-1}}}\right)\nonumber\\
&\qquad\qquad\times h_{i_{1}}\ldots h_{i_{k-1}}h_{-i_{1}-\ldots-i_{k-1}}V(g_{k})\;.
\end{align}
Finally, we find for $w(t)$
\begin{align}
w_{k}(t)=\sum_{g\in\mathcal{G}_{k}}\frac{h(t)}{S(g)}\sum_{i_{1},\ldots,i_{k-1}\in\mathbb{Z}}&\left(\frac{1}{h_{i_{1}}t^{i_{1}}}+\ldots+\frac{1}{h_{i_{k-1}}t^{i_{k-1}}}+\frac{1}{h_{-i_{1}-\ldots-i_{k-1}}t^{-i_{1}-\ldots-i_{k-1}}}\right)\nonumber\\
&\qquad\qquad\times h_{i_{1}}\ldots h_{i_{k-1}}h_{-i_{1}-\ldots-i_{k-1}}V(g_{k})\;.
\end{align}
We emphasize that the choice of the particular rooted version of $g$ that we used here (\textit{i.e.} $g_{k}$) is completely arbitrary: any rooting of $g$ could have been chosen.
\end{subsection}

\begin{subsection}{Explicit expression for \texorpdfstring{$\alpha(t)$}{alpha(t)}}
In order to calculate $E(\gamma)$, we need to extract the negative powers in $t$ of $w(t)$ to find $\alpha(t)$. Thus, we only need to keep the negative powers of $h(t)/t^{z}=(1-t)^{L}/t^{n+z}$. We will use the notation $[\cdots]_{(-)}$ for the negative powers in $t$ of a sum of powers of $t$. We have
\begin{equation}
\left[\frac{h(t)}{t^{z}}\right]_{(-)}=\sum_{i=0}^{n+z-1}\C{L}{i}(-1)^{i}t^{i-n-z}\;.
\end{equation}
It is convenient for the calculation of the generating function of the cumulants of the current to write this as a series in $(1-t)$ instead of a series in $t$. We have
\begin{equation}
\left[\frac{h(t)}{t^{z}}\right]_{(-)}=\sum_{i=0}^{n+z-1}\sum_{l=0}^{\infty}\C{L}{i}\C{i-n-z}{l}(-1)^{i+l}(1-t)^{l}\;.
\end{equation}
The sum over $i$ can be performed in the previous expression by considering the term of order $0$ in $u$ in the product of the two formal series
\begin{align}
&\frac{(1-u)^{L}}{u^{n+z-1}}=\sum_{i=0}^{L}\C{L}{i}(-1)^{i}u^{i-n-z+1}\\
&\frac{1}{(1-u)^{l+1}}=\sum_{j=0}^{\infty}\C{-j-1}{l}(-1)^{l}u^{j}\;.
\end{align}
This gives the binomial sum
\begin{equation}
\sum_{i=0}^{n+z-1}\C{L}{i}\C{i-n-z}{l}(-1)^{i+l}=\C{L-l-1}{n+z-1}(-1)^{n+z-1}\;,
\end{equation}
and we find
\begin{equation}
\left[\frac{h(t)}{t^{z}}\right]_{(-)}=\sum_{l=0}^{\infty}\C{L-l-1}{n+z-1}(-1)^{n+z-1}(1-t)^{l}\;.
\end{equation}
Using this and
\begin{equation}
h_{z}=\C{L}{n+z}(-1)^{n+z}\;,
\end{equation}
we have finally found the following expression for $\alpha(t)$:
\begin{align}
\alpha(t)=2\sum_{k=0}^{\infty}\sum_{l=0}^{\infty}(-1)^{kn}C^{k}(1-t)^{l}\sum_{g\in\mathcal{G}_{k}}\frac{1}{S(g)}\sum_{i_{1},\ldots,i_{k-1}\in\mathbb{Z}}&\left(-\frac{\C{L-l-1}{n+i_{1}-1}}{\C{L}{n+i_{1}}}-\ldots-\frac{\C{L-l-1}{n+i_{k}-1}}{\C{L}{n+i_{k-1}}}-\frac{\C{L-l-1}{n-1-i_{1}-\ldots-i_{k-1}}}{\C{L}{n-i_{1}-\ldots-i_{k-1}}}\right)\nonumber\\
&\times\C{L}{n+i_{1}}\ldots\C{L}{n+i_{k-1}}\C{L}{n-i_{1}-\ldots-i_{k-1}}V(g_{k})\;.
\end{align}
The global factor $2$ comes from the factor $2$ in the definition (\ref{w[alpha,beta]}) of $w(t)$.\\\indent
We note that the previous equation can be expressed in terms of the tree functions defined in section \ref{Section functions trees}. Indeed, $V(g_{k})$ is equal to $V_{\xi_{\lambda}}(\theta(g))$ if the map $\theta$ from $\mathcal{G}$ to $\widetilde{\mathcal{G}}$ of section \ref{Section functions trees} is defined such that the choice of the elementary root in $\theta(g)$ is the same as in $g_{k}$, and if all the arrows are pointing away from the elementary root. With such a map $\theta$, the product of binomial coefficients in the previous expression is also equal to $U_{\varphi_{l},\eta}(\theta(g)^{*})$, the functions $\varphi_{l}$ and $\eta$ being defined in equations (\ref{phi(z)}) and (\ref{eta(z)}). Thus, we can write for $\alpha(t)$ the following expression
\begin{equation}
\alpha(t)=2\sum_{k=0}^{\infty}\sum_{l=0}^{\infty}(-1)^{kn}C^{k}(1-t)^{l}\sum_{g\in\mathcal{G}_{k}}\frac{1}{S(g)}\sum_{i_{1},\ldots,i_{k-1}\in\mathbb{Z}}U_{\varphi_{l},\eta}(\theta(g)^{*})V_{\xi_{\lambda}}(\theta(g))\;.
\end{equation}
Introducing the function $W_{\varphi}^{\eta,\xi}$ defined in (\ref{W(g) tree 2}) and using the expression (\ref{alpha[Q]}) of $\alpha(t)$ in terms of the polynomial $Q$, we finally obtain
\begin{equation}
\log\left(\frac{x^{n}Q(t/x)}{Q(t)}\right)=-2\sum_{k=1}^{\infty}\sum_{l=0}^{\infty}\left(\frac{B}{2}\right)^{k}(1-t)^{l}\sum_{g\in\mathcal{G}_{k}}\frac{W_{\varphi_{l}}^{\eta,\xi_{\lambda}}(g)}{S(g)}\;,
\end{equation}
where $B$ is given in terms of the parameter $C$ defined in (\ref{C[gamma,Q(0)]}) by
\begin{equation}
B=2(-1)^{n}\C{L}{n}C\;.
\end{equation}
\end{subsection}

\end{section}

\begin{section}{Proof of the explicit expression of the generating function \texorpdfstring{$E(\gamma)$}{E(gamma)}}
\label{Appendix E(gamma) explicit Lagrange}
In this appendix, we prove the explicit expression (\ref{E(gamma) forests}) for the generating function of the cumulants of the current $E(\gamma)$. We show that this expression, which involves a sum over forest sets, is a consequence of the parametric expression (\ref{gamma(B) trees}) and (\ref{E(gamma) trees}) of $E(\gamma)$ involving sums over tree sets.

\begin{subsection}{Elimination of \texorpdfstring{$B$}{B} in the parametric expression of \texorpdfstring{$E(\gamma)$}{E(gamma)}}
The equations (\ref{gamma(B) trees}) and (\ref{E(gamma) trees}) express respectively $\gamma$ and $E(\gamma)$ as formal series in a parameter $B$. These equations are of the form
\begin{equation}
\label{gamma(B) formal series}
\gamma=f(B)=\sum_{k=1}^{\infty}f_{k}B^{k}\;,
\end{equation}
and
\begin{equation}
\label{E(gamma) formal series}
\frac{E(\gamma)-J\gamma}{p}=g(B)=\sum_{k=2}^{\infty}g_{k}B^{k}\;.
\end{equation}
We want to invert the relation (\ref{gamma(B) formal series}) to obtain an expression for $B$ in terms of $\gamma$, and then insert this expression into (\ref{E(gamma) formal series}) to obtain an explicit expression in $\gamma$ of $E(\gamma)$. The inversion of the relation (\ref{gamma(B) formal series}) between $B$ and $\gamma$ can be performed using the Lagrange inversion formula (see \textit{e.g.} \cite{W05.1,FS09.1}). This formula implies that, for a formal series $g$ such as the one defined in (\ref{E(gamma) formal series}), we have
\begin{equation}
[g(B)]_{(\gamma^{r})}=\frac{1}{r}\left[\frac{g'(B)}{f(B)^{r}}\right]_{(B^{-1})}\;,
\end{equation}
where $[\cdots]_{(\gamma^{r})}$ and $[\cdots]_{(B^{-1})}$ stand respectively for the coefficient of the term in $\gamma^{r}$ and the one of the term in $B^{-1}$ of the expression $\cdots$. Using the expansion (\ref{E(gamma) formal series}) of $g(B)$ into powers of $B$, we obtain
\begin{equation}
[g(B)]_{(\gamma^{r})}=\sum_{k=2}^{\infty}\frac{k\times g_{k}}{r}\left[\frac{B^{k-1}}{f(B)^{r}}\right]_{(B^{-1})}=\sum_{k=2}^{\infty}\frac{k\times g_{k}}{r}\left[\frac{1}{f(B)^{r}}\right]_{(B^{-k})}\;.
\end{equation}
From (\ref{E(gamma) formal series}), we find for $E(\gamma)$
\begin{equation}
\frac{E(\gamma)-J\gamma}{p}=g(B)=\sum_{r=2}^{\infty}[g(B)]_{(\gamma^{r})}\gamma^{r}=\sum_{r=2}^{\infty}\sum_{k=2}^{\infty}\frac{k\times g_{k}}{r}\left[\frac{1}{f(B)^{r}}\right]_{(B^{-k})}\gamma^{r}\;.
\end{equation}
We need to expand $f(B)^{-r}$ into powers of $B$ for $r$ strictly positive integer. By the multinomial theorem (with a negative exponent), we have
\begin{align}
\frac{1}{f(B)^{r}}&=\frac{1}{(f_{1}B)^{r}\left(1+\frac{f_{2}}{f_{1}}B+\frac{f_{3}}{f_{1}}B^{2}+\ldots\right)^{r}}\\
&=(f_{1}B)^{-r}\sum_{b_{1},b_{2},\ldots\in\mathbb{N}}\C{r-1+b_{1}+b_{2}+\ldots}{r-1,b_{1},b_{2},\ldots}\prod_{j=1}^{\infty}\left(-\frac{f_{j+1}B^{j}}{f_{1}}\right)^{b_{j}}\;.\nonumber
\end{align}
The term in $B^{-k}$ of $f(B)^{-r}$ is then
\begin{equation}
\left[\frac{1}{f(B)^{r}}\right]_{(B^{-k})}=(f_{1})^{-r}\sum_{\substack{b_{1},b_{2},\ldots\in\mathbb{N}\\b_{1}+2b_{2}+3b_{3}+\ldots=r-k}}\frac{(r-1+b_{1}+b_{2}+\ldots)!}{(r-1)!}\prod_{j=1}^{\infty}\frac{1}{b_{j}!}\left(-\frac{f_{j+1}}{f_{1}}\right)^{b_{j}}\;,
\end{equation}
and we find for $E(\gamma)$ (shifting $k$ of $1$)
\begin{equation}
\frac{E(\gamma)-J\gamma}{p}=\sum_{r=2}^{\infty}\frac{1}{r!}\left(\frac{\gamma}{f_{1}}\right)^{r}\sum_{k=1}^{r-1}(k+1)g_{k+1}\!\!\!\!\sum_{\substack{b_{1},b_{2},\ldots\in\mathbb{N}\\b_{1}+2b_{2}+\ldots=r-k-1}}\!\!\!\!(r-1+b_{1}+b_{2}+\ldots)!\prod_{j=1}^{\infty}\frac{1}{b_{j}!}\left(-\frac{f_{j+1}}{f_{1}}\right)^{b_{j}}\;.
\end{equation}
We must now replace in the previous equation the coefficients $f_{j}$ and $g_{k+1}$ by their respective values given by (\ref{gamma(B) trees}) and (\ref{E(gamma) trees}). We have in particular $f_{1}=-1/L$. We finally find, putting together all the powers of $2$:
\begin{align}
\label{E(gamma) explicit trees}
\frac{E(\gamma)-J\gamma}{p}=&\frac{2(1-x)}{L(L-1)}\sum_{r=2}^{\infty}\left(-\frac{L\gamma}{2}\right)^{r}\sum_{k=1}^{r-1}\frac{k+1}{r!}\left(\sum_{g\in\mathcal{G}_{k+1}}\frac{W_{z\mapsto z^{2}}^{\eta,\xi_{\lambda}}(g)}{S(g)}\right)\\
&\qquad\qquad\times\sum_{\substack{b_{1},b_{2},\ldots\in\mathbb{N}\\b_{1}+2b_{2}+\ldots=r-k-1}}\frac{(r-1+b_{1}+b_{2}+\ldots)!}{\prod\limits_{j=1}^{\infty}(-1)^{b_{j}}b_{j}!}\prod_{j=1}^{\infty}\left(\sum_{g\in\mathcal{G}_{j+1}}\frac{W_{z\mapsto 1}^{\eta,\xi_{\lambda}}(g)}{S(g)}\right)^{b_{j}}\;.\nonumber
\end{align}
We have obtained an expression of $E(\gamma)$ in terms of tuples of trees. We now want to replace these tuples of trees by forests.
\end{subsection}

\begin{subsection}{Expression of \texorpdfstring{$E(\gamma)$}{E(gamma)} as a sum over forests}
Equation (\ref{E(gamma) explicit trees}) involves a sum over $b_{1}$ trees of size $2$, \ldots, $b_{k-1}$ trees of size $k$, $1+b_{k}$ trees of size $k+1$, $b_{k+1}$ trees of size $k+2$, \ldots\ We note $f$ the $m$-tuple (ordered set of $m$ trees) constituted by these trees, ordered by increasing size. We have $m=\sum_{j=1}^{\infty}(b_{j}+\delta_{j,k})$. In equation (\ref{E(gamma) explicit trees}), we thus sum over all the $m$-tuples of trees $f$ such that the $b_{1}$ first trees of $f$ are of size $2$, the $b_{2}$ following of size $3$, \ldots\ The sum over the $m$-tuples $f$ can be replaced by a sum over the forests $h$ containing $b_{1}$ trees of size $2$, \ldots, $b_{k-1}$ trees of size $k$, $1+b_{k}$ trees of size $k+1$, $b_{k+1}$ trees of size $k+2$, \ldots\ There, we must take into account the fact that several $m$-tuples of trees will correspond to the same forest: two $m$-tuples correspond to the same forest if and only if they are made of the same trees but not in the same order. We consider the action on the set of the $m$-tuples of trees $f$ of the subgroup of the group of permutations of $m$ elements keeping the size of the trees in increasing order in the $m$-tuples. The number of $m$-tuples of trees corresponding to the same forest as a given $m$-tuple $f$ is equal to the cardinal of the orbit of $f$ under this group action. Thus, we have
\begin{equation}
\operatorname{card}\operatorname{Orb}(f)=\frac{\prod\limits_{j=1}^{\infty}(b_{j}+\delta_{j,k})!}{P_{f}(h)}\;,
\end{equation}
where the product of the factorials of the $b_{j}+\delta_{j,k}$ is the cardinal of the group of the permutations leaving the sizes of the trees in increasing order in the $m$-tuple $f$, while $P_{f}(h)$, defined after equation (\ref{Sf(h)}), is equal to the cardinal of the stabilizer of $f$ under the group action. Calling $g_{0}$ one of the trees of size $k+1$, we obtain
\begin{align}
\frac{E(\gamma)-J\gamma}{p}=&\frac{2(1-x)}{L(L-1)}\sum_{r=2}^{\infty}\left(-\frac{L\gamma}{2}\right)^{r}\sum_{k=1}^{r-1}\frac{k+1}{r!}\left(\sum_{g\in\mathcal{G}_{k+1}}\frac{W_{z\mapsto z^{2}}^{\eta,\xi_{\lambda}}(g)}{S(g)}\right)\\
&\qquad\times\sum_{\substack{b_{1},b_{2},\ldots\in\mathbb{N}\\b_{1}+2b_{2}+\ldots=r-k-1}}\frac{(b_{k}+1)(r-1+b_{1}+b_{2}+\ldots)!}{\prod\limits_{j=1}^{\infty}(-1)^{b_{j}}}\nonumber\\
&\qquad\qquad\times\sum_{\substack{h\in\mathcal{H},\\\text{$\forall j$ $b_{j}+\delta_{j,k}$ trees}\\\text{of size $j+1$}}}\frac{1}{P_{f}(h)}\,\frac{W_{z\mapsto z^{2}}^{\eta,\xi_{\lambda}}(g_{0})}{S(g_{0})}\prod_{\substack{g\in h\\|g|=k+1\\\text{and $g\neq g_{0}$}}}\frac{W_{z\mapsto 1}^{\eta,\xi_{\lambda}}(g)}{S(g)}\prod_{\substack{g\in h\\|g|\neq k+1}}\frac{W_{z\mapsto 1}^{\eta,\xi_{\lambda}}(g)}{S(g)}\;.\nonumber
\end{align}
We now want to put $W_{z\mapsto z^{2}}^{\eta,\xi_{\lambda}}(g_{0})$ together with the product for $g$ of size $k+1$ ($g\neq g_{0}$) of the $W_{z\mapsto 1}^{\eta,\xi_{\lambda}}(g)$. We use the fact that, by the definition (\ref{W(h) forest 2}) of $W_{\varphi}^{\eta,\xi}$ acting on a forest, we have
\begin{align}
&\sum_{\substack{h=\{g_{0},\ldots,g_{b_{k}}\}\in\mathcal{H}\\|g_{0}|=\ldots=|g_{b_{k}}|=k+1}}\frac{W_{z\mapsto z^{2}}^{\eta,\xi_{\lambda}}(g_{0})}{S(g_{0})}\frac{W_{z\mapsto 1}^{\eta,\xi_{\lambda}}(g_{1})}{S(g_{1})}\ldots\frac{W_{z\mapsto 1}^{\eta,\xi_{\lambda}}(g_{b_{k}})}{S(g_{b_{k}})}\nonumber\\
&=\sum_{i_{1}^{(0)},\ldots,i_{k}^{(0)}\in\mathbb{Z}}\ldots\sum_{i_{1}^{(b_{k})},\ldots,i_{k}^{(b_{k})}\in\mathbb{Z}}\frac{(k+1)^{b_{k}}}{b_{k}+1}\left(\left(i_{1}^{(0)}\right)^{2}+\ldots+\left(i_{k}^{(0)}\right)^{2}+\left(-i_{1}^{(0)}-\ldots-i_{k}^{(0)}\right)^{2}+\ldots\right.\nonumber\\
&\qquad\qquad\qquad\qquad\qquad\qquad\qquad\qquad\qquad\left.+\left(i_{1}^{(b_{k})}\right)^{2}+\ldots+\left(i_{k}^{(b_{k})}\right)^{2}+\left(-i_{1}^{(b_{k})}-\ldots-i_{k}^{(b_{k})}\right)^{2}\right)\nonumber\\
&\qquad\quad\times\eta\left(i_{1}^{(0)}\right)\ldots\eta\left(i_{k}^{(0)}\right)\ldots\eta\left(i_{1}^{(b_{k})}\right)\ldots\eta\left(i_{k}^{(b_{k})}\right)\times\sum_{\substack{h=\{g_{0},\ldots,g_{b_{k}}\}\in\mathcal{H}\\|g_{0}|=\ldots=|g_{b_{k}}|}}\frac{V_{\xi_{\lambda}}(\theta(g_{0}))}{S(g_{0})}\ldots\frac{V_{\xi_{\lambda}}(\theta(g_{b_{k}}))}{S(g_{b_{k}})}\nonumber\\
&=\frac{(k+1)^{b_{k}}}{b_{k}+1}\frac{W_{z\mapsto z^{2}}^{\eta,\xi_{\lambda}}(\{g_{0},\ldots,g_{b_{k}}\})}{S(g_{0})\times\ldots\times S(g_{b_{k}})}\;,
\end{align}
with the notation $\{g_{0},\ldots,g_{b_{k}}\}$ for the forest made of the trees $g_{0}$, \ldots, $g_{b_{k}}$. The factor $(k+1)^{b_{k}}$ is the product of the factors $k+1$ of all the $W_{z\mapsto 1}^{\eta,\xi_{\lambda}}(g_{j})$, while the factor $1/(b_{k}+1)$ comes from the symmetrization of $z\mapsto z^{2}$ over all the trees of the forest $\{g_{0},\ldots,g_{b_{k}}\}$. We obtain
\begin{align}
\frac{E(\gamma)-J\gamma}{p}=&\frac{2(1-x)}{L(L-1)}\sum_{r=2}^{\infty}\frac{1}{r!}\left(-\frac{L\gamma}{2}\right)^{r}\sum_{k=1}^{r-1}\sum_{\substack{b_{1},b_{2},\ldots\in\mathbb{N}\\b_{1}+2b_{2}+\ldots=r-k-1}}\!\!\!\!\!\!\!\!\frac{(k+1)^{b_{k}+1}\times(r-1+b_{1}+b_{2}+\ldots)!}{\prod\limits_{j=1}^{\infty}(-1)^{b_{j}}}\nonumber\\
&\qquad\qquad\times\sum_{\substack{h\in\mathcal{H},\\\text{$\forall j$ $b_{j}+\delta_{j,k}$ trees}\\\text{of size $j+1$}}}\frac{W_{z\mapsto z^{2}}^{\eta,\xi_{\lambda}}(\{g\in h,|g|=k+1\})\times\prod\limits_{\substack{g\in h\\|g|\neq k+1}}W_{z\mapsto 1}^{\eta,\xi_{\lambda}}(g)}{P_{f}(h)\prod\limits_{g\in h}S(g)}\;.
\end{align}
We now change $b_{k}+1$ into $b_{k}$ so that for all $j$, we have exactly $b_{j}$ trees of size $j+1$ (even for $j=k$). Then, we can put the sum over $k$ inside the sum over the forests. We can write
\begin{align}
&\frac{E(\gamma)-J\gamma}{p}=\frac{2(1-x)}{L(L-1)}\sum_{r=2}^{\infty}\frac{1}{r!}\left(-\frac{L\gamma}{2}\right)^{r}\times\sum_{\substack{b_{1},b_{2},\ldots\in\mathbb{N}\\b_{1}+2b_{2}+\ldots=r-1}}\frac{(r-2+b_{1}+b_{2}+\ldots)!}{(-1)\times\prod\limits_{j=1}^{\infty}(-1)^{b_{j}}}\nonumber\\
&\qquad\times\sum_{\substack{h\in\mathcal{H},\\\text{$\forall j$ $b_{j}$ trees}\\\text{of size $j+1$}}}\frac{\sum\limits_{k=1}^{r-1}\openone_{b_{k}>0}\times(k+1)^{b_{k}}\times W_{z\mapsto z^{2}}^{\eta,\xi_{\lambda}}(\{g\in h,|g|=k+1\})\times\prod\limits_{\substack{g\in h\\|g|\neq k+1}}W_{z\mapsto 1}^{\eta,\xi_{\lambda}}(g)}{P_{f}(h)\prod\limits_{g\in h}S(g)}\;.
\end{align}
The number of trees in the forest $\{g\in h,|g|=k+1\}$ being equal to $b_{k}$, $W_{z\mapsto z^{2}}^{\eta,\xi_{\lambda}}(\{g\in h,|g|=k+1\})$ is equal to zero if $b_{k}=0$. The constraint $\openone_{b_{k}>0}$ can thus be forgotten. By definition of $W_{\varphi}^{\eta,\xi}$ for the trees and the forests, we have then
\begin{equation}
\sum_{k=1}^{r-1}(k+1)^{b_{k}}\times W_{z\mapsto z^{2}}^{\eta,\xi_{\lambda}}(\{g\in h,|g|=k+1\})\!\!\prod_{\substack{g\in h\\|g|\neq k+1}}\!\!W_{z\mapsto 1}^{\eta,\xi_{\lambda}}(g)=\left(\prod_{j=1}^{\infty}(j+1)^{b_{j}}\right)W_{z\mapsto z^{2}}^{\eta,\xi_{\lambda}}(\{g\in h\})\;.
\end{equation}
For $j\neq k$, the factor $(j+1)^{b_{j}}$ comes from the $W_{z\mapsto 1}^{\eta,\xi_{\lambda}}(g)$ for $g$ one of the $b_{j}$ trees of size $j+1$. We have then
\begin{align}
&\frac{E(\gamma)-J\gamma}{p}=\\
&\qquad-\frac{2(1-x)}{L(L-1)}\sum_{r=2}^{\infty}\frac{1}{r!}\left(-\frac{L\gamma}{2}\right)^{r}\sum_{\substack{b_{1},b_{2},\ldots\in\mathbb{N}\\b_{1}+2b_{2}+\ldots=r-1}}\sum_{\substack{h\in\mathcal{H},\\\text{$\forall j$ $b_{j}$ trees}\\\text{of size $j+1$}}}\frac{W_{z\mapsto z^{2}}^{\eta,\xi_{\lambda}}(h)}{\frac{P_{f}(h)}{(r-2+b_{1}+b_{2}+\ldots)!}\prod\limits_{j=1}^{\infty}\frac{(-1)^{b_{j}}}{(j+1)^{b_{j}}}\prod\limits_{g\in h}S(g)}\;.\nonumber
\end{align}
We observe that $r-1=b_{1}+2b_{2}+\ldots$ is equal to $|h|-\overline{h}$ and that $b_{1}+b_{2}+\ldots$ is equal to the number $\overline{h}$ of trees in $h$. It implies that $(r-2+b_{1}+b_{2}+\ldots)!$ is equal to $(|h|-1)!$. We also note that the $j+1$ are the sizes of the trees $g\in h$. We can write
\begin{equation}
\frac{P_{f}(h)}{(r-2+b_{1}+b_{2}+\ldots)!}\prod\limits_{j=1}^{\infty}\frac{(-1)^{b_{j}}}{(j+1)^{b_{j}}}\prod\limits_{g\in h}S(g)=\frac{P_{f}(h)(-1)^{\overline{h}}}{(|h|-1)!}\prod\limits_{g\in h}\frac{S(g)}{|g|}=S_{f}(h)\;.
\end{equation}
We have recovered the forest symmetry factor of $h$ defined in (\ref{Sf(h)}). We can remove the sum over the $b_{j}$ and sum directly over the forests $h\in\mathcal{H}_{r-1}$. We finally obtain the explicit expression (\ref{E(gamma) forests}) for $E(\gamma)$:
\begin{equation}
\frac{E(\gamma)-J\gamma}{p}=-\frac{2(1-x)}{L(L-1)}\sum_{r=2}^{\infty}\frac{(-1)^{r}L^{r}\gamma^{r}}{2^{r}r!}\sum_{h\in\mathcal{H}_{r-1}}\frac{W_{z\mapsto z^{2}}^{\eta,\xi_{\lambda}}(h)}{S_{f}(h)}\;.
\end{equation}
\end{subsection}

\end{section}

\end{document}